\documentclass[10pt, conference, compsocconf]{IEEEtran}
\ifCLASSINFOpdf
  \usepackage[pdftex]{graphicx}
\else
  \usepackage[dvips]{graphicx}
\fi
\usepackage[cmex10]{amsmath}
\usepackage{amssymb}

\usepackage{picinpar}

\usepackage[font=footnotesize,caption=false]{subfig}
\newtheorem{theorem}{Theorem}
\newtheorem{lemma}[theorem]{Lemma}

\newtheorem{example}{Example}

\newcommand{\singi}{singleton}

\newcommand{\singis}{singletons}

\newcommand{\singiadj}{singleton}

\newif\ifAppendix
\newif\ifDraft

\Appendixtrue

\Draftfalse

\newif\ifLong
\ifAppendix
\Longfalse
\else
\Longfalse
\fi

\ifAppendix
\newcommand{\RefAppendix}{the appendix}
\else
\newcommand{\RefAppendix}{the full version of this paper~\cite{extended}}
\fi

\ifDraft

\fi

\newcommand{\Naturals}{\mathbb{N}}

\newcommand{\Reals}{\mathbb{R}}
\newcommand{\RealsPos}{\mathbb{R}_{> 0}}
\newcommand{\RealsNonNeg}{\mathbb{R}_{\geq 0}}

\newcommand{\True}{\mathbf{true}}
\newcommand{\False}{\mathbf{false}}

\newcommand{\Tuple}[1]{{\langle{#1}\rangle}}

\newcommand{\Setdef}[2]{\left\{{#1}\mid{#2}\right\}}
\newcommand{\Set}[1]{\left\{{#1}\right\}}
\newcommand{\Def}{\mathrel{{:}{=}}}
\newcommand{\bnfor}{\mathrel{|}}

\newcommand{\ZIMITL}{\ensuremath{\textup{MITL}_{0,\infty}}}
\newcommand{\MITL}{\ensuremath{\textup{MITL}}}

\newcommand{\AP}{p}
\newcommand{\APs}{\mathit{AP}}

\newcommand{\Const}{n}

\newcommand{\Formula}{\phi}
\newcommand{\AnotherFormula}{\psi}
\newcommand{\LSF}{\alpha}
\newcommand{\RSF}{\beta}

\newcommand{\U}{\mathrel{\textup{\bf U}}}

\newcommand{\R}{\mathrel{\textup{\bf R}}}

\newcommand{\G}{\mathop{\textup{\bf G}}}

\newcommand{\AnyTemporalOp}{\mathrel{\textup{\bf {Op}}}}

\newcommand{\SU}{\mathrel{\textup{\bf U}^\textup{s}}}
\newcommand{\SUI}[1]{\mathrel{\textup{\bf U}^\textup{s}_{#1}}}
\newcommand{\SR}{\mathrel{\textup{\bf R}^\textup{s}}}
\newcommand{\SRI}[1]{\mathrel{\textup{\bf R}^\textup{s}_{#1}}}
\newcommand{\SF}{\mathop{\textup{\bf F}^\textup{s}}}
\newcommand{\SFI}[1]{\mathop{\textup{\bf F}^\textup{s}_{#1}}}
\newcommand{\SG}{\mathop{\textup{\bf G}^\textup{s}}}
\newcommand{\SGI}[1]{\mathop{\textup{\bf G}^\textup{s}_{#1}}}

\newcommand{\AnyIneq}{\mathbin{\bowtie}}
\newcommand{\AnyLower}{\mathbin{\triangleright}}
\newcommand{\AnyUpper}{\mathbin{\triangleleft}}
\newcommand{\Implies}{\Rightarrow}
\newcommand{\Iff}{\Leftrightarrow}
\newcommand{\Equiv}{\equiv}

\newcommand{\Strict}[1]{\bar{#1}}
\newcommand{\Index}{i}
\newcommand{\AnotherIndex}{j}
\newcommand{\AThirdIndex}{k}
\newcommand{\AFourthIndex}{m}

\newcommand{\Point}[2]{{(#1,#2)}}
\newcommand{\Earlier}{\prec}
\newcommand{\Timepoints}[1]{T(#1)}

\newcommand{\LaterPoints}[3]{T_+(#1,\Point{#2}{#3})}

\newcommand{\Clocks}{X}
\newcommand{\ClocksNext}{X'}
\newcommand{\Clock}{x}
\newcommand{\ClockNext}{x'}
\newcommand{\ClockValuation}{\nu}
\newcommand{\AnotherClockValuation}{\nu'}
\newcommand{\Valuation}{v}
\newcommand{\Diff}{\delta}
\newcommand{\Clockcons}[1]{\mathcal{C}(#1)}
\newcommand{\Acons}{C}
\newcommand{\GDef}{\mathrel{::=}}
\newcommand{\Valusat}{\models}

\newcommand{\NonClocks}{Z}
\newcommand{\NonClocksNext}{\Next{Z}}
\newcommand{\NonClock}{z}
\newcommand{\NonClockNext}{\Next{z}}

\newcommand{\Ta}{\Tuple{\Locs, \Init, \Clocks, \Edges, \Invs}}
\newcommand{\Ata}{\mathcal{A}}
\newcommand{\Locs}{L}
\newcommand{\Loc}{l}
\newcommand{\AnotherLoc}{l'}
\newcommand{\Init}{l_\textup{init}}
\newcommand{\Invs}{I}

\newcommand{\Edges}{E}
\newcommand{\Guard}{g}
\newcommand{\Resets}{R}

\newcommand{\TATrace}{\pi} 

\newcommand{\EStep}[1]{\xrightarrow{#1}}

\newcommand{\Bound}{k}
\newcommand{\Enc}[1]{{\left|\!\left[{#1}\right]\!\right|}}
\newcommand{\EncNext}[1]{{\left|\!\left[{#1}\right]\!\right|'}}

\newcommand{\At}{\mathit{at}}

\newcommand{\LeftOpen}{\mathit{lefto}}

\newcommand{\Open}{\mathit{open}}

\newcommand{\Reset}[1]{r_{#1}}

\newcommand{\AtStep}[2]{{#2^{[#1]}}}

\newcommand{\Next}[1]{#1'}

\newcommand{\AnotherSt}{t}

\newcommand{\Models}{\models}
\newcommand{\ModelsNot}{\not\models}

\newcommand{\Delay}{\delta}
\newcommand{\DelayNext}{\delta'}

\newcommand{\XLoc}[1]{l_{#1}}
\newcommand{\XClockA}{c_1}
\newcommand{\XClockB}{c_2}

\newcommand{\ClockMax}[1]{\mathrm{m}_{#1}}
\newcommand{\AnotherClock}{y}

\newcommand{\LB}{T}
\newcommand{\UB}{T'}
\newcommand{\Time}{t}

\newcommand{\ITr}{\sigma}
\newcommand{\IntervalI}[1]{I_{#1}}
\newcommand{\ValuI}[1]{v_{#1}}
\newcommand{\Offset}{t}
\newcommand{\AnotherOffset}{u}
\newcommand{\ThirdOffset}{v}
\newcommand{\FirstOffset}{t_1}
\newcommand{\SecondOffset}{t_2}
\newcommand{\ITrSuffix}[3]{{#1}^{(#2,#3)}}

\newcommand{\Left}{l}
\newcommand{\Right}{r}
\newcommand{\C}{c}
\newcommand{\D}{\delta}

\newcommand{\EncInit}{\mathcal{I}}
\newcommand{\EncTr}{\mathcal{T}}
\newcommand{\EncInv}{\mathcal{INV}}
\newcommand{\EncFair}{\mathcal{F}}

\newcommand{\EReset}{R_\C}
\newcommand{\DelayReset}{{D_\C}}
\newcommand{\Timing}{T_\C}
\newcommand{\Obligation}{\mathit{oblig}}

\newcommand{\RightOpen}{\mathit{righto}}

\newcommand{\TRight}{\hat\Right}

\newcommand{\IState}[2]{\ensuremath{\left\langle\!\begin{smallmatrix}#1\\#2\end{smallmatrix}\!\right\rangle}}
\newcommand{\IStateC}[2]{\IState{[#1,#1]}{#2}}
\newcommand{\IStateO}[3]{\ensuremath{\left\langle\!\begin{smallmatrix}(#1,#2)\\#3\end{smallmatrix}\!\right\rangle}}

\newcommand{\CInt}[1]{{\lfloor{#1}\rfloor}}
\newcommand{\CFrac}[1]{\mathop{\operatorname{fract}}(#1)}
\newcommand{\REquiv}{\approx}

\newcommand{\SameRegion}[1]{E_{#1,\Bound}}
\newcommand{\Loop}[1]{\mathit{loop}^{[#1]}}

\newcommand{\TSshort}{transition system}
\newcommand{\TSlong}{symbolic transition system with clock-like variables}
\newcommand{\TStitle}{Symbolic Transition Systems with Clock-like Variables}

\newcommand{\mitlopsec}[1]{\subsubsection{#1}}

\newcommand{\Val}{Val}
\newcommand{\Reg}{Reg}

\begin{document}

%
\title{Bounded Model Checking of an MITL Fragment for Timed Automata}


\author{\IEEEauthorblockN{Roland Kindermann, Tommi Junttila and Ilkka Niemel{\"a}}
\IEEEauthorblockA{Department of Information and Computer Science\\
Aalto University\\
P.O.Box 15400, FI-00076 Aalto, Finland\\
Email: \{Roland.Kindermann,Tommi.Junttila,Ilkka.Niemela\}@aalto.fi}
}


%


\maketitle

\begin{abstract}
Timed automata (TAs) are a common formalism for modeling timed
systems.
Bounded model checking (BMC) is a verification method that searches
for runs violating a property using a SAT or SMT solver.
MITL is a real-time extension of the linear time logic LTL. Originally,
MITL was defined for traces of non-overlapping time intervals rather than the
``super-dense'' time traces allowing for intervals overlapping in single points that are employed by the nowadays common semantics of timed
automata.
In this paper we extend the semantics of a fragment of MITL to super-dense time traces and devise a bounded model checking encoding for the fragment.
We prove correctness and completeness in the sense that using a sufficiently large bound a counter-example to any given non-holding property can be found.
We have implemented the proposed bounded model checking approach and experimentally studied the efficiency and scalability of the implementation. 
\end{abstract}


%
\IEEEpeerreviewmaketitle

\section{Introduction}

Fully-automated verification has many industrial applications.
A particularly interesting and challenging setting for the use of verification are systems for which timing aspects are of high importance like safety instrumented systems or communication protocols.
In this paper, we study verification in a setting where \emph{both} the \emph{system} and the \emph{specification} contain quantitative timing aspects, allowing not only to specify, e.g., that a certain situation will eventually lead to a reaction but also that the reaction will happen within a certain amount of time. Allowing such timing aspects to be part of both the specification and  the system adds an additional challenge.

Timed automata~\cite{AlurDill:TCS1994} are a widely employed formalism for the representation of finite state systems augmented with real-valued clocks.
Timed automata have been studied for two decades and various tools for the verification of timed automata exist. Most existing verification techniques and tools, like the model checker Uppaal~\cite{Uppaaltutorial:2004}, however do not support quantitative specifications on the timing of events. We feel that the ability to state, e.g., that a certain condition triggers a reaction within a certain amount of time provides a clear improvement over being able only to specify that a reaction will eventually occur.
For specifications, we use the  linear time logic \ZIMITL\ \cite{AlurFederHenzinger:JACM1996}, an extension adding lower and upper time bounds to the popular logic LTL.

Industrial size systems often have a huge discrete state space in addition to the infinite state space of timing-related parts of the system. We feel that fully symbolic verification is a key to tackling large discrete state spaces. We, thus, provide a translation of a pair of a timed automaton representing a system and a \ZIMITL\ formula into a symbolic \TSshort\ that can serve as a foundation for various symbolic verification methods. It is proven that the translated system has a trace if and only if the original timed automaton has a trace satisfying the formula.
We, furthermore, demonstrate how to employ the translation for SMT-based bounded model checking using the region-abstraction for timed automata \cite{AlurDill:TCS1994}. We show completeness of the approach and prove the applicability of the region abstraction to the \TSshort .
Finally, we evaluate the scalability of the approach and the cost for checking specifications containing timing experimentally.

\ZIMITL\ is a fragment of the logic MITL \cite{AlurFederHenzinger:JACM1996} for which the question whether or not a given timed automaton has a trace satisfying or violating a given formula is PSPACE complete~\cite{AlurFederHenzinger:JACM1996}. Previously, a verification approach for \ZIMITL\ specifications was introduced in \cite{AlurFederHenzinger:JACM1996} and improved upon in~\cite{DBLP:conf/formats/MalerNP06_long}. At this point, however, there are to our best knowledge no implementations or results of experiments using these methods available. Additionally, a major difference between the techniques described in \cite{AlurFederHenzinger:JACM1996,DBLP:conf/formats/MalerNP06_long} and our approach lies in the precise semantics of timed automata used. While previous approaches use dense-time semantics, we extend \ZIMITL\ to super-dense time. Although dense and super-dense time semantics of timed automata are often used interchangeably in the literature (and in fact do not differ in any important fashion when, e.g., verifying reachability constraints), we will show that equivalences between \ZIMITL\ formulas fundamental to the techniques in \cite{AlurFederHenzinger:JACM1996,DBLP:conf/formats/MalerNP06_long} do not hold anymore when using dense-time semantics.

\section{Timed Automata}

We first give basic definitions for timed automata
(see e.g.\ \cite{AlurDill:TCS1994,Alur:CAV1999,BengtssonYi:2003}).
For simplicity,
we use basic timed automata in the theoretical parts of the paper.
However,
in practice (and the experimental part of the paper)
one usually defines a network of timed automata
that can also have (shared and local) finite domain non-clock variables
manipulated on the edges.
The symbolic bounded model checking encodings presented later in the paper can
be extended to handle both of these features:
see, e.g.,~\cite{Sorea:ENTCS2002,DBLP:conf/forte/AudemardCKS02}
on how to handle synchronization in a network of timed automata.
%
Alternatively,
one can specify timed systems with a symbolic formalism~\cite{KindermannJunttilaNiemela:ACSD2011}.

Let $\Clocks$ be a set of real-valued \emph{clock variables}.
%
A
\emph{clock valuation} $\ClockValuation$ is a function
$\ClockValuation : \Clocks \to \RealsNonNeg$.
For $\Diff \in \RealsNonNeg$ we define the valuation
$\ClockValuation + \Diff$ by
$\forall \Clock \in \Clocks:
 (\ClockValuation+\Diff)(\Clock) = \ClockValuation(\Clock)+\Diff$.
The set of \emph{clock constraints} over $\Clocks$, $\Clockcons{\Clocks}$,
is defined by the grammar
$\Acons \GDef {\True \mid {\Clock \AnyIneq \Const} \mid {\Acons \land \Acons}}$
where $\Clock \in \Clocks$,
${\AnyIneq} \in \Set{<, \leq, =, \geq, >}$ and
$\Const \in \Naturals$.
A valuation $\ClockValuation$ satisfies $\Acons\in\Clockcons{\Clocks}$,
denoted by $\ClockValuation \Valusat \Acons$, if it evaluates $\Acons$ to true.

A \emph{timed automaton} (TA)
is a tuple $\Ta$ where
\begin{itemize}
\item
  $\Locs$ is a finite set of \emph{locations},
\item
  $\Init \in \Locs$ is the \emph{initial location} of the automaton,
\item
  $\Clocks$ is a finite set of real-valued \emph{clock variables},
\item
  $\Edges \subseteq {\Locs \times \Clockcons{\Clocks} \times 2^{\Clocks} \times \Locs}$ is a finite set of edges,
  each edge $\Tuple{l,g,R,l'}\in\Edges$ specifying a \emph{guard} $g$ and
  a set $R$ of \emph{clocks to be reset},
  and
\item
  $\Invs : \Locs \to \Clockcons{\Clocks}$
  assigns an \emph{invariant} to each location.
\end{itemize}

\iftrue
\begin{figure}[tb]
  \centering
  \includegraphics[scale=.55]{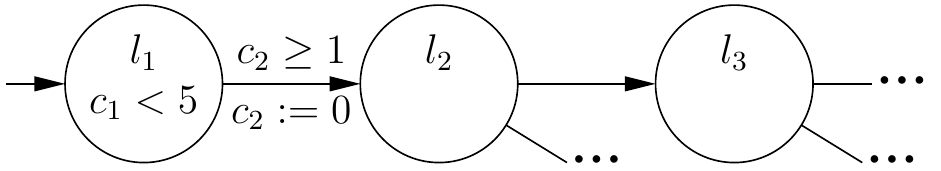}
  \caption{A timed automaton}
  \label{Fig:TA}
\end{figure}
\else
\begin{figwindow}[1,l,%
\makebox[55mm][c]{\includegraphics[width=50mm]{TA}},%
{A timed automaton
\label{Fig:TA}}
]
\fi
As an example,
Figure~\ref{Fig:TA} shows a part of a timed automaton with
locations $\XLoc{1}$, $\XLoc{2}, ...$, and
two clocks $\XClockA$ and $\XClockB$.
The initial location is $\XLoc{1}$, having the invariant $\XClockA < 5$.
The invariant of the location $\XLoc{2}$ is $\True$.
The edge from $\XLoc1$ to $\XLoc2$ has the guard $\XClockB \ge 1$
and the reset set $\Set{\XClockB}$.
The guard of the edge from $\XLoc2$ to $\XLoc3$ is $\True$ and
its reset set is empty.
\iftrue
\else
\end{figwindow}
\fi

A \emph{state} of a timed automaton $\Ata = \Ta$ is  a pair
$\Tuple{\Loc, \ClockValuation}$,
where $\Loc \in \Locs$ is a location
and
$\ClockValuation$ is a clock valuation over $\Clocks$.
A state $\Tuple{\Loc, \ClockValuation}$ is
(i)~\emph{initial} if $\Loc = \Init$ and
$\ClockValuation(\Clock) = 0$ for each $\Clock \in \Clocks$,
and
(ii)~\emph{valid} if $\ClockValuation \Valusat \Invs(\Loc)$.
Let $\Tuple{\Loc, \ClockValuation}$ and $\Tuple{\AnotherLoc, \AnotherClockValuation}$ be
states of $\Ata$.
There is a \emph{time elapse step} of $\Delay \in \RealsPos$ time units
from $\Tuple{\Loc, \ClockValuation}$ to $\Tuple{\AnotherLoc, \AnotherClockValuation}$,
denoted by $\Tuple{\Loc, \ClockValuation} \EStep{\Delay} \Tuple{\AnotherLoc, \AnotherClockValuation}$,
if
(i)~$\Loc = \AnotherLoc$,
(ii)~$\AnotherClockValuation = \ClockValuation+\Delay$,
and
(iii)~$\Tuple{\AnotherLoc, \AnotherClockValuation}$ is a valid state.
Intuitively,
there is a time elapse step from a state to another
if the second state can be reached from the first one by
letting $\Delay$ amount of time pass.
There is a \emph{discrete step} from  $\Tuple{\Loc, \ClockValuation}$ to
$\Tuple{\AnotherLoc, \AnotherClockValuation}$,
denoted by $\Tuple{\Loc, \ClockValuation} \EStep{0} \Tuple{\AnotherLoc, \AnotherClockValuation}$,
if there is an edge $\Tuple{\Loc, \Guard, \Resets, \AnotherLoc} \in \Edges$
such that
(i)~$\ClockValuation \Valusat \Guard$,
(ii)~$\Tuple{\AnotherLoc, \AnotherClockValuation}$ is a valid state,
and
(iii)~$\AnotherClockValuation(\Clock) = 0$ for all $\Clock \in \Resets$ and
$\AnotherClockValuation(\Clock) = \ClockValuation(\Clock)$ for all $\Clock \in {\Clocks \setminus \Resets}$.
That is, discrete steps can be used to change the current location 
as long as the guard and the target location invariant are satisfied.
A discrete step resets some clocks and leaves the other's values unchanged,
i.e., a discrete step does not take any time.
%

A \emph{run} of $\Ata$ is an infinite sequence of states 
$\TATrace=\Tuple{\Loc_0, \ClockValuation_0} \EStep{\Delay_{0}} \Tuple{\Loc_1, \ClockValuation_1}  \EStep{\Delay_{1}} \ldots$,
such that
(i) $\Tuple{\Loc_0, \ClockValuation_0}$ is valid and initial,
and
(ii) $\Tuple{\Loc_\Index, \ClockValuation_\Index} \EStep{\Delay_{\Index}} \Tuple{\Loc_{\Index+1}, \ClockValuation_{\Index+1}}$ with some $\Delay_{\Index} \in \Reals$
for each consecutive pair of states.
%
%
E.g.,
the automaton in Figure~\ref{Fig:TA}
has a run
$\Tuple{\XLoc1,(0,0)} \EStep{3.5} \Tuple{\XLoc1,(3.5,3.5)} \EStep{0} \Tuple{\XLoc2,(3.5,0.0)} \EStep{0} \Tuple{\XLoc3,(3.5,0.0)} \EStep{1.1} \Tuple{\XLoc3,(4.6,1.1)} \ldots$
where each clock valuation $\Set{\XClockA \mapsto v, \XClockB \mapsto w}$ is
abbreviated with $(v,w)$.
%
A run is
\emph{non-zeno} if the total amount $\sum_{i=0}^\infty \Delay_{i}$ of time passed in the run is infinite.
In the rest of the paper, we will only consider non-zeno runs.

Observe that on timed automata runs,
the automaton can visit multiple locations without time elapsing in between.
For instance,
at the time point 3.5 in the run given above, 
the automaton is after the first time elapse step in location $\XLoc1$,
then after the first discrete step in location $\XLoc2$,
and
finally after the second discrete step in location $\XLoc3$.
These kind of ``super-dense'' runs
differ from the dense runs that can be represented with ``signals'',
i.e.~by mapping each time point in $\RealsNonNeg$ to a single value.
%
%
As we will see in the next section, 
considering super-dense timed automata runs
complicates model checking as, e.g.,
we cannot get rid of the timed until operator
in the way we would if dense runs were used.

Note that previous papers on timed automata use both dense (e.g.~\cite{AlurDill:TCS1994}) and super-dense time (e.g.~\cite{Alur:CAV1999}), often without addressing the different semantics. 
From a practical perspective, super-dense runs appear paradox, as they permit multiple successive events to happen with no time passing in between. An alternative way of interpreting super-dense time, however, is that the amount of time in between events is just too small to be of interest and is, thus, abstracted away.
We also take the fact that Uppaal \cite{Uppaaltutorial:2004}, arguably the most successful timed model checker, not only allows for super-dense time traces but actually even makes it possible to \emph{enforce} super-dense behaviors by marking locations as ``urgent'' or ``committed'' as a strong indication that there is an interest in super-dense traces in practice.

\section{The Logic \ZIMITL\ for super-dense time}

Next,
 we describe the syntax and semantics of \ZIMITL{} formulas 
%
over ``super-dense timed traces''
which, as discussed in Sect.~\ref{Sect:TA2TT},
can represent timed automata runs.

%
%
\subsection{Syntax and Semantics}

Assuming a set $\APs$ of atomic propositions,
the syntax of \ZIMITL{} formula follows that in~\cite{AlurFederHenzinger:JACM1996},
and
is defined by the BNF grammar
$
  \Formula
  \GDef
  \AP \bnfor
  \Formula \bnfor
  \neg\Formula \bnfor
  \Formula \land \Formula \bnfor
  \Formula \lor  \Formula \bnfor
  \Formula \SUI{\AnyIneq\Const} \Formula
  \bnfor
  \Formula \SRI{\AnyIneq\Const} \Formula
$
where $\AP$ ranges over $\APs$,
$\Const$ ranges over $\Naturals$,
and
${\AnyIneq}$ ranges over $\Set{{<}, {\leq}, {\geq}, {>}}$.
Intuitively,
a strict timed until formula $\Formula \SUI{\AnyIneq\Const} \AnotherFormula$
states that
$\Formula$ holds in all later time points until
$\AnotherFormula$ holds at a time point $\Offset$ satisfying the timing constraint, i.e. $\Offset \AnyIneq \Const$.
Rational time constraints could be allowed in the temporal operators
without influencing the expressivity of the logic
(see \cite{AlurFederHenzinger:JACM1996} for \MITL{} on dense traces).
We define the
usual abbreviations:
$\True \Equiv (\AP \lor {\neg \AP})$,
$\False \Equiv {\neg\True}$,
${\SFI{\AnyIneq\Const}\Formula} \Equiv {\True \SUI{\AnyIneq\Const} \Formula}$,
and
${\SGI{\AnyIneq\Const}\Formula} \Equiv {\False \SRI{\AnyIneq\Const} \Formula}$.

We now define the semantics of \ZIMITL{} over ``super-dense'' timed traces,
and
then later show the correspondence of timed automata runs to such traces.
A \emph{super-dense timed trace} over a set of atomic propositions $\APs$
is an infinite sequence
$\ITr = \Tuple{\IntervalI{0},\ValuI{0}} \Tuple{\IntervalI{1},\ValuI{1}} \ldots$,
where
\begin{itemize}
\item
  each $\ValuI{\Index}$ is a subset of $\APs$,
\item
  each $\IntervalI{\Index}$ is either an open interval $(\LB_\Index,\UB_\Index)$
  or
  a \singi\ $[\LB_\Index,\LB_\Index]$
  with
  $0 \le \LB_\Index < \UB_\Index$
  and
  $\LB_\Index,\UB_\Index \in \RealsNonNeg$,
\item
  $\IntervalI{0} = [0,0]$,
\item
  for each $\Index \in \Naturals$ it holds that
  (i)
  $\IntervalI{\Index} = (\LB_\Index,\UB_\Index)$ implies
  $\IntervalI{\Index+1} = [\UB_\Index,\UB_\Index]$,
  and
  (ii)
  $\IntervalI{\Index} = [\LB_\Index,\LB_\Index]$
  implies either
  $\IntervalI{\Index+1} = [\LB_\Index,\LB_\Index]$ or
  $\IntervalI{\Index+1} = (\LB_\Index,\UB_{\Index+1})$;
  and
\item
  every 
  $\Time\in \RealsNonNeg$ is contained in at least one $\IntervalI{\Index}$.
\end{itemize}
For each trace element $\Tuple{\IntervalI{\Index},\ValuI{\Index}}$,
 equivalently written as $\IState{\IntervalI{\Index}}{\ValuI{\Index}}$,
the interpretation is that the atomic propositions in $\ValuI{\Index}$
hold in all the time points in the interval $\IntervalI{\Index}$.
As consecutive \singis\ are allowed,
it is possible for an atomic proposition to change its value
an
arbitrary finite number of times at a given time point.
This is required to capture timed automata traces containing two or more successive discrete steps and differentiates super-dense timed traces from dense ones.
In the semantics part we could have allowed
general intervals;
however, 
our constructions depend on discriminating the end points of left/right-closed intervals and thus we use this normal form already here.
A \emph{dense timed trace} is a super-dense timed trace with no consecutive
\singis\ (i.e., every time point $\Time\in \RealsNonNeg$ occurs
in exactly one $\IntervalI{\Index}$).

The set of all \emph{points} in a trace $\ITr$ is defined by
$\Timepoints{\ITr} = \Setdef{\Point{\Index}{\Offset}}{\Index\in\Naturals,\Offset\in\IntervalI{\Index}}$.
Two points,
$\Point{\Index}{\Offset},\Point{\Index'}{\Offset'}\in\Timepoints{\ITr}$,
are ordered with the ``earlier'' relation $\Earlier$ defined by
$
 \Point{\Index}{\Offset}\Earlier\Point{\Index'}{\Offset'}
 \Iff
 {{\Index < \Index'} \lor ({\Index = \Index'} \land {\Offset < \Offset'})}
$
and
the set of all points later than $\Point{\Index}{\Offset}$
is defined by
$
 \LaterPoints{\ITr}{\Index}{\Offset}
 \Def
 \Setdef{\Point{\Index'}{\Offset'}\in\Timepoints{\ITr}}
        {\Point{\Index}{\Offset}\Earlier\Point{\Index'}{\Offset'}}
$.

\newcommand{\ITrAt}[2]{\sigma^{\Point{#1}{#2}}}

Given a super-dense timed trace
$\ITr$ 
over $\APs$,
a formula $\Formula$ over $\APs$,
and
a point $\Point{\Index}{\Offset}$ in $\ITr$,
we define the satisfies relation $\ITr^\Point{\Index}{\Offset} \Models \Formula$
iteratively as follows:
\begin{itemize}
\item
  $\ITrAt{\Index}{\Offset} \Models \AP$
  iff $\AP \in \ValuI{\Index}$,
  where $\AP$ is an atomic proposition.
\item
  $\ITrAt{\Index}{\Offset} \Models {\neg \Formula}$
  iff
  $\ITrAt{\Index}{\Offset} \Models {\Formula}$ does not hold.
\item
  $\ITrAt{\Index}{\Offset} \Models (\Formula \land \AnotherFormula)$
  iff
  $\ITrAt{\Index}{\Offset} \Models \Formula$ and
  $\ITrAt{\Index}{\Offset} \Models \AnotherFormula$.
\item
  $\ITrAt{\Index}{\Offset} \Models (\Formula \lor \AnotherFormula)$
  iff
  $\ITrAt{\Index}{\Offset} \Models \Formula$ or
  $\ITrAt{\Index}{\Offset} \Models \AnotherFormula$.
\item
  $\ITrAt{\Index}{\Offset} \Models (\Formula \SUI{\AnyIneq\Const} \AnotherFormula)$
  iff
  $\exists \Point{\Index'}{\Offset'} \in \LaterPoints{\ITr}{\Index}{\Offset} :
   ({\Offset' - \Offset} \AnyIneq \Const) \land 
   (\ITrAt{\Index'}{\Offset'} \Models \AnotherFormula)
    \land
    \big(\forall \Point{\Index''}{\Offset''} \in \LaterPoints{\ITr}{\Index}{\Offset}:
         \Point{\Index''}{\Offset''} \Earlier \Point{\Index'}{\Offset'}
         \Implies
         (\ITrAt{\Index''}{\Offset''} \Models \Formula)\big)$
\item
  $\ITrAt{\Index}{\Offset} \Models (\Formula \SRI{\AnyIneq\Const} \AnotherFormula)$
  iff
  $\forall \Point{\Index'}{\Offset'} \in \LaterPoints{\ITr}{\Index}{\Offset}:
   \big(({\Offset'-\Offset} \AnyIneq \Const) \land 
        \neg(\ITrAt{\Index'}{\Offset'} \Models \AnotherFormula)\big)
   \Implies
   \big(\exists \Point{\Index''}{\Offset''} \in \LaterPoints{\ITr}{\Index}{\Offset}:
    \Point{\Index''}{\Offset''}\Earlier\Point{\Index'}{\Offset'}
    \land
    (\ITrAt{\Index''}{\Offset''} \Models \Formula)\big)$
\end{itemize}
For any formula $\Formula$,
we abbreviate $\ITrAt{0}{0} \Models \Formula$ with $\ITr \Models \Formula$.
\begin{example}\label{Ex:semantics}
Consider the super-dense timed trace
$\ITr =
  \IStateC{0}{\emptyset}
  \IStateO{0}{4}{\Set{p}}
  \IStateC{4}{\Set{p}}
  \IStateC{4}{\Set{q}}
  \IStateC{4}{\emptyset}
  \ldots$.
Now $\ITr \Models {p \SUI{\le 4} q}$ as
$\ITrAt{3}{4} \Models q$
and
$\ITrAt{\Index}{\Offset} \Models p$
for all $0 < \Index < 3$ and $0 \le \Offset \le 4$.
As an another example,
$\ITr \Models \SFI{\le 3}((\SGI{\le 1} p) \land (\SFI{< 2}q))$
also holds because
(i) $\ITrAt{1}{\Offset} \Models {\SGI{\le 1} p}$ for all $0 \le \Offset < 3$,
and
(ii) $\ITrAt{1}{\Offset} \Models {\SFI{< 2} q}$ for all $2 < \Offset < 4$.
\end{example}

As illustrated in Ex.~\ref{Ex:semantics},
neither $\Formula$ nor $\AnotherFormula$ need to hold in the current point
in order to satisfy $\Formula\SUI{\AnyIneq\Const}\AnotherFormula$.
Conversely,
$\Formula\SUI{\AnyUpper\Const}\AnotherFormula$ with $\AnyUpper\in\Set{{<},{\leq}}$ does not necessarily hold even if $\AnotherFormula$ holds in the first state:
e.g.,
$\IStateC{0}{\Set{q}} \IStateO{0}{3}{\emptyset} ...$
does not satisfy $p \SUI{<2} q$.
As \cite{AlurFederHenzinger:JACM1996} observes,
the reason for this slightly unintuitive semantics is that they allow
expressing formulas that would not be expressible if more intuitive semantics
where the current point in time is relevant for the timed until operator as well were used.
On the other hand,
expressing that $\Formula$ holds from the current point in time on until $\AnotherFormula$ holds can be done using the formula
$\AnotherFormula \lor {(\Formula \land (\Formula\SUI{\AnyIneq\Const}\AnotherFormula))}$.

We can define the ``untimed versions'' of the temporal operators with
${\SF \Formula} \Equiv {\SFI{\ge 0} \Formula}$,
${\SG \Formula} \Equiv {\SGI{\ge 0} \Formula}$,
${\Formula \SU \AnotherFormula} \Equiv {\Formula \SUI{\ge 0} \AnotherFormula}$,
and
${\Formula \SR \AnotherFormula} \Equiv {\Formula \SRI{\ge 0} \AnotherFormula}$.
An easily made misconception is that the time-aspect of a timed trace is
irrelevant when evaluating ``untimed'' operators, i.e.,
that they could be evaluated on $\omega$-words obtained 
when
removing intervals from a trace;
this is not the case.
In fact, even when not taking the ``only in the future'' part of the semantics,
illustrated in the previous example, into account,
considering the sets of propositions only is not sufficient.
%
As an example,
the formula $p \SU q$ is satisfied on
$
\IStateC{0}{\Set{p}}
\IStateO{0}{2}{\Set{p}}
\IStateC{2}{\Set{q}}
\ldots
$
but not on
$
\IStateC{0}{\Set{p}}
\IStateO{0}{2}{\Set{p}}
\IStateC{2}{\Set{p}}
\IStateO{2}{3.5}{\Set{q}}
\ldots
$.
The issue in the second trace is that
as the interval on which $q$ holds is an open one,
any point in it has a previous point at which only $q$, but not $p$, holds.
%
This illustrates that even for the ``untimed'' versions of the operators,
timing is relevant.

Observe that with super-dense timed traces we cannot get rid of
the timed until operator $\SUI{\AnyIneq\Const}$ by using the
``timed until is redundant'' theorem of~\cite{DBLP:conf/formats/MalerNP06_long},
vital for the transducer construction presented there.
That is,
$\Formula \SUI{\ge\Const} \AnotherFormula$
is \emph{not} equivalent to
$(\SGI{\le\Const}(\Formula \U \AnotherFormula)) \land {\SFI{\ge\Const}\AnotherFormula}$
in our setting.\footnote{Here, $\U$ is the non-strict until operator, i.e. $\Formula\U\AnotherFormula\Def\AnotherFormula\lor(\Formula\land(\Formula\SU\AnotherFormula))$}
For example,
in the trace
$\ITr =
  \IStateC{0}{\Set{p}}
  \IStateO{0}{2}{\Set{p}}
  \IStateC{2}{\Set{p}}
  \IStateC{2}{\Set{q}}
  \IStateC{2}{\emptyset}
  \ldots$
we have $\ITr \Models {p \SUI{\ge 2} q}$ but
$\ITr \ModelsNot (\SGI{\le 2}(p \U q)) \land {\SFI{\ge 2} q}$
as
$\ITrAt{4}{2} \ModelsNot {p \U q}$.
Likewise, the corresponding equivalences used in~\cite{AlurFederHenzinger:JACM1996} do not hold when using super-dense time, e.g. $p\SUI{\geq 2}q$ is \emph{not} equivalent to $\SGI{<2} p\land\SGI{\leq 2} (q \lor(p \land(p \SU p)))$ which can be demonstrated by the exact same trace.

Similarly,
it is not possible to
use the classic LTL equality
${\Formula \R \AnotherFormula} \Equiv
 (\G \AnotherFormula) \lor (\AnotherFormula \U (\Formula \land \AnotherFormula))$
to handle timed release operator by means of the other operators in our setting:
e.g.,
when
$\ITr =
  \IStateC{0}{\emptyset}
  \IStateO{0}{2}{\Set{\AnotherFormula}}
  \IStateC{2}{\Set{\AnotherFormula}}
  \IStateO{2}{4}{\Set{\Formula}}
  \ldots$
we have
$\ITr \Models {\Formula \SRI{\le 3} \AnotherFormula}$
but
$\ITr \ModelsNot \SGI{\le 3} \AnotherFormula$
and
$\ITr \ModelsNot \AnotherFormula \SUI{\le 3} (\Formula \land \AnotherFormula)$.

One can verify that the usual dualities hold
for the operators:
$\neg\neg\Formula \Equiv \Formula$,
$\neg(\Formula \lor \AnotherFormula) \Equiv
 {(\neg\Formula) \land (\neg\AnotherFormula)}$,
$\neg(\Formula \land \AnotherFormula) \Equiv
 {(\neg\Formula) \lor (\neg\AnotherFormula)}$,
${\neg(\Formula\SUI{\AnyIneq\Const}\AnotherFormula)} \Equiv
 {(\neg\Formula)\SRI{\AnyIneq\Const}(\neg\AnotherFormula)}$,
and
${\neg(\Formula\SRI{\AnyIneq\Const}\AnotherFormula)} \Equiv
  {(\neg\Formula)\SUI{\AnyIneq\Const}(\neg\AnotherFormula)}$.
These allow us to transform a formula into
\emph{positive normal form} in which negations only appear in front of
atomic propositions.
From now on, we assume that all formulas are in positive normal form.

%
%
\subsection{Trace Refinement and Fineness}

To perform model checking of \ZIMITL{} formulas, we do
not want the values of sub-formulas to change during open
intervals. We next formalize this and show how it can be
achieved by means of trace refinement; the definitions and
results here are extended from those in Sect.~2 of \cite{AlurFederHenzinger:JACM1996}.

\newcommand{\TrRef}{\preceq}

A trace $\ITr'$ is a \emph{refinement} of a trace $\ITr$,
denoted by $\ITr' \TrRef \ITr$,
if it can be obtained 
by replacing each open interval $\IStateO{T_i}{T'_i}{v_i}$ in the trace $\ITr$
with a sequence of intervals
$\IStateO{T_{i,0}}{T_{i,1}}{v_i}
 \IStateC{T_{i,1}}{v_i} \IStateO{T_{i,1}}{T_{i,2}}{v_i} \ldots
  \IStateO{T_{i,k-1}}{T_{i,k}}{v_i}$ 
of $2k-1$ consecutive, non-overlapping intervals with
$k \ge 1$, $T_{i,0} = T_i$. and $T_{i,k} = T'_i$.
Naturally,
if $\Formula$ is a \ZIMITL{} formula and $\ITr'$ is a refinement of $\ITr$,
then
$\ITr' \Models \Formula$ iff $\ITr \Models \Formula$.
%

Taking an arbitrary trace $\ITr$,
it may happen that the value of a compound 
sub-formula changes 
within
an open interval. 
To capture the desired case when this does not happen,
we
call 
$\ITr$ \emph{fine for a formula $\Formula$}
(or \emph{$\Formula$-fine})
if
for each sub-formula $\AnotherFormula$ of $\Formula$
(including $\Formula$ itself),
for each interval $\IntervalI{\Index}$ in $\ITr$,
and
for all $\Offset, \Offset' \in \IntervalI{\Index}$,
it holds that
$\ITrSuffix{\ITr}{\Index}{\Offset} \Models \AnotherFormula$ iff
$\ITrSuffix{\ITr}{\Index}{\Offset'} \Models \AnotherFormula$.
%

\begin{example}
The following super-dense timed trace
$\ITr =
  \IStateC{0}{\Set{p}}
  \IStateO{0}{4.1}{\Set{p}}
  \IStateC{4.1}{\Set{p}}
  \IStateC{4.1}{\Set{q}}
  \IStateC{4.1}{\emptyset}
  \ldots$
is not fine for $\SGI{\le 1} p$ as, e.g., 
(i) $\ITrSuffix{\ITr}{1}{t} \Models \SGI{\le 1} p$ for all $0 \le t < 3.1$
but
(ii) $\ITrSuffix{\ITr}{1}{t} \ModelsNot \SGI{\le 1} p$ for all $3.1 \le t < 4.1$.
We can make the beginning of the trace $\SGI{\le 1} p$-fine
by refining it to
$  \IStateC{0}{\Set{p}}
  \IStateO{0}{3.1}{\Set{p}}
  \IStateC{3.1}{\Set{p}}
  \IStateO{3.1}{4.1}{\Set{p}}
  \IStateC{4.1}{\Set{p}}
  \ldots$.
\end{example}

By definition,
every trace $\ITr$ is fine for each atomic proposition $\AP \in \APs$.
Furthermore, if $\ITr$ is $\Formula$-fine and $\AnotherFormula$-fine,
then it is also fine for
$\neg\Formula$,
$\Formula \land \AnotherFormula$,
and
$\Formula \lor \AnotherFormula$.
For temporal operators $\SUI{\AnyIneq\Const}$ and $\SRI{\AnyIneq\Const}$,
we have the following lemma stating that their values can change only once
during an open interval
given the trace is fine for the sub-formulas:
\newcommand{\SplitIntervalText }{
  If a trace $\ITr$ is fine for $\Formula$ and $\AnotherFormula$,
  $\Index \in \Naturals$,
  $\Offset,\AnotherOffset \in \IntervalI{\Index}$,
  ${\AnyUpper} \in \Set{<,\le}$, and
  ${\AnyLower} \in \Set{\ge,>}$,
  then
  \begin{itemize}
  \item
    if
    $\ITrSuffix{\ITr}{\Index}{\Offset} \Models {\Formula \SUI{\AnyUpper\Const} \AnotherFormula}$
    and
    $\AnotherOffset \ge \Offset$,
    then
    $\ITrSuffix{\ITr}{\Index}{\AnotherOffset} \Models {\Formula \SUI{\AnyUpper\Const} \AnotherFormula}$;
  \item
    if
    $\ITrSuffix{\ITr}{\Index}{\Offset} \Models {\Formula \SUI{\AnyLower\Const} \AnotherFormula}$
    and $\AnotherOffset \le \Offset$,
    then
    $\ITrSuffix{\ITr}{\Index}{\AnotherOffset} \Models {\Formula \SUI{\AnyLower\Const} \AnotherFormula}$;
  \item
    if
    $\ITrSuffix{\ITr}{\Index}{\Offset} \Models {\Formula \SRI{\AnyUpper\Const} \AnotherFormula}$
    and $\AnotherOffset \le \Offset$,
    then
    $\ITrSuffix{\ITr}{\Index}{\AnotherOffset} \Models {\Formula \SRI{\AnyUpper\Const} \AnotherFormula}$;
  \item
    if 
    $\ITrSuffix{\ITr}{\Index}{\Offset} \Models {\Formula \SRI{\AnyLower\Const} \AnotherFormula}$
    and $\AnotherOffset \ge \Offset$,
    then
    $\ITrSuffix{\ITr}{\Index}{\AnotherOffset} \Models {\Formula \SRI{\AnyLower\Const} \AnotherFormula}$.
  \end{itemize}}
\begin{lemma}\label{Lemma:SplitInterval}
\SplitIntervalText
\end{lemma}
Thus,
if $\ITr$ is fine for two formulas,
it can be made fine for their compound by
splitting each open interval at most once.

\newcommand{\ExistenceOfFineRefinementText}{
  Let $\Formula$ be a \ZIMITL{} formula and $\ITr$ a trace.
  There is a refinement $\ITr'$ of $\ITr$ that is $\Formula$-fine.
  Such a refinement can be obtained by splitting each open interval in
  $\ITr$ into at most $2^K$ new open intervals and $2^K-1$ \singis ,
  where $K$ is the number of timed until and release operators in $\Formula$.
}
\begin{lemma}\label{Lemma:ExistenceOfFineRefinement}
\ExistenceOfFineRefinementText
\end{lemma}

%
%
\subsection{Timed Automata Runs as Super-Dense Timed Traces}
\label{Sect:TA2TT}

We now describe the relationship between timed automata
runs and super-dense timed traces. In our theory part, when
model checking timed automata with \ZIMITL{}, we assume
that the atomic propositions only concern locations of the
automaton. That is, they are of form ``$@\XLoc{i}$'', where $\XLoc{i}$ is a
location in the automaton. Of course, in the practice when
compositions of timed automata with discrete local variables
are handled, the atomic propositions can be more complex.
However, we do assume that the atomic propositions do not
change their values during the time elapse steps.

Consider a run
$\TATrace =
 \Tuple{\Loc_0, \ClockValuation_0} \EStep{\Delay_{0}}
 \Tuple{\Loc_1, \ClockValuation_1} \EStep{\Delay_{1}} ...$
of a timed automaton $\Ata$.
For each 
$\Tuple{\Loc_i,\ClockValuation_i}$ in 
$\TATrace$
let $t_i = \sum_{j=0}^{i-1}\Delay_{j}$ be the cumulative time spent in
the run before the state,
i.e.\ $t_i$ is ``the time when the state occurs in 
$\TATrace$''.
Thus, at the time point $t_i$ the automaton is in the state $\Tuple{\Loc_i,\ClockValuation_i}$
and
we shall have $\IStateC{t_i}{\Set{@\Loc_i}}$ in the corresponding timed trace.
The time elapse steps in the run produce the missing open 
intervals:
when
$\Tuple{\Loc_i, \ClockValuation_i} \EStep{\Delay_{i}}
 \Tuple{\Loc_{i+1}, \ClockValuation_{i+1}}$
with $\Delay_{i} > 0$ (and thus $\Loc_i = \Loc_{i+1}$),
then 
an
open interval element
$\IStateO{t_i}{t_{i+1}}{\Set{@\Loc_i}}$
lies
in
between 
$\IStateC{t_i}{\Set{@\Loc_i}}$ and
$\IStateC{t_{i+1}}{\Set{@\Loc_i}}$
in the timed trace.

\begin{example}
The run
$\Tuple{\XLoc1,(0,0)} \EStep{3.5} \Tuple{\XLoc1,(3.5,3.5)} \EStep{0} \Tuple{\XLoc2,(3.5,0)} \EStep{0} \Tuple{\XLoc3,(3.5,0)} \EStep{1.1} \Tuple{\XLoc3,(4.6,1.1)} \ldots$
of the automaton in Figure~\ref{Fig:TA}
corresponds to the trace
{\small
$\ITr =
\IStateC{0}{\Set{@\XLoc1}}
\IStateO{0}{3.5}{\Set{@\XLoc1}}
\IStateC{3.5}{\Set{@\XLoc1}}
\IStateC{3.5}{\Set{@\XLoc2}}
\IStateC{3.5}{\Set{@\XLoc3}}
\IStateO{3.5}{4.6}{\Set{@\XLoc3}}
\ldots$
}
\end{example}

Recall that we will need to consider certain refinements of timed traces when
model checking with $\ZIMITL$ formulas.
All the refinements of a timed trace produced by a timed automata run can
be produced by other runs of the same automaton.
That is,
considering
 a trace coming from a run
$\TATrace =
 \Tuple{\Loc_0, \ClockValuation_0} \EStep{\Delay_{0}}
 \Tuple{\Loc_1, \ClockValuation_1} \EStep{\Delay_{1}} ...$
of a timed automaton,
each 
refinement
can be obtained by 
considering the corresponding run $\TATrace'$
where each time elapse step
$\Tuple{\Loc_\Index, \ClockValuation_\Index} \EStep{\Delay_\Index} \Tuple{\Loc_{\Index+1}, \ClockValuation_{\Index+1}}$ in $\TATrace$,
with $\Delay_\Index > 0$ and $\Loc_{\Index+1} = \Loc_{\Index}$,
is split into a sequence
$\Tuple{\Loc_\Index, \ClockValuation_\Index} \EStep{\Delay_{\Index,1}} \Tuple{\Loc_\Index, \ClockValuation_{\Index,1}} \EStep{\Delay_{\Index,2}}  ... \EStep{\Delay_{\Index,k}} \Tuple{\Loc_\Index, \ClockValuation_{\Index,k}}$
of time elapse steps
such that $\sum_{1 \le j \le k}\Delay_{\Index,j} = \Delay_\Index$
(and thus $\ClockValuation_{\Index,k} = \ClockValuation_{\Index+1}$).

\newcommand{\xplscale}{.62}

\section{Symbolic Encoding of Timed Traces}

\label{sect:sett}

We now describe how to \emph{symbolically} represent systems producing super-dense timed traces.
The symbolical representation intended not as a replacement for timed automata but 
as a foundation for their symbolic verification, i.e. it is intended for use in the ``back-end'' of the verification tool and not as a modeling language. 
After the formalism is introduced, it will be shown how timed automata can be represented in this framework.
The next section will then address the question of how to encode $\ZIMITL$ formulas in this framework so that they are symbolically evaluated.
Finally, in Sect.~\ref{Sect:BMC} it will be demonstrated how finite versions of
these encodings can be obtained by using region abstraction,
allowing us to perform actual symbolic model checking of $\ZIMITL$ formulas on timed automata.

\newcommand{\TraceOf}[1]{\mathop{\operatorname{trace}}(#1)}

\subsection{\TStitle}

\label{ss:ts}

In the following, we use standard concepts of propositional and first-order logics, and assume that the formulas are interpreted modulo some background theory such as linear arithmetics
(see e.g.~\cite{Handbook:SMT} and the references therein).
Given a set of typed variables,
a valuation $\Valuation$ over the set is a function that assigns each variable in the set a value in the domain of the variable.
We use $\Valuation \Models \phi$ to denote that $\Valuation$ evaluates
a quantifier-free formula $\phi$ over the set to true.

\newcommand{\STTS}{\mathcal{S}}
\newcommand{\STTSf}[1]{\mathcal{S}_{#1}}
\newcommand{\St}{s}
\newcommand{\Stp}{s'}
\newcommand{\StI}[1]{s_{#1}}
\newcommand{\EncAPs}{\widehat{\APs}}
\newcommand{\EncAP}{\hat{\AP}}

A \emph{\TSlong}, for brevity simply referred to as a \TSshort\ for the remainder of the paper,
over a set $\APs$ of atomic propositions
is a tuple
$\Tuple{\NonClocks,\Clocks,\EncInit,\EncInv,\EncTr,\EncFair,\EncAPs}$,
where
\begin{itemize}
\item
  $\NonClocks = \Set{\NonClock_1,\ldots,\NonClock_n}$ is a set of typed \emph{non-clock variables},
$\NonClocksNext = \Set{\NonClockNext_1,\ldots,\NonClockNext_n}$ being
  their \emph{next-state versions},
\item
  $\Clocks = \Set{\Clock_1,\ldots,\Clock_m}$ is a set of non-negative real-valued \emph{clock variables},
  $\ClocksNext = \Set{\ClockNext_1,\ldots,\ClockNext_m}$ again being their
  \emph{next-state versions},
\item
  $\EncInit$ is the \emph{initial state formula} over $\NonClocks \cup \Clocks$,
\item
  $\EncInv$ is the \emph{state invariant formula} over $\NonClocks \cup \Clocks$,
\item
  $\EncTr$ is the \emph{transition relation formula}
  over $\NonClocks \cup \Clocks \cup \Set{\Diff} \cup \NonClocksNext \cup \ClocksNext$,
  with a real-valued \emph{duration variable} $\Diff$,
\item
  $\EncFair$ is a finite set of \emph{fairness formulas} over $\NonClocks$,
  and
\item
  $\EncAPs$ associates each atomic proposition $\AP \in \APs$
  with a corresponding formula $\EncAP$ over $\NonClocks$.
\end{itemize}
To ensure that the clock variables are used properly,
we require that all the atoms in all the formulas in the system
follow these rules:
(i) if a non-clock variable in $\NonClocks$ or in $\NonClocksNext$ occurs in the atom, then none of the variables in $\Clocks \cup \ClocksNext \cup \Set{\Delay}$ occur in it,
and
(ii) if a variable in $\Clocks \cup \ClocksNext \cup \Set{\delta}$ occurs in it, then it is of the forms $\ClockNext = 0$, $\ClockNext = \Clock + \delta$, $\Clock \AnyIneq \Const$, $\Clock + \delta \AnyIneq \Const$, or $\Delay \AnyIneq 0$ where $\AnyIneq \in \Set{{<},{\le},{=},{\ge},{>}}$, $\Clock,\ClockNext\in\Clocks$ and $\Const \in \Naturals$.
Furthermore,
for all valuations $\tau$ over $\NonClocks \cup \Clocks \cup \Set{\Delay} \cup \NonClocksNext \cup \ClocksNext$ such that $\tau \Models \EncTr$,
it must hold that $\tau(\Delay) \ge 0$ and
for each clock $\Clock \in \Clocks$ either $\tau(\ClockNext)=0$ or $\tau(\ClockNext)=\tau(\Clock)+\tau(\Delay)$.

\newcommand{\TracesOf}[1]{\mathop{\operatorname{traces}}(#1)}
\newcommand{\STTSRun}{\tau}

A \emph{state} of 
the system
now is
a valuation $\St$ over $\NonClocks \cup \Clocks$ and
a \emph{run} an infinite sequence $\St_0 \EStep{\Delay_0} \St_1 \EStep{\Delay_1} \St_2 \ldots$ such that
\begin{itemize}
\item
  $\Delay_0 = 0$
  and for all $\Index \in \Naturals$ we have
  $\Delay_\Index \ge 0$,
  $\St_\Index(\Clock) \ge 0$ when $\Clock \in \Clocks$,
  and
  ${\Delay_\Index>0} \Implies {\Delay_{\Index+1}=0}$,
\item
  $\St_0 \Models \EncInit$ and $\St_\Index \Models \EncInv$ holds for all $\Index \in \Naturals$,
\item
   for all $\Index \in \Naturals$ it holds that
  $\Setdef{y\mapsto \St_\Index(y)}{y\in\NonClocks\cup\Clocks}\cup\Set{\Delay \mapsto \Delay_\Index} \cup \Setdef{y' \mapsto \St_{\Index+1}(y)}{y \in \NonClocks\cup\Clocks} \Models \EncTr$,
  and
\item
  for each $f \in \EncFair$,
  there are infinitely many states $\St$ in the run for which $\St \Models f$ holds.
\end{itemize}
A run $\STTSRun = \St_0 \EStep{\Delay_0} \St_1 \EStep{\Delay_1} \St_2 \EStep{\Delay_2} \ldots$
represents the super-dense timed trace
$\TraceOf{\STTSRun} =
 \Tuple{\IntervalI{0},\ValuI{0}}
 \Tuple{\IntervalI{1},\ValuI{1}}
 \Tuple{\IntervalI{2},\ValuI{2}} \ldots$ over $\APs$
where for each $\Index \in \Naturals$,
\begin{itemize}
\item
  $\ValuI{\Index} = \Setdef{\AP \in \APs}{\St_\Index \Models \EncAP}$,
  and
\item
  letting $t_\Index = \sum_{\AnotherIndex=0}^{\Index-1}\Delay_\AnotherIndex$,
  (i)
  if $\Delay_\Index = 0$,
  then $\IntervalI{\Index} = [t_\Index,t_\Index]$,
  and
  (ii)
  if $\Delay_\Index > 0$,
  then $\IntervalI{\Index} = (t_\Index,t_\Index+\Delay_\Index)$.
\end{itemize}
The set of all traces of a \TSshort\ $\STTS$ is $\TracesOf{\STTS} = \Setdef{\TraceOf{\STTSRun}}{\text{$\STTSRun$ is a run of $\STTS$}}$.
The \TSshort\ $\STTS$ is \emph{refinement-admitting}
if $\ITr \in \TracesOf{\STTS}$ implies $\ITr' \in \TracesOf{\STTS}$ for all the refinements $\ITr'$ of $\ITr$.

\subsection{Encoding Timed Automata Traces}

Recall the correspondence between timed automata runs and traces
discussed in Sect.~\ref{Sect:TA2TT}.
Given a timed automaton $\Ata = \Ta$,
we can encode it as a \TSshort\ 
$\STTSf{\Ata} = \Tuple{\NonClocks,\Clocks,\EncInit,\EncInv,\EncTr,\emptyset,\EncAPs}$,
where\footnote{Strictly,
  the atoms $\DelayNext = 0$ and $\DelayNext > 0$ are not allowed in $\EncTr$;
  this can be handled by adding new Boolean variables
  $\underline{\Delay = 0}$ and $\underline{\Delay > 0}$ in $\NonClocks$,
  forcing $\underline{\Delay = 0} \Implies (\Delay=0)$ and
  $\underline{\Delay > 0} \Implies (\Delay>0)$ in $\EncTr$,
  and then
  using $\underline{\Delay = 0}'$ instead of $\DelayNext = 0$
  and $\underline{\Delay > 0}'$ instead of $\DelayNext > 0$
  in the rest of $\EncTr$.}
\begin{itemize}
\item
  $\NonClocks=\Set{\At}$, where $\At$ is a variable with the domain $\Locs$,
\item
  $\EncInit \Def (\At=\Init) \land \bigwedge_{\Clock\in\Clocks}(\Clock=0)$,
\item
  $\EncInv \Def \bigwedge_{\Loc\in\Locs} (\At=\Loc) \Implies \Invs(\Loc)$
\item
$\begin{array}[t]{@{}r@{ }c@{ }l}
    \EncTr & \Def & \big((\Diff=0\land\Diff'=0) \Implies
    \bigvee_{\Tuple{\Loc,\Guard,\Reset,\Loc'}\in\Edges}
     \At{=}\Loc \land \At'{=}\Loc'
     \\
     & & {}
     \quad\quad \land \Guard \land
           (\bigwedge_{\Clock\in\Resets} \Clock'=0) \land
           (\bigwedge_{\Clock\in\Clocks\setminus\Resets} \Clock'=\Clock) \big)\\
& \land & \big((\Diff>0\lor\Diff'>0) {\Implies}
        (\At'{=}\At \land \bigwedge_{\Clock\in\Clocks}
        \ClockNext{=}\Clock{+}\Delay )\big)
        \\
& \land & \big(\Delay=0\lor\DelayNext=0\big)
\end{array}$
(Recall that $\Diff$ special real-valued \emph{duration variable}) 
\item
  $\EncAPs$ associates each atomic proposition $@\Loc$, where $\Loc \in \Locs$, with the formula $(\At = \Loc)$.
\end{itemize}
Now $\TracesOf{\STTSf{\Ata}}$ is exactly the set of super-dense timed traces corresponding to the runs of the automaton $\Ata$.
Every state of $\STTSf{\Ata}$ corresponds to a time interval in the timed trace of $\Ata$. Thus, there are three types of transitions 
encoded in  $\EncTr$.
Firstly, a \singi -to-\singi\ transition, corresponding to a discrete transition of $\Ata$, occurs when $\Diff$ and $\Diff'$ are both zero. 
Secondly, a \singi -to-open transition occurs when the $\Diff$ is zero and $\Diff'$ non-zero. On such a transition, all variables remain unchanged. Hence, the clocks values correspond to the left bound of the interval.
Thirdly, on a open-to-\singi\ transition ($\Diff>0$ and $\Diff'=0$) 
the clock variables are updated according to the length of the open interval.

Due to the ``repetition of time elapse steps'' property of timed automata discussed in Sect.~\ref{Sect:TA2TT},
the \TSshort\ $\STTSf{\Ata}$ is also refinement-admitting.

%
%
%
\section{Symbolic Encoding of $\ZIMITL$ formulas}

\label{sect:seozf}

Let $\STTS = \Tuple{\NonClocks,\Clocks,\EncInit,\EncInv,\EncTr,\EncFair,\EncAPs}$ be a \TSshort\ over $\APs$ encoding some timed system producing super-dense timed traces.
We now augment $\STTS$ with new variables and constraints
so that $\ZIMITL$ formulas over $\APs$ are symbolically evaluated
in the runs of the \TSshort s.
We say that the resulting \TSshort\ $\STTSf{\Formula} = \langle\NonClocks\cup\NonClocks_\Formula,\Clocks\cup\Clocks_\Formula,\EncInit\land\EncInit_\Formula,\EncInv,\EncTr\land\EncTr_\Formula,\EncFair\cup\EncFair_\Formula,\EncAPs\rangle$ over $\APs$
encodes $\Formula$
if $\NonClocks_\Formula$ includes a Boolean variable $\Enc{\AnotherFormula}$
for each sub-formula $\AnotherFormula$ of $\Formula$ (including $\Formula$ itself).
Furthermore,
we require two conditions on such encodings.

First,
we want to make sure that the encoding $\STTSf{\Formula}$ is \emph{sound}
in the following senses:
\newcommand{\SoundDef}{
\begin{itemize}
\item
  all the traces of $\STTS$ (i.e, projections of runs to the atomic propositions) are preserved:
  $\TracesOf{\STTSf{\Formula}} = \TracesOf{\STTS}$ 
\item
  when $\Formula$ is holds in a state,
  then it holds in the corresponding interval:
  for each run $\STTSRun = \St_0 \St_1 
  \ldots$ of $\STTSf{\Formula}$
  with $\TraceOf{\STTSRun} = \ITr = \Tuple{\IntervalI{0},\ValuI{0}} \Tuple{\IntervalI{1},\ValuI{1}} 
  \ldots$,
  and
   each $\Index \in \Naturals$,
  $\St_{\Index}(\Enc{\Formula}) = \True$
  implies
  $\forall \Offset \in \IntervalI{\Index} : \ITrAt{\Index}{\Offset} \Models \Formula$.
\end{itemize}
}
\SoundDef
For fine traces we want to faithfully capture the cases when a formula
holds on some interval.
To this end, we say that the encoding $\STTSf{\Formula}$ is \emph{complete}
if for every $\Formula$-fine trace $\ITr = \Tuple{\IntervalI{0},\ValuI{0}} \Tuple{\IntervalI{1},\ValuI{1}} \Tuple{\IntervalI{2},\ValuI{2}} \ldots$
 in $\TracesOf{\STTS}$,
there is a run $\STTSRun = \St_0 \St_1 \St_2 \ldots$ in $\STTSf{\Formula}$
  such that
  $\TraceOf{\STTSRun} = \ITr$
  and
  for all points $\Point{\Index}{\Offset}$ in $\ITr$
  it holds that
  $\ITrAt{\Index}{\Offset} \Models \Formula$
  implies
  $\ValuI{\Index}(\Enc{\Formula}) = \True$.

Therefore,
our model checking task ``Does a refinement-admitting \TSshort\ $\STTS$ have a run corresponding to a trace $\ITr$ with $\ITr \Models \Formula$?''
is reduced to the problem of
deciding whether $\STTSf{\Formula}$ has a run $\St_0 \St_1 \St_2 \ldots$ with $\St_0(\Enc{\Formula}) = \True$.

%
%
\subsection{Encoding Propositional Subformulas}

Let $\STTS = \Tuple{\NonClocks,\Clocks,\EncInit,\EncInv,\EncTr,\EncFair,\EncAPs}$ be a \TSshort\ over $\APs$.
\ifLong
For the atomic formulas $\Formula$ of forms $\AP$ and $\neg\AP$,
it is possible to make a \TSshort\  $\STTSf{\Formula} = \Tuple{\NonClocks\cup\Set{\Enc{\AP}},\Clocks,\EncInit,\EncInv,\EncTr\land\EncTr_\Formula,\EncFair,\EncAPs}$ encoding $\Formula$
as follows:
\begin{itemize}
\item
  if $\Formula = \AP$,
  then $\EncTr_\Formula = (\Enc{\Formula} \Iff \EncAP)$.
\item
  if $\Formula = \neg\AP$,
  then $\EncTr_\Formula = (\Enc{\Formula} \Iff {\neg\EncAP})$.
\end{itemize}
\else
For the atomic formulas $\Formula$ of forms $\AP$ and $\neg\AP$,
it is possible to make a \TSshort\  $\STTSf{\Formula} = \Tuple{\NonClocks\cup\Set{\Enc{\AP}},\Clocks,\EncInit,\EncInv,\EncTr\land\EncTr_\Formula,\EncFair,\EncAPs}$ encoding $\Formula$
by (i) defining $\EncTr_\Formula \Def (\Enc{\Formula} \Iff \EncAP)$ if $\Formula = \AP$
and
(ii)
$\EncTr_\Formula \Def (\Enc{\Formula} \Iff {\neg\EncAP})$
if 
$\Formula = \neg\AP$.
\fi
Similarly,
assuming that $\Formula$ is either of form $\LSF \land \RSF$ or $\LSF \lor \RSF$
for some $\ZIMITL$ formulas $\LSF$ and $\RSF$,
and
that $\STTS$ encodes both $\LSF$ and $\RSF$,
we can make a \TSshort\  $\STTSf{\Formula} = \Tuple{\NonClocks\cup\Set{\Enc{\AP}},\Clocks,\EncInit,\EncInv,\EncTr\land\EncTr_\Formula,\EncFair,\EncAPs}$
encoding $\Formula$ as follows:
\ifLong
\begin{itemize}
\item
  if $\Formula = {\LSF \lor \RSF}$,
  then $\EncTr_\Formula = (\Enc{\Formula} \Iff (\Enc{\LSF} \lor \Enc{\RSF}))$,
  and
\item
  if $\Formula = {\LSF \land \RSF}$,
  then $\EncTr_\Formula = (\Enc{\Formula} \Iff (\Enc{\LSF} \land \Enc{\RSF}))$.
\end{itemize}
\else
(i)
  if $\Formula = {\LSF \lor \RSF}$,
  then $\EncTr_\Formula \Def (\Enc{\Formula} \Iff (\Enc{\LSF} \lor \Enc{\RSF}))$,
  and,
(ii)
  if $\Formula = {\LSF \land \RSF}$,
  then $\EncTr_\Formula \Def (\Enc{\Formula} \Iff (\Enc{\LSF} \land \Enc{\RSF}))$.
\fi
 
The lemmas for the soundness and completeness of the encodings are given in
Sect.~\ref{Sect:SoundnessAndCompleteness}.

\subsection{Encoding \ZIMITL operators}

In the following sub-sections,
we present encodings for the other $\ZIMITL$ operators.
In each encoding, we may introduce some new non-clock and clock variables
such as $\C$ and $\LeftOpen$;
these variables are ``local'' to the encoded subformula $\AnotherFormula$ and
not used elsewhere, we do not subscript them (e.g.\ $\C$ really means $\C_\AnotherFormula$) for the sake of readability.
We also introduce new transition relation constraints (i.e.\ conjuncts in $\EncTr_\AnotherFormula$), initial state constraints and fairness conditions.
We will use $\Open$ as a shorthand for $(\Delay > 0)$.

%
%
\newcommand{\OUT}{MUU}
\newcommand{\OUTnext}{MUU}
\renewcommand{\OUT}{\Left \SUI{\AnyUpper\Const} \Right}

\mitlopsec{Encoding $\Left \SUI{\AnyUpper\Const} \Right$ and $\Left \SRI{\AnyUpper\Const} \Right$ with ${\AnyUpper} \in \Set{<,\le}$}

These operators can be expressed with simpler ones by using the following lemma (proven in \RefAppendix ):
\newcommand{\UbeqText}{
  $\ITrAt{\Index}{\Offset} \models {\Formula \SUI{\AnyUpper\Const} \AnotherFormula}$ iff
  $\ITrAt{\Index}{\Offset} \models (\SFI{\AnyUpper\Const}\AnotherFormula) \land (\Formula\SU\AnotherFormula)$
  for all
  $\Index \in \Naturals$,
  $\Offset \in \IntervalI{\Index}$,
  ${\AnyUpper} \in \Set{{<},{\leq}}$,
  and $\Const \in \Naturals$.}
\begin{lemma}\label{lem:ubeq}
\UbeqText

\end{lemma}
Using the $ \SUI{\AnyUpper\Const}$ / $ \SRI{\AnyUpper\Const}$ duality, we can now also express ${\Formula \SRI{\AnyUpper\Const} \AnotherFormula}$
as
$(\SGI{\AnyUpper\Const}\AnotherFormula) \land (\Formula\SR\AnotherFormula)$.

\newcommand{\LEFT}{\Enc{\Left}}
\newcommand{\LEFTnext}{\EncNext{\Left}}
\newcommand{\RIGHT}{\Enc{\Right}}
\newcommand{\RIGHTnext}{\EncNext{\Right}}

\mitlopsec{Encoding $\Left\SU\Right$}
\label{s:enc.su}
\renewcommand{\OUT}{\Enc{\Left \SU \Right}}
\renewcommand{\OUTnext}{\EncNext{\Left \SU \Right}}
\renewcommand{\LEFT}{\Enc{\Left}}
\renewcommand{\LEFTnext}{\EncNext{\Left}}
\renewcommand{\RIGHT}{\Enc{\Right}}
\renewcommand{\RIGHTnext}{\EncNext{\Right}}

We encode ``untimed'' until formulas $\Left\SU\Right$
essentially like in the traditional LTL case~\cite{BiereEtAl:LMCS2006} but must consider open intervals and \singis\ separately.

Assume $\Left\SU\Right$ holds on the current interval.
If that interval is open,
$\Left$ and one of the following hold:
(i) $\Right$ holds on the current interval, (ii) $\Right$ holds on the next interval (which is
a \singi ), or (iii) $\Left$ holds on the next interval and
$\Left\SU\Right$ is satisfied as well. This is captured by the following 
constraint:
\begin{equation}
  \label{m:enc-u-open}
  {\OUT{\land}\Open}
  \Implies
  {\LEFT \land (\RIGHT \lor \RIGHTnext \lor (\LEFTnext \land \OUTnext))}
\end{equation}

If, in contrast, the current interval is a \singi ,
then there are two possibilities:
(i) the next interval is a \singi\ and $\Right$ holds,
or
(ii)
both $\Left$ and $\Left\SU\Right$ hold on the next interval:
\begin{equation}
  \label{m:enc-u-singular}
  {\OUT {\land} {\neg\Open}}
  \Implies
  {(\neg\Open' {\land} \RIGHTnext) \lor (\LEFTnext {\land} \OUTnext)}
\end{equation}

\ifLong
Finally,
as in the traditional LTL encoding,
we must add the following fairness condition in order to avoid
the case where $\OUT$ and $\LEFT$ are $\True$ on all intervals starting from some point but $\Right$ does not hold at any future time point:
\begin{equation}
\EncFair_{\Left \SU \Right} = \Set{\neg\OUT \lor \RIGHT}
\end{equation}
\else
Finally,
as in the traditional LTL encoding,
we must add a fairness condition in order to avoid
the case where $\OUT$ and $\LEFT$ are $\True$ on all intervals starting from some point but $\Right$ does not hold at any future time point, i.e. $\EncFair_{\Left \SU \Right} = \Set{\neg\OUT \lor \RIGHT}$.
\fi

\begin{figure}[tb]
  \centering
  \includegraphics[scale=\xplscale]{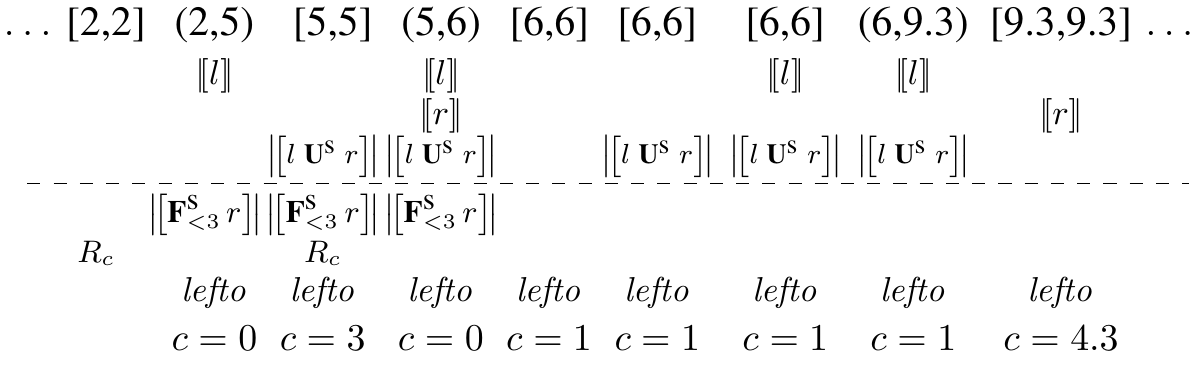}
  \caption{Encoding $\Left \SU \Right$ and $\SFI{<3}\Right$}
  \label{Fig:EncEx1}
\end{figure}

\begin{example}
  Figure~\ref{Fig:EncEx1} illustrates
  an evaluation of the encoding variables
  on a trace (ignore the text below the dashed line for now).
  Note that $\Enc{\Left \SU \Right}$ is (correctly) evaluated to $\True$
  on the second $[6,6]$-interval
  despite $\Left$ not holding.
\end{example}

%
%
\mitlopsec{Encoding $\SFI{\leq 0}\Right$}
\renewcommand{\OUT}{\Enc{\SFI{\leq 0}\Right}}
\renewcommand{\OUTnext}{\EncNext{\SFI{\leq 0}\Right}}

A formula $\SFI{\leq 0}\Right$ holding
requires a future interval at which $\Right$ holds and
which can be reached without any time passing.
Thus, $\SFI{\leq 0}\Right$
is satisfied only on a
\singi\
 where
the next interval is a \singi\ as well and (i)
$\Right$ or (ii) $\SFI{\leq 0}\Right$ holds on the next interval:
\begin{equation}
  \label{eq:f.leqz}
  \OUT
  \Implies
  {{\neg\Open} \land {\neg\Open'} \land (\RIGHTnext \lor \OUTnext)}
\end{equation}
No fairness conditions are needed
as the non-zenoness requirement always guarantees a future open interval.

%
%
\mitlopsec{Encoding $\SFI{\AnyUpper\Const}\Right$ with $\Const > 0$}
\renewcommand{\OUT}{\Enc{\SFI{\AnyUpper\Const}\Right}}
\renewcommand{\OUTnext}{\EncNext{\SFI{\AnyUpper\Const}\Right}}

In the encoding of $\SFI{\AnyUpper\Const}$, we first add
the constraints for $\SU$
replacing $\Left$ by $\True$.
\begin{eqnarray}
  {\OUT \land \Open}
  & \Implies &
  {\RIGHT \lor \RIGHTnext \lor \OUTnext}
  \label{eq:enc.f.ub.open}
  \\
  {\OUT \land {\neg\Open}}
  & \Implies &
  {\RIGHTnext \lor \OUTnext}
  \label{eq:enc.f.ub.singular}
\end{eqnarray}

Next, we observe that for encoding timing related aspect, it is sufficient to at any point remember the earliest interval at which $\SFI{\AnyUpper\Const} \Right$ holds and after which $\Right$ has not held yet.
If $\Right$ is encountered in time for the earliest such interval, then interval where $\Right$ holds is close enough to any later interval where $\OUT$ holds as well.
Correspondingly, 
we use a real-valued (clock-like) auxiliary variable $\C$ and a boolean auxiliary variable $\LeftOpen$
to remember the time passed since and type of the earliest interval on which  $\OUT$ held and after which we have not seen $\RIGHT$.
\ifLong
The correct values in the first interval are forced by the initial state formula
\begin{equation}
  \EncInit_{\SFI{\AnyUpper\Const} \Right}
  \Def
  {{\C = 0} \land {\neg\LeftOpen}}
  \label{eq:enc.f.ub.initial}
\end{equation}
\else
The correct values in the first interval are forced by the initial state formula
  $\EncInit_{\SFI{\AnyUpper\Const} \Right}
  \Def
  {{\C = 0} \land {\neg\LeftOpen}}$.
\fi
To update $\C$ and $\LeftOpen$, we define the shorthand $\EReset$ to be $\True$
 when we
have not seen $\OUT$ without seeing $\Right$ afterwards or $\Right$ holds on an open current or an arbitrary next interval.
\begin{equation}
  \EReset
  \Def
  (\neg\OUT \lor (\Open\land\RIGHT)\lor\RIGHTnext) \land \OUTnext
\end{equation}
We then
(i) reset $\C$ and $\LeftOpen$ on the next interval if $\EReset$ holds on the current interval,
and
(ii) update $\C$ and leave $\LeftOpen$ unchanged if $\EReset$ does not hold.
\begin{eqnarray}
  \EReset
  & \Implies &
  {\C'=0} \land (\LeftOpen' \Iff \Open')
  \label{eq:enc.f.ub.clk.reset}
  \\
  \neg\EReset
  & \Implies &
  {\C'=\C+\Delay} \land (\LeftOpen'\Iff\LeftOpen)
  \label{eq:enc.f.ub.clk.update}
\end{eqnarray}

We introduce a shorthand $\Timing$ (defined below)
such that $\Timing$
holds if
for each point on the interval where we reset $\C$ 
there is a point on the \emph{next} interval that satisfies the $\AnyUpper\Const$ constraint.
We then require that $\OUT$ being $\True$,
and $\Right$ being $\False$ or
the current interval being a \singi\ implies that $\Timing$ holds.
\begin{equation}
  (\OUT \land \neg(\RIGHT \land \Open)) \Implies \Timing
  \label{eq:enc.f.ub.timing}
\end{equation}

\ifLong
In the case of $\SFI{<\Const}\Right$, we define $\Timing$ as follows
\begin{equation}
  \Timing
  \Def
  {\C+\D < \Const} \lor (\LeftOpen \land {\C+\D\leq\Const})
\end{equation}
and in the case of $\SFI{\leq\Const}\Right$ by
\begin{equation}
  \Timing
  \Def
  {\C+\D < \Const} \lor ((\neg\Open'\lor\LeftOpen)\land {\C+\D\leq\Const})
\end{equation}
\else
In the case of $\SFI{<\Const}\Right$, we define
$
  \Timing
  \Def
  {\C+\D < \Const} \lor (\LeftOpen \land {\C+\D\leq\Const})
$
and in the case of $\SFI{\leq\Const}\Right$ we define
$
  \Timing
  \Def
  {\C+\D < \Const} \lor ((\neg\Open'\lor\LeftOpen)\land {\C+\D\leq\Const})
$.
\fi

\begin{example}
  An evaluation of the encoding variables
  is shown (below the dashed line) in Figure~\ref{Fig:EncEx1}.
  Especially,
  observe that $\Enc{\SFI{>3}\Right}$ is \emph{not} evaluated to true
  on the interval $(6,9.3)$ although $\SFI{>3}\Right$ holds on \emph{some} points in the interval: we are interested in \emph{sound} encodings and
  $\Enc{\SFI{>3}\Right}$ does not hold on \emph{all} the points in the interval.
\end{example}

%
%
\mitlopsec{Encoding $\Left\SUI{\AnyLower\Const}\Right$ with $\AnyLower \in \Set{\ge,>}$}
\renewcommand{\OUT}{\Enc{\Left\SUI{\AnyLower\Const}\Right}}
\renewcommand{\OUTnext}{\EncNext{\Left\SUI{\AnyLower\Const}\Right}}

To encode $\Left\SUI{\AnyLower\Const}\Right$,
we define shorthands $\Timing$ and $\TRight$.
$\Timing$ will later be defined so that $\Timing$ holds iff for every previous point at which $\OUT$ held there is a point on the current interval that satisfies the $\AnyLower\Const$ timing constraint.
\ifLong
We, then, define $\TRight$ as follows:
\begin{equation}
  \TRight
  \Def
  {\RIGHT \land \Timing}
  \label{m:enc.u.lb.def.rhat}
\end{equation}
\else
We, then, define $
  \TRight
  \Def
  {\RIGHT \land \Timing}
$.
\fi
Next, we add a boolean ``obligation'' variable $\Obligation$ to remember when we need to see $\TRight$ at a future point.
Whenever $\OUT$ is $\True$, we also require $\Obligation$ to be $\True$.
\begin{equation}
  \label{m:enc.u.lb.out.obl}
  \OUT\Implies\Obligation
\end{equation}
In case $\Const>0$, we additionally require $\Obligation$ and $\Left$ to hold on the next interval.
\begin{equation}
  \label{m:enc.u.lb.out.xobl}
  \OUT\Implies(\Obligation'\land\Left')
\end{equation}
Next, we add constraints similar to
those
for the
$\SU$-operator but with $\OUT$ and $\RIGHT$ replaced by $\Obligation$ and $\TRight$.
\iffalse 
\begin{equation}
  (\Obligation\land\Open)
  \Implies
  (\LEFT\land(\TRight\lor\TRight'\lor(\LEFTnext\land\Obligation')))
  \label{m:enc.u.lb.obl.open}
\end{equation}
\begin{equation}
  (\Obligation\wedge\neg\Open)
  \Implies
  ((\neg\Open'\land\TRight')\lor(\LEFTnext\land\Obligation'))
  \label{m:enc.u.lb.obl.singular}
\end{equation}
\else 
\begin{eqnarray}
  (\Obligation\land\Open)
  \Implies
  (\LEFT\land(\TRight\lor\TRight'\lor(\LEFTnext\land\Obligation')))
&&
  \label{m:enc.u.lb.obl.open}
\\
  (\Obligation\wedge\neg\Open)
  \Implies
  ((\neg\Open'\land\TRight')\lor(\LEFTnext\land\Obligation'))
&&
  \label{m:enc.u.lb.obl.singular}
\end{eqnarray}
\fi 

We want to determine whether the $\AnyLower\Const$ constraint holds for \emph{all} previous points at which $\OUT$ holds.
We, thus, use a
real-valued 
variable $\C$ and
a boolean variable $\RightOpen$
to measure the time since the most recent corresponding 
interval.
We, thus, reset $\C$ to zero and use $\RightOpen$ to remember the type of the \emph{current} interval whenever $\OUT$ holds. Otherwise, we update $\C$ and $\RightOpen$ as before.
\begin{eqnarray}
  \OUT
  &\Implies&
  {\C'=0} \land (\RightOpen'\Iff\Open)
  \label{m:enc.u.lb.clk.reset}
  \\
  {\neg\OUT}
  &\Implies&
  {\C'=\C+\D} \land (\RightOpen'\Iff\RightOpen)
  \label{m:enc.u.lb.clk.update}
\end{eqnarray}

\ifLong
Next, in case
$\Left\SUI{>\Const}\Right$, we define
\begin{equation}
\label{m:enc.u.lb.def.timing.strict}
\Timing\Def\C+\D > \Const \lor (\RightOpen\land\C+\D \geq \Const)
\end{equation}
and in case $\Left\SUI{\geq\Const}\Right$
\begin{equation}
\label{m:enc.u.lb.def.timing.nonstrict}
\Timing\Def\C+\D > \Const \lor ((\RightOpen\lor\neg\Open)\land\C+\D \geq \Const)
\end{equation}
\else
Next, in case
$\Left\SUI{>\Const}\Right$, we define
$\Timing\Def\C+\D > \Const \lor (\RightOpen\land\C+\D \geq \Const)$
and in case $\Left\SUI{\geq\Const}\Right$ we define $\Timing\Def\C+\D > \Const \lor ((\RightOpen\lor\neg\Open)\land\C+\D \geq \Const)$.
\fi

\ifLong
Finally,
as for the untimed $\SU$-operator,
we need a fairness condition to prevent a situation where $\Obligation$ holds globally but $\Right$ never holds.
\begin{equation}
  \label{eq:enc.u.lb.fairness}
  \EncFair_{\Left \SUI{\AnyLower\Const} \Right}
  \Def
  \Set{{\neg\Obligation} \lor \RIGHT}
\end{equation}
\else
Finally,
as for the untimed $\SU$-operator,
we need a fairness condition to prevent a situation where $\Obligation$ holds globally but $\Right$ never holds. We define $\EncFair_{\Left \SUI{\AnyLower\Const} \Right} \Def \Set{{\neg\Obligation} \lor \RIGHT}$.
\fi
Note that, here, we use $\RIGHT$, not $\TRight$. For instance, when $\Left\SUI{\AnyLower\Const}\Right$ and $\Right$ hold globally, there may never be a point where $\Timing$ is $\True$ and thus $\TRight$ always stays $\False$.

\begin{figure}[tb]
  \centering
  \includegraphics[scale=\xplscale]{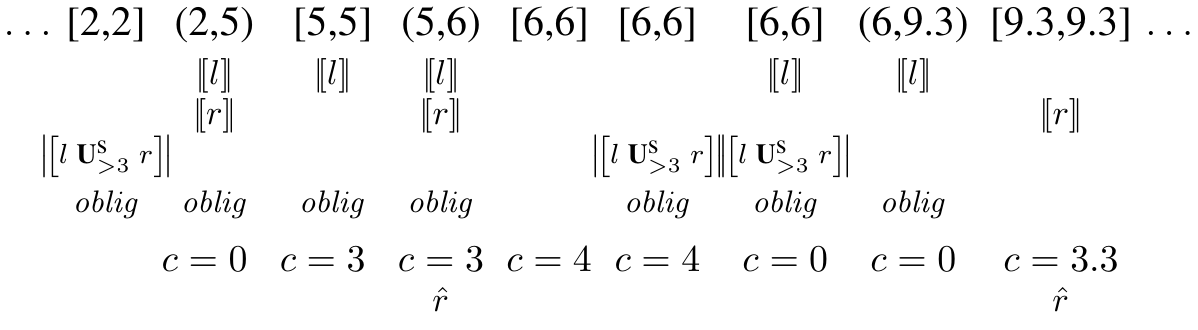}
  \caption{Encoding $\Left \SUI{>3} \Right$}
  \label{Fig:EncEx2}
\end{figure}

\begin{example}
  Figure \ref{Fig:EncEx2} illustrates how the encoding variables
  of $\Left \SUI{>3} \Right$
  variables could be evaluated on a trace.
  Again,
  $\Enc{\Left \SUI{>3} \Right}$ is not true on the interval $(6,9.3)$
  because $\Left \SUI{>3} \Right$ holds only on some points on it but not on all.
\end{example}

%
%
\mitlopsec{Encoding $\Left\SR\Right$}
\renewcommand{\OUT}{\Enc{\Left\SR\Right}}
\renewcommand{\OUTnext}{\EncNext{\Left\SR\Right}}

For encoding $\Left\SR\Right$,
we
use an auxiliary boolean variable $\Obligation$.
Intuitively, $\Obligation$ being $\True$ means that before seeing any point at which $\RIGHT$ is $\False$, we need to see a point where $\LEFT$ is $\True$.

We require $\Obligation$ to hold on the current interval when $\OUT$ holds on an open interval and on the next interval when $\OUT$ holds on a \singi .
\begin{eqnarray}
  (\OUT \land \Open) & \Implies & \Obligation
  \label{eq:r.unt.out-open}
  \\
  \label{eq:r.unt.out-sing}
  (\OUT\land\neg\Open) & \Implies & \Obligation'
\end{eqnarray}
The obligation to see $\Left$ before $\neg\Right$ remains active
until
 $\Left$ holds:
\begin{equation}
  \Obligation
  \Implies
  (\LEFT \lor \Obligation')
  \label{eq:r.unt.consec}
\end{equation}
As a final constraint, $\Right$ needs to hold on all intervals where the obligation is $\True$, with the exception of open intervals on which $\Left$ holds, leading to
\begin{equation}
  \Obligation
  \Implies
  ((\Open\land\LEFT) \lor \RIGHT)
  \label{eq:r.unt.obl-r}
\end{equation}

%
%
\mitlopsec{Encoding $\SGI{\leq 0}\Right$}
\renewcommand{\OUT}{\Enc{\SGI{\leq 0}\Right}}
\renewcommand{\OUTnext}{\EncNext{\SGI{\leq 0}\Right}}

$\SGI{\leq 0}\Right$ trivially holds
 when the current or the next interval is open.
Furthermore, $\SGI{\leq 0}\Right$ holds when both current and next interval are \singis\ and $\Right$ and $\SGI{\leq 0}\Right$ hold on the next interval.
\begin{equation}
  \OUT
  \Implies
  (\Open \lor \Open' \lor (\RIGHTnext \land \OUTnext))
  \label{eq:g.leqz}
\end{equation}

%
%
\mitlopsec{Encoding $\SGI{\AnyUpper\Const}\Right$ with $\Const > 0$}
\renewcommand{\OUT}{\Enc{\SGI{\AnyUpper\Const}\Right}}
\renewcommand{\OUTnext}{\EncNext{\SGI{\AnyUpper\Const}\Right}}

First,
we require that $\Right$ holds on all
open intervals on which $\OUT$ holds. Furthermore, we will later define a shorthand $\Timing$ to hold whenever there is an interval on which $\OUT$ held sufficiently shortly in the past to still require $\Right$ to hold, resulting in
\begin{equation}
  ((\OUT \land \Open) \vee \Timing)
  \Implies
  \RIGHT
  \label{eq:g.ub.timing}
\end{equation}
Like in the $\SUI{\AnyLower}$ encoding, we use a real-valued variable $\C$ and a boolean variable $\RightOpen$ to measure time from the most recent interval at which $\OUT$ held.
\begin{eqnarray}
  \OUT
  &\Implies&
  {\C'=0} \land (\RightOpen'\Iff\Open)
  \label{eq:g.ub.clk1}
  \\
  {\neg\OUT}
  &\Implies&
  {\C'=\C+\D} \land (\RightOpen'\Iff\RightOpen)
  \label{eq:g.ub.clk2}
\end{eqnarray}
\ifLong
Now, in the case of $\SGI{<\Const}$ we define
\begin{equation}
  \Timing
  \Def
  {\C<\Const}
  \label{eq:g.ub.timing-strict}
\end{equation}
and
in the case of $\SGI{\leq\Const}$
\begin{equation}
  \Timing
  \Def
  {\C<\Const} \lor ({\C\leq\Const} \land {\neg\Open} \land {\neg\RightOpen})
  \label{eq:g.ub.timing-nonstrict}
\end{equation}
\else
Now, in the case of $\SGI{<\Const}$ we define
$
  \Timing
  \Def
  {\C<\Const}
$
and
for $\SGI{\leq\Const}$
we define
$
  \Timing
  \Def
  {\C<\Const} \lor ({\C\leq\Const} \land {\neg\Open} \land {\neg\RightOpen})
$
\fi

%
%
\mitlopsec{Encoding $\Left\SRI{\AnyLower\Const}\Right$}
\renewcommand{\OUT}{\Enc{\Left\SRI{\AnyLower\Const}\Right}}
\renewcommand{\OUTnext}{\EncNext{\Left\SRI{\AnyLower\Const}\Right}}
For encoding the lower bound until operators, we use a boolean variable $\Obligation$ and the same update rules as for the untimed $\SR$ operator.
\begin{eqnarray}
\label{eq:nr.lb.left.open}
(\OUT\wedge\Open) & \Implies & \Obligation \\
\label{eq:nr.lb.left.singular}
(\OUT\wedge\neg\Open) & \Implies & \Obligation' \\
\label{eq:nr.lb.obligation.update}
\Obligation & \Implies & (\LEFT\lor\Obligation')
\end{eqnarray}
We add a modified version of Constraint~\ref{eq:r.unt.consec} and use a shorthand $\Timing$ (defined later) to identify intervals that contain time points $\AnyLower\Const$ from a point where $\OUT$ holds.
\begin{equation}
\label{eq:nr.lb.timing}
(\Obligation\wedge\Timing) \Implies ((\LEFT\wedge\Open)\lor\RIGHT)
\end{equation}
Next, we add a constraint for intervals of length $>\Const$. On such an interval, $\Left$ or $\Right$ has to hold if $\OUT$ holds.
\begin{equation}
\label{eq:nr.lb.first}
(\OUT\land\D>\Const) \Implies (\LEFT\lor\RIGHT)
\end{equation}

\begin{figure}[tb]
  \centering
  \includegraphics[scale=\xplscale]{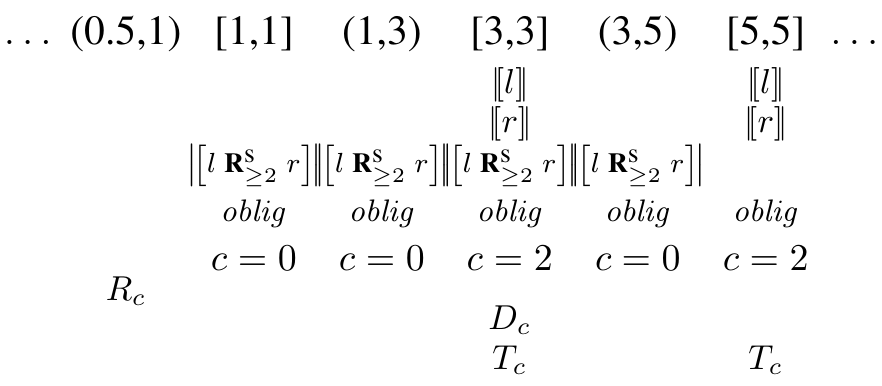}
  \caption{Encoding $\Left \SRI{\geq 2}\Right$.
  }
  \label{Fig:EncEx3}
\end{figure}

For encoding $\SRI{\AnyLower\Const}$, we use an auxiliary real-valued variable $\C$ and a boolean variable $\LeftOpen$ to measure the time passed since
the earliest interval at which $\OUT$ holds and whose obligation to see $\Left$ before $\Right$ is still active. 
This is, in principle, similar to the $\SFI{\AnyUpper\Const}$ encoding except for a special case illustrated in Figure~\ref{Fig:EncEx3}.
Here, on the fourth interval $\C$ and $\LeftOpen$ are needed for two purposes: to measure the time passed since the second interval (which introduced a still open obligation) and to start measuring time since the current interval (which introduces a fresh obligation as $\LEFT$ holds satisfying the previous obligation).
\ifLong
We define a shorthand $\DelayReset$ to captures precisely this situation and will later delay resetting $\C$ by one step whenever $\DelayReset$ holds.
\begin{equation}
\label{eq:nr.lb.delay}
\DelayReset\Def(\neg\Open\land\Obligation\land\Left\land\OUT)
\end{equation}
\else
We define a shorthand $\DelayReset\Def(\neg\Open\land\Obligation\land\Left\land\OUT)$ to captures precisely this situation and will later delay resetting $\C$ by one step whenever $\DelayReset$ holds.
\fi
Otherwise, $\C$ needs to be reset on the next interval if $\OUT$ holds on that
interval and
(i)
if there is an open obligation it is satisfied on the current interval
and
(ii) the current interval is not a \singi\ on which $\OUT$ holds, i.e. does not add an obligation to the next
\iffalse
interval.
\begin{equation}
\EReset\Def\OUT'\land(\neg\Obligation\lor\Left)\land(\Open\lor\neg\OUT)
\end{equation}
\else
interval, i.e.
$
\EReset\Def\OUT'\land(\neg\Obligation\lor\Left)\land(\Open\lor\neg\OUT)
$.
\fi

As said before,
we delay resetting $\C$ and $\LeftOpen$ by one interval when $\DelayReset$ holds, i.e. set $\C$ to 0 and $\LeftOpen$ to $\False$.
\begin{equation}
\label{eq:nr.lb.clk.delayed}
\DelayReset\Implies(\C'=0\wedge\neg\LeftOpen')
\end{equation}
When $\EReset$ holds, $\C$ and $\LeftOpen$ are reset as for the $\SFI{\AnyUpper\Const}$ operator and when neither holds we update them as usual:
\begin{eqnarray}
\label{eq:nr.lb.clk.reset}
&\EReset\Implies(\C'=0\wedge(\LeftOpen'\Iff\Open'))&\\
\label{eq:nr.lb.clk.update}
&(\neg\EReset\land\neg\DelayReset)\Implies(\C'=\C+\D\wedge(\LeftOpen'\Iff\LeftOpen))&
\end{eqnarray}
We set the initial values of $\C$ and $\LeftOpen$ to correspond measuring time from the initial
\iffalse
interval:
\begin{equation}
\label{eq:nr.lb.initial}
  \EncInit_{\Left\SRI{\AnyLower\Const}\RIGHT}
  \Def
	\C=0\land\neg\LeftOpen
\end{equation}
\else
interval, i.e.
$
  \EncInit_{\Left\SRI{\AnyLower\Const}\RIGHT}
  \Def
	\C=0\land\neg\LeftOpen
$.
\fi

Finally, we define $\Timing$ to hold precisely if there is a point on the current interval that is $\AnyLower\Const$ time units away from a point belonging to the interval at which we started measuring time.
\ifLong
In the case of $\SRI{>\Const}$, we define:
\begin{equation}
\label{eq:nr.lb.timing.strict}
\Timing\Def\C+\D>\Const
\end{equation}
and in case $\SRI{\geq\Const}$
\begin{equation}
\label{eq:nr.lb.timing.nonstrict}
\Timing\Def\C+\D>\Const\lor(\neg\LeftOpen\land\neg\Open\land\C+\D\geq\Const))
\end{equation}
\else
In the case of $\SRI{>\Const}$, we define
$
\Timing\Def\C+\D>\Const
$
and for $\SRI{\geq\Const}$ we define
$
\Timing\Def\C+\D>\Const\lor(\neg\LeftOpen\land\neg\Open\land\C+\D\geq\Const))
$.
\fi

%
%
\subsection{Soundness and Completeness of the Encodings}
\label{Sect:SoundnessAndCompleteness}

The encoding just given is sound and complete in the sense defined by the following lemmas which are  proven in \RefAppendix .
\newcommand{\CenfFText}{
The \TSshort\ 
  $\STTSf{\AP}$ is a sound encoding for $\AP$ and
  $\STTSf{\neg\AP}$ is a sound encoding for $\neg\AP$.
  If a \TSshort\  $\STTS$ over $\APs$ is a sound encoding of $\LSF$ and $\RSF$,
  then
  the \TSshort\  $\STTSf{\AnyTemporalOp \LSF}$ over $\APs$ is a sound encoding of $\AnyTemporalOp \LSF$ for each ${\AnyTemporalOp} \in \Set{{\SFI{\leq 0}},{\SFI{<\Const}},{\SFI{\leq\Const}},{\SGI{\leq0}},{\SGI{<\Const}},{\SGI{\leq\Const}}}$,
  and
  $\STTSf{\LSF \AnyTemporalOp \RSF}$
  is a sound encoding of $\LSF \AnyTemporalOp \RSF$ for each ${\AnyTemporalOp} \in \Set{{\land},{\lor},{\SU},{\SUI{\geq\Const}},{\SUI{>\Const}},{\SR},{\SRI{\geq\Const}},{\SRI{>\Const}}}$.
}
\begin{lemma}\label{Lemma:constraints_enforce_formula}
\CenfFText
\end{lemma}

\newcommand{\FenfCText}{
  The \TSshort\ 
  $\STTSf{\AP}$ is a complete encoding for $\AP$,
  $\STTSf{\neg\AP}$ is a complete encoding for $\neg\AP$.
  If a \TSshort\  $\STTS$ over $\APs$ is a complete encoding of $\LSF$ and $\RSF$,
  then
  the \TSshort\  $\STTSf{\AnyTemporalOp \LSF}$ over $\APs$ is a complete encoding of $\AnyTemporalOp \LSF$ for each ${\AnyTemporalOp} \in \Set{{\SFI{\leq 0}},{\SFI{<\Const}},{\SFI{\leq\Const}},{\SGI{\leq0}},{\SGI{<\Const}},{\SGI{\leq\Const}}}$,
  and
  $\STTSf{\LSF \AnyTemporalOp \RSF}$
  is a complete encoding of $\LSF \AnyTemporalOp \RSF$ for each ${\AnyTemporalOp} \in \Set{{\land},{\lor},{\SU},{\SUI{\geq\Const}},{\SUI{>\Const}},{\SR},{\SRI{\geq\Const}},{\SRI{>\Const}}}$.
}

\begin{lemma}\label{Lemma:formula_implies_constraints}
\FenfCText
\end{lemma}

\section{Bounded Model Checking}
\label{Sect:BMC}

Naturally, one cannot directly handle infinite formula representations
capturing infinite runs with SMT solvers.
Thus in bounded model checking (\emph{BMC}) one considers finite representations, i.e.\ looping, lasso-shaped paths only.
We show that,
by using region abstraction~\cite{AlurDill:TCS1994},
we can indeed capture all runs that satisfy a $\ZIMITL$ formula with
such finite representations.
For this we must assume that the domains of all the non-clock variables in $\NonClocks$ are finite.

Assume a \TSshort\ 
$\Tuple{\NonClocks,\Clocks,\EncInit,\EncInv,\EncTr,\EncFair,\EncAPs}$
over a set $\APs$ of atomic propositions.
%
%
For each clock $\Clock \in \Clocks$,
let $\ClockMax{\Clock}$ be the largest constant $\Const$
occurring in atoms of forms $\Clock \AnyIneq \Const$ and
$\Clock + \delta \AnyIneq \Const$ in $\EncInit$, $\EncInv$, and $\EncTr$.
Two states,
$\St$ and $\AnotherSt$ (i.e. valuations over $\NonClocks\cup\Clocks$ as defined in Sect.~\ref{ss:ts}),
belong to the same equivalence class called \emph{region},
denoted by $\St \REquiv \AnotherSt$,
if
(i) $\St(\NonClock) = \AnotherSt(\NonClock)$ for each non-clock variable $\NonClock \in \NonClocks$,
and
(ii) for all clocks $\Clock, \AnotherClock \in \Clocks$
\begin{enumerate}
\item
  either
  (a) $\CInt{\St(\Clock)} = \CInt{\AnotherSt(\Clock)}$
  or
  (b)
  $\St(\Clock) > \ClockMax{\Clock}$ and
  $\AnotherSt(\Clock) > \ClockMax{\Clock}$;
\item
  if $\St(\Clock) \leq \ClockMax{\Clock}$,
  then
  $\CFrac{\St(\Clock)} = 0$ iff
  $\CFrac{\AnotherSt(\Clock)} = 0$,
  where $\CFrac{i}$ denotes the fractional part of $i$;
   and
\item
  if
  $\St(\Clock) \leq \ClockMax{\Clock}$ and
  $\St(\AnotherClock) \leq \ClockMax{\AnotherClock}$,
  then 
  $\CFrac{\St(\Clock)} \leq \CFrac{\St(\AnotherClock)}$
  iff
  $\CFrac{\AnotherSt(\Clock)} \leq \CFrac{\AnotherSt(\AnotherClock)}$.
\end{enumerate}
%
%
%
Next, we will apply the bisimulation property of regions introduced in~\cite{AlurDill:TCS1994} to transition systems.
\begin{lemma}
\label{lem:sddfuiet}
  Assume two states, $\St$ and $\AnotherSt$,
  such that $\St \REquiv \AnotherSt$.
  It holds that
  (i) $\St \Models \EncInit$ iff $\AnotherSt \Models \EncInit$, and
  (ii) $\St \Models \EncInv$ iff $\AnotherSt \Models \EncInv$.
  Furthermore,
  if there is
  a $\Delay_\St \in \RealsNonNeg$ and
  a state $\St'$
  such that
  ${\St \cup \Set{\Delay \mapsto \Delay_\St} \cup \Setdef{y' \mapsto \St'(y)}{y \in {\Clocks \cup \NonClocks}}} \Models \EncTr$,
  then
  there is a $\Delay_\AnotherSt \in \RealsNonNeg$
  and a state $\AnotherSt'$
  such that
  ${\AnotherSt \cup \Set{\Delay \mapsto \Delay_\AnotherSt} \cup \Setdef{y' \mapsto \AnotherSt'(y)}{y \in {\Clocks \cup \NonClocks}}} \Models \EncTr$
  and
  $\St' \REquiv \AnotherSt'$.
\end{lemma}
Lemma~\ref{lem:sddfuiet} 
is proven in \RefAppendix .

When the domains of the non-clock variables are finite, as we have assumed,
the set of equivalence classes induced by $\REquiv$ is finite, too.
In this case
we can prove, in a similar fashion as the corresponding lemma in \cite{KindermannJunttilaNiemela:FORTE2012},
that all runs of a \TSshort\  also have corresponding runs whose
projections on the equivalences classes induced by $\REquiv$ are
lasso-shaped looping runs:
\begin{lemma}
\label{lem:weiht}
Let $\Val$ be the set of all valuations over $\NonClocks$ and $\Reg$ the set of clock regions.
  If the \TSshort\ $\STTS$ has an arbitrary infinite run starting in some state $\St_0$, then it also has a run
  run
  $\STTSRun = \St_0 \EStep{\Delay_0} \St_1 \EStep{\Delay_1} \St_2 \EStep{\Delay_2} \ldots$ such that
  for some $\Index,\Bound\in\Naturals$ with $0 \leq \Index \le \Bound \le (\left|\Clocks\right|+\left|\EncFair\right|+2)\cdot\left|\Val\right|\cdot\left|\Reg\right|$ and for every $\AnotherIndex$ with $\AnotherIndex\geq\Index$ we have
  $\St_{\AnotherIndex} \REquiv \St_{\AnotherIndex+\Bound-\Index+1}$.
\end{lemma}

Intuitively, Lemma~\ref{lem:weiht} states that if $\STTS$ has a run starting in a given state, then $\STTS$ 
has a run starting in the same state that begins to loop \emph{through the same regions} 
after a finite prefix. E.g., if $\Index=7$ and $\Bound=10$, then $\St_7\REquiv\St_{11}\REquiv\St_{15}\REquiv\St_{19}\ldots$ and $\St_8\REquiv\St_{12}\REquiv\St_{16}\REquiv\St_{20}\ldots$.
In particular, Lemma~\ref{lem:weiht} implies that if we are interested in whether $\STTS$ has any run at all, it is sufficient to search for runs that are lasso-shaped under the region abstraction.
%
Such runs
can be captured with finite bounded model checking encodings. 
Given a formula $\AnotherFormula$ over $\NonClocks \cup \Clocks \cup \Set{\Delay} \cup \NonClocksNext \cup \ClocksNext$ and an index $\Index \in \Naturals$,
let $\AtStep{\Index}{\AnotherFormula}$ be the the formula over
$\Setdef{\AtStep{\Index}{y}}{y \in \NonClocks \cup \Clocks \cup \Set{\Delay}} \cup \Setdef{\AtStep{\Index+1}{y}}{y \in \NonClocks \cup \Clocks}$
obtained by
replacing each variable $y \in \NonClocks \cup \Clocks \cup \Set{\Delay}$ 
with the 
variable $\AtStep{\Index}{y}$ and
each $y' \in \NonClocksNext \cup \ClocksNext$ with
the 
variable $\AtStep{\Index+1}{y}$.
E.g.,
$\AtStep{3}{((\ClockNext=\Clock+\Delay) \land \neg\AP)}$ is
$(\AtStep{4}{\Clock}=\AtStep{3}{\Clock}+\AtStep{3}{\Delay}) \land \neg\AtStep{3}{\AP}$).
Now the bounded model checking encoding for 
bound $\Bound$ is:
$$
\begin{array}{r@{\ }c@{\ }l}
\Enc{\STTS,\Bound}
&{\Def}&
{\AtStep{0}{\Delay} = 0} \land
\AtStep{0}{\EncInit} \land
\bigwedge_{0 \le \AnotherIndex \le \Bound} \AtStep{\AnotherIndex}{\EncInv} \land {}
\\
&&
\bigwedge_{0 \le \AnotherIndex < \Bound} \AtStep{\AnotherIndex}{\EncTr}
\land
\bigwedge_{1 \le \AnotherIndex \leq \Bound} (\Loop{\AnotherIndex} \Implies \SameRegion{\AnotherIndex})
\land {}
\\
&&
\bigwedge_{0 \le \AnotherIndex < \Bound} ({\AtStep{\AnotherIndex}{\Delay}> 0} \Implies {\AtStep{\AnotherIndex+1}{\Delay}=0})
\land {}
\\
&&
\mathit{Fair}_\Bound \land \mathit{NonZeno}_\Bound
\land\bigvee_{1\le\AnotherIndex\le\Bound} \Loop{\AnotherIndex}
\end{array}
$$
where
(i)
$\SameRegion{\AnotherIndex}$ is a formula evaluating to true
if state $\AnotherIndex-1$ and
state $\Bound$ (i.e. the valuations of the variables with superscripts $\AnotherIndex-1$ and $\Bound$, respectively) are in the same region
(see \cite{KindermannJunttilaNiemela:FORTE2012} for different ways to implement this),
and
(ii)
$\mathit{Fair}_\Bound$ and $\mathit{NonZeno}_\Bound$ are constraints
forcing that the fairness formulas are holding in the loop and
that sufficiently much time passes in the loop to unroll it to a non-zeno run
(again, see \cite{KindermannJunttilaNiemela:FORTE2012}).
Intuitively, the conjuncts of $\Enc{\STTS,\Bound}$ encode the following: (a) the first interval is a \singi\ and satisfies the initial constraint, (b) all intervals satisfy the invariant and all pairs of successive states the transition relation, (c) if some $\Loop{\AnotherIndex}$ holds then state $\AnotherIndex-1$ and state $\Bound$ are in the same region, (d)~there are no two successive open intervals, (e) the fairness formulas are satisfied within the looping part of the trace, (f) the trace is non-zeno and (g) at least one $\Loop{\AnotherIndex}$ is true, meaning that the trace is ``looping under region abstraction''.

Now,
if we wish to find out whether a \TSshort\ $\STTS$ has a run corresponding to a trace $\ITr$ such that $\ITr \Models \Formula$ for a $\ZIMITL$ formula $\Formula$,
we can check whether 
$\Enc{\STTSf{\Formula},\Bound} \land \AtStep{0}{\Enc{\Formula}}$
is satisfiable for some $0 < \Bound \le (\left|\Clocks\right|+\left|\EncFair\right|+2)\cdot\left|\Val\right|\cdot\left|\Reg\right|$. 
This upper bound is
very
large and, in practice, much lower bounds are often used (and sufficient for finding traces). Then, however, the possibility remains that a trace exists despite none being found with the bound used.

\section{Experimental Evaluation}

We have studied the feasibility of the BMC encoding developed in this paper experimentally. We have devised a straightforward implementation of the approach following
 the encoding scheme given in Sect.~\ref{sect:sett} and~\ref{sect:seozf}. With experiments on a class of models we (i) show that it is possible to develop relatively efficient implementations of the approach, (ii) demonstrate that the approach scales reasonably and (iii) are able to estimate the “cost of timing” by comparing the verification of properties using timed operators both to
verifying \ZIMITL\ properties that do not use timing constraints and region-based LTL BMC~\cite{KindermannJunttilaNiemela:FORTE2012,BiereEtAl:TACAS1999}.

\begin{figure}[tp]
\centering
\subfloat[Non-holding property]{
\includegraphics[width=.223\textwidth]{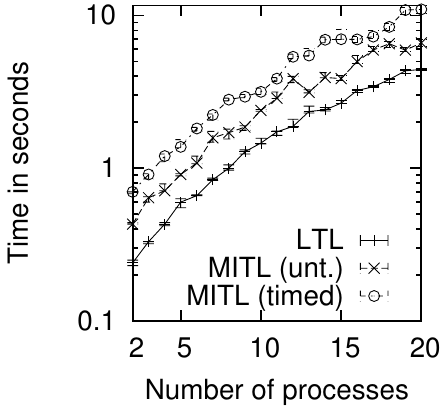}
\label{fig:exp-non}
}%
\subfloat[Holding property]{
\includegraphics[width=.223\textwidth]{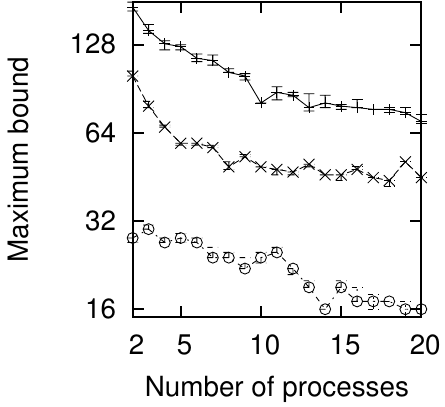}
\label{fig:exp-hold}
}
\caption{Experimental results}
\end{figure}

As a model for the experimentation we used the Fischer mutual exclusion protocol
with two to 20 agents. This protocol is  commonly used for the evaluation of timed verification approaches. The encoding used for the experiments is based on a model that comes with the model checker Uppaal~\cite{Uppaaltutorial:2004} which also uses super-dense time.
We checked one property that holds (``requesting state leads to waiting state eventually'') and one that does not (``there is no trace visiting the critical section and the non-critical section infinitely often'').\footnote{Here, we search for counter-examples, i.e. encode $\neg\Formula$ instead of $\Formula$.} Each property was checked in three variants: as an LTL property using the approach from~\cite{KindermannJunttilaNiemela:FORTE2012}, as the corresponding MITL property (only untimed operators) and with timing constraints added.
Both MITL BMC and LTL BMC were used in an incremental fashion, i.e. bounds are increased starting with bound one until a counter-example is found
and constraints are shared by successive SMT solver calls where possible.
All experiments were run under Linux on Intel Xeon X5650 CPUs limiting memory to 4 GB and CPU time to 20 minutes. As an SMT solver, Yices~\cite{DBLP:conf/cav/DutertreM06} version 1.0.37 was used. All plots report minimum, maximum and median over 11 executions. The implementation and the benchmark used are available on the first author's website.

Figure~\ref{fig:exp-non} shows the time needed for finding a counter-example to the non-holding property.
No timeouts were encountered, even when using the  timed MITL properties.
Figures~\ref{fig:exp-hold} shows the maximum bound reached within 20 minutes when checking the holding property. The bounds reached for the timed property are significantly lower than the bounds reached for the LTL property with the untimed MITL BMC bounds lying between.
While there is both a cost for using the MITL framework for an untimed property and an additional cost for adding timing constraints,
checking timed constraints using MITL BMC is  certainly feasible.
The performance could be further improved using
well-known optimization techniques
e.g. by adding the possibility
for finite counter-examples~\cite{BiereEtAl:LMCS2006}, a technique used in the
LTL BMC implementation used for the experiments. When
verifying properties without timing constraints, using LTL
BMC, however, is advisable not only because of the better
performance but also because a lower bound is needed to
find a trace as open intervals are irrelevant for LTL formulas.

\section{Conclusions}

In this paper, we extend the linear time logic \ZIMITL\ to super-dense time semantics. We devise a method to encode both a timed automaton and a \ZIMITL\ formula as a symbolic transition system.  The encoding provides a foundation for different kinds of fully symbolic verification methods. Soundness and completeness of the encoding are proven in \RefAppendix. 
Furthermore, we demonstrate how the encoding can be employed for bounded model checking (BMC) using the well-known region abstraction. We have implemented the approach. An experimental evaluation of the BMC approach
indicated that a reasonably efficient implementation is feasible.

\section*{Acknowledgements}
This work has been financially supported by the Academy of Finland  under project 128050 and under the Finnish Centre of Excellence in Computational Inference (COIN).





\bibliographystyle{IEEEtran}
\bibliography{paper}

\ifAppendix
\clearpage
\appendix
\newenvironment{relemma}[1]{\renewcommand{\thetheorem}{#1}\begin{lemma}}{\end{lemma}}
\newenvironment{retheorem}[1]{\renewcommand{\thetheorem}{#1}\begin{theorem}}{\end{theorem}}

%
%
%

%
%
\subsection{Duality of until and release operators}

\begin{lemma}\label{Lemma:Duality}
  For any trace $\ITr=\Tuple{\IntervalI{0},\ValuI{0}},\Tuple{\IntervalI{1},\ValuI{1}},\ldots$ over $\APs$, \ZIMITL\ formulas $\Formula$ and $\AnotherFormula$ over $\APs$, $\Index\in\Naturals$, $\Offset\in\IntervalI{\Index}$ it holds that
  $\ITrAt{\Index}{\Offset} \Models (\Formula \SUI{\AnyIneq\Const} \AnotherFormula)$
  iff
  $\ITrAt{\Index}{\Offset} \Models \neg(\neg\Formula \SRI{\AnyIneq\Const} \neg\AnotherFormula)$
\end{lemma}
\begin{IEEEproof}
 $
  \ITrAt{\Index}{\Offset} \Models \neg({\neg\Formula} \SRI{\AnyIneq\Const} {\neg\AnotherFormula}) $
  if and only if (by definition) \\
$  \neg\forall \Point{\Index'}{\Offset'} \in \LaterPoints{\ITr}{\Index}{\Offset}: \big(({\Offset'-\Offset} \AnyIneq \Const) \land \neg(\ITrAt{\Index'}{\Offset'} \Models {\neg\AnotherFormula})\big)
   \Implies \big(\exists \Point{\Index''}{\Offset''} \in \LaterPoints{\ITr}{\Index}{\Offset} : \Point{\Index''}{\Offset''}\Earlier\Point{\Index'}{\Offset'} \land (\ITrAt{\Index''}{\Offset''} \Models \neg\Formula)\big) $ \\
  if and only if (pushing negations inside)
  \\
  $\exists \Point{\Index'}{\Offset'} \in \LaterPoints{\ITr}{\Index}{\Offset} :
  \big(({\Offset'-\Offset} \AnyIneq \Const) \land 
  \neg (\ITrAt{\Index'}{\Offset'} \Models \neg \AnotherFormula)\big)
  \land
  \big(\forall \Point{\Index''}{\Offset''} \in \LaterPoints{\ITr}{\Index}{\Offset}:
  \Point{\Index''}{\Offset''}\Earlier\Point{\Index'}{\Offset'}
  \Implies
  \neg(\ITrAt{\Index''}{\Offset''} \Models \neg\Formula)\big)$
  \\
  if and only if (replacing
    $\neg (\ITrAt{\Index'}{\Offset'} \Models \neg \cdot)\big)$
  by
    $(\ITrAt{\Index'}{\Offset'} \Models \cdot)\big)$
  \\
  $\exists \Point{\Index'}{\Offset'} \in \LaterPoints{\ITr}{\Index}{\Offset}:
  ({\Offset'-\Offset} \AnyIneq \Const)
   \land 
   (\ITrAt{\Index'}{\Offset'} \Models \AnotherFormula)\big)
   \land
   \big(\forall \Point{\Index''}{\Offset''} \in \LaterPoints{\ITr}{\Index}{\Offset}:
   \Point{\Index''}{\Offset''}\Earlier\Point{\Index'}{\Offset'}
   \Implies
   (\ITrAt{\Index''}{\Offset''} \Models \Formula)\big)$
  \\
  if and only if (by definition)
  $\ITrAt{\Index}{\Offset} \Models (\Formula \SUI{\AnyIneq\Const} \AnotherFormula)$.
\end{IEEEproof}

\begin{lemma}\label{Lemma:Duality2}
  For any trace $\ITr=\Tuple{\IntervalI{0},\ValuI{0}},\Tuple{\IntervalI{1},\ValuI{1}},\ldots$ over $\APs$, \ZIMITL\ formulas $\Formula$ and $\AnotherFormula$ over $\APs$, $\Index\in\Naturals$, $\Offset\in\IntervalI{\Index}$ it holds that
  $\ITrAt{\Index}{\Offset} \Models (\Formula \SRI{\AnyIneq\Const} \AnotherFormula)$
  iff
  $\ITrAt{\Index}{\Offset} \Models \neg(\neg\Formula \SUI{\AnyIneq\Const} \neg\AnotherFormula)$
\end{lemma}
\begin{IEEEproof}
  $\ITrAt{\Index}{\Offset} \Models \neg(\neg\Formula \SUI{\AnyIneq\Const} \neg\AnotherFormula)$
  iff (by Lemma~\ref{Lemma:Duality})
  $\ITrAt{\Index}{\Offset} \Models \neg(\neg(\neg\neg\Formula \SRI{\AnyIneq\Const} \neg\neg\AnotherFormula))$
  iff (double negations)
  $\ITrAt{\Index}{\Offset} \Models (\Formula \SRI{\AnyIneq\Const} \AnotherFormula)$.
\end{IEEEproof}

%
%
\subsection{Proof of Lemma~\ref{Lemma:SplitInterval}}

\begin{relemma}{\ref{Lemma:SplitInterval}}
\SplitIntervalText
\end{relemma}
\begin{IEEEproof}
  If $\IntervalI{\Index}$ is a \singi , then the lemma holds trivially.
  Thus, assume that $\IntervalI{\Index}$ is an open interval.
  We have the following four cases.
  \begin{itemize}
  \item
    Assume that $\ITrAt{\Index}{\Offset} \Models {\Formula \SUI{\AnyUpper\Const} \AnotherFormula}$.
    Thus there exists a $\Point{\Index'}{\Offset'} \in \LaterPoints{\ITr}{\Index}{\Offset}$ such that
    $({\Offset' - \Offset} \AnyUpper \Const) \land 
    (\ITrAt{\Index'}{\Offset'} \Models \AnotherFormula)
    \land
    \big(\forall \Point{\Index''}{\Offset''} \in \LaterPoints{\ITr}{\Index}{\Offset},
    \Point{\Index''}{\Offset''} \Earlier \Point{\Index'}{\Offset'}
    \Implies
    (\ITrAt{\Index''}{\Offset''} \Models \Formula)\big)$.
    Let $\AnotherOffset \ge \Offset$ with $\AnotherOffset \in \IntervalI{\Index}$.
    Now $\LaterPoints{\ITr}{\Index}{\AnotherOffset} \subseteq \LaterPoints{\ITr}{\Index}{\Offset}$.

    If $\Index' > \Index$ or
    ${\Index' = \Index} \land {\AnotherOffset < \Offset'}$,
    then 
    $({\Offset' - \AnotherOffset} \AnyUpper \Const) \land 
     (\ITrAt{\Index'}{\Offset'} \Models \AnotherFormula)
     \land
     \big(\forall \Point{\Index''}{\Offset''} \in \LaterPoints{\ITr}{\Index}{\AnotherOffset},
     \Point{\Index''}{\Offset''} \Earlier \Point{\Index'}{\Offset'}
     \Implies
     (\ITrAt{\Index''}{\Offset''} \Models \Formula)\big)$,
    implying
    $\ITrSuffix{\ITr}{\Index}{\AnotherOffset} \Models {\Formula \SUI{\AnyUpper\Const} \AnotherFormula}$ irrespective whether $\ITr$ is fine for $\Formula$ and $\AnotherFormula$ or not.

    If $\Index' = \Index$ and $\AnotherOffset \ge \Offset'$,
    then
    there is a $\AnotherOffset' > \AnotherOffset$ with
    $\AnotherOffset' \in \IntervalI{\Index}$ and
    ${\AnotherOffset' - \AnotherOffset} \AnyUpper \Const$
    as $\IntervalI{\Index}$ is an open interval.
    As $\ITr$ is fine for $\AnotherFormula$ and 
    $\ITrAt{\Index}{\Offset'} \Models \AnotherFormula$,
    it holds that
    $\ITrAt{\Index}{\AnotherOffset'} \Models \AnotherFormula$ as well.
    As
    $\forall \Point{\Index}{\Offset''} \in \LaterPoints{\ITr}{\Index}{\Offset},
    \Point{\Index}{\Offset''} \Earlier \Point{\Index}{\Offset'}
    \Implies
    (\ITrAt{\Index}{\Offset''} \Models \Formula)$,
    there is at least one $\Offset < \Offset'' < \Offset'$,
    and
    $\ITr$ is fine for $\Formula$,
    we have
    $\forall \Point{\Index}{\AnotherOffset''} \in \LaterPoints{\ITr}{\Index}{\AnotherOffset},
    \Point{\Index}{\AnotherOffset''} \Earlier \Point{\Index}{\AnotherOffset'}
    \Implies
    (\ITrAt{\Index}{\AnotherOffset''} \Models \Formula)$.    
    Therefore, $\ITrSuffix{\ITr}{\Index}{\AnotherOffset} \Models {\Formula \SUI{\AnyUpper\Const} \AnotherFormula}$.

  \item
    Assume that
    $\ITrAt{\Index}{\Offset} \Models
     {\Formula \SUI{\AnyLower\Const} \AnotherFormula}$.
    Thus there exists a $\Point{\Index'}{\Offset'} \in \LaterPoints{\ITr}{\Index}{\Offset}$ such that
    $({\Offset' - \Offset} \AnyLower \Const) \land 
    (\ITrAt{\Index'}{\Offset'} \Models \AnotherFormula)
    \land
    \big(\forall \Point{\Index''}{\Offset''} \in \LaterPoints{\ITr}{\Index}{\Offset},
    \Point{\Index''}{\Offset''} \Earlier \Point{\Index'}{\Offset'}
    \Implies
    (\ITrAt{\Index''}{\Offset''} \Models \Formula)\big)$.
    Let $\AnotherOffset \le \Offset$ with $\AnotherOffset \in \IntervalI{\Index}$.
    Thus ${\Offset' - \AnotherOffset} \AnyLower \Const$.
    Because
    (i)
    $\forall \Point{\Index''}{\Offset''} \in \LaterPoints{\ITr}{\Index}{\Offset},
    \Point{\Index''}{\Offset''} \Earlier \Point{\Index'}{\Offset'}
    \Implies
    (\ITrAt{\Index''}{\Offset''} \Models \Formula)$,
    (ii)
    there is at least one $\Offset < \Offset'' < \Offset'$ with $\Offset'' \in \IntervalI{\Index}$ as $\IntervalI{\Index}$ is open,
    and
    (iii)
    $\ITr$ is fine for $\Formula$,
    we have
    $\forall \Point{\Index''}{\Offset''} \in \LaterPoints{\ITr}{\Index}{\AnotherOffset},
    \Point{\Index''}{\Offset''} \Earlier \Point{\Index'}{\Offset'}
    \Implies
    (\ITrAt{\Index''}{\Offset''} \Models \Formula)$.    
    Therefore, $\ITrSuffix{\ITr}{\Index}{\AnotherOffset} \Models {\Formula \SUI{\AnyUpper\Const} \AnotherFormula}$.

  \item
    Assume that
    $\ITrSuffix{\ITr}{\Index}{\Offset} \Models {\Formula \SRI{\AnyUpper\Const} \AnotherFormula}$.
    Thus
    $\forall \Point{\Index'}{\Offset'} \in \LaterPoints{\ITr}{\Index}{\Offset},
   \big(({\Offset'-\Offset} \AnyUpper \Const) \land 
   \neg(\ITrAt{\Index'}{\Offset'} \Models \AnotherFormula)\big)
   \Implies
   \big(\exists \Point{\Index''}{\Offset''} \in \LaterPoints{\ITr}{\Index}{\Offset},
   \Point{\Index''}{\Offset''}\Earlier\Point{\Index'}{\Offset'}
   \land
   (\ITrAt{\Index''}{\Offset''} \Models \Formula)\big)$.
   Let $\AnotherOffset \le \Offset$ with $\AnotherOffset \in \IntervalI{\Index}$.

   Suppose that
   $({\AnotherOffset'-\Offset'} \AnyUpper \Const) \land 
    \neg(\ITrAt{\AnotherIndex'}{\AnotherOffset'} \Models \AnotherFormula)$
   for some $\Point{\AnotherIndex}{\AnotherOffset'} \in \LaterPoints{\ITr}{\Index}{\AnotherOffset}$.
   If $\Point{\Index}{\Offset} \Earlier \Point{\AnotherIndex'}{\AnotherOffset'}$,
   then 
   $\exists \Point{\Index''}{\Offset''} \in \LaterPoints{\ITr}{\Index}{\Offset},
   \Point{\Index''}{\Offset''}\Earlier\Point{\AnotherIndex'}{\AnotherOffset'}
   \land
   (\ITrAt{\Index''}{\Offset''} \Models \Formula)$.
   On the other hand,
   if $\Point{\AnotherIndex'}{\AnotherOffset'} = \Point{\Index}{\Offset}$
   or
   $\Point{\AnotherIndex'}{\AnotherOffset'} \Earlier \Point{\Index}{\Offset}$,
   then
   $\AnotherIndex' = \Index$,
   $\neg(\ITrAt{\Index}{\ThirdOffset} \Models \AnotherFormula)$
   for all $\ThirdOffset \in \IntervalI{\Index}$ as $\ITr$ is fine for $\AnotherFormula$,
   there is a $\Point{\Index}{\Offset''} \in \LaterPoints{\ITr}{\Index}{\Offset},
   \Point{\Index}{\Offset} \Earlier \Point{\Index}{\Offset''}
   \land
   (\ITrAt{\Index}{\Offset''} \Models \Formula)$
   as $\IntervalI{\Index}$ is open,
   $\ITrAt{\Index}{\ThirdOffset} \Models \Formula$
   for all $\ThirdOffset \in \IntervalI{\Index}$ as $\ITr$ is fine for $\Formula$,
   and
   $\ITrSuffix{\ITr}{\Index}{\AnotherOffset} \Models {\Formula \SRI{\AnyUpper\Const} \AnotherFormula}$.

 \item
   Assume that
   $\ITrSuffix{\ITr}{\Index}{t} \Models
    {\Formula \SRI{\AnyLower\Const} \AnotherFormula}$.
   Thus
   $\forall \Point{\Index'}{\Offset'} \in \LaterPoints{\ITr}{\Index}{\Offset},
    \big(({\Offset'-\Offset} \AnyLower \Const) \land 
         \neg(\ITrAt{\Index'}{\Offset'} \Models \AnotherFormula)\big)
    \Implies
    \big(\exists \Point{\Index''}{\Offset''} \in \LaterPoints{\ITr}{\Index}{\Offset},
         \Point{\Index''}{\Offset''}\Earlier\Point{\Index'}{\Offset'}
         \land
         (\ITrAt{\Index''}{\Offset''} \Models \Formula)\big)$.
    Let $\AnotherOffset \ge \Offset$ with $\AnotherOffset \in \IntervalI{\Index}$.
    Suppose that 
    $({\AnotherOffset'-\AnotherOffset} \AnyLower \Const) \land 
     \neg(\ITrAt{\AnotherIndex'}{\AnotherOffset'} \Models \AnotherFormula)$
    for some
    $\Point{\AnotherIndex'}{\AnotherOffset'} \in \LaterPoints{\ITr}{\Index}{\AnotherOffset}$.
    As ${\AnotherOffset'-\Offset}\AnyLower \Const$,
    there exists a
    $\Point{\Index''}{\Offset''} \in \LaterPoints{\ITr}{\Index}{\Offset}$
    such that
    $\Point{\Index''}{\Offset''}\Earlier\Point{\Index'}{\Offset'}
     \land
     (\ITrAt{\Index''}{\Offset''} \Models \Formula)\big)$.
    If $\Point{\AnotherIndex'}{\AnotherOffset'} \Earlier \Point{\Index''}{\Offset''}$, we are done.
    On the other hand,
    if 
    $\Point{\Index''}{\Offset''} = \Point{\AnotherIndex'}{\AnotherOffset'}$ or
    $\Point{\Index''}{\Offset''} \Earlier \Point{\AnotherIndex'}{\AnotherOffset'}$,
    then 
    $\ITrAt{\Index}{\ThirdOffset} \Models \Formula$ for all $\ThirdOffset \in \IntervalI{\Index}$ as $\ITr$ is fine for $\Formula$,
    there exists a $\Point{\Index}{\AnotherIndex''}$ such that
    $\Point{\Index}{\AnotherIndex} \Earlier \Point{\Index}{\AnotherIndex''} \Earlier \Point{\AnotherIndex'}{\AnotherOffset'}$
    as $\IntervalI{\Index}$ is open,
    and
    thus
    $\ITrSuffix{\ITr}{\Index}{\AnotherOffset} \Models {\Formula \SRI{\AnyLower\Const} \AnotherFormula}$.
  \end{itemize}
\end{IEEEproof}

%
%
\subsection{Proof of Lemma~\ref{Lemma:ExistenceOfFineRefinement}}

\begin{relemma}{\ref{Lemma:ExistenceOfFineRefinement}}
\ExistenceOfFineRefinementText
\end{relemma}
\begin{IEEEproof}
  Let $[\Formula_1,...,\Formula_n]$ be a list containing all the
  sub-formulas of $\Formula$
  so that the sub-formulas of a sub-formula $\Formula_i$ are listed before
  $\Formula_i$.
  Thus $\Formula_1$ is an atomic proposition and $\Formula_n = \Formula$.

  We now construct a trace $\ITr_i$ for each $1 \le i \le n$
  such that
  $\ITr_i$ is fine for all sub-formulas $\Formula_j$ with $1 \le j \le i$.

  If $\Formula_i$ is an atomic proposition or
  of forms $\neg\Formula_j$, $\Formula_j \land \Formula_k$,
  or $\Formula_j \lor \Formula_k$ with $j,k < i$,
  then $\ITr_i = \ITr_{i-1}$ is fine for $\Formula_i$ as well.
  
  If $\Formula_i$ is an until or release formula 
  of forms
  $\Formula_j \SUI{\AnyIneq\Const} \Formula_k$ or
  $\Formula_j \SRI{\AnyIneq\Const} \Formula_k$,
  then
  by (i) recalling that $\ITr_{i-1}$ is fine for $\Formula_j$ and $\Formula_k$
  (ii) applying Lemma~\ref{Lemma:SplitInterval},
  we obtain a $\Formula_i$-fine trace $\ITr_i$
  by splitting each open interval in $\ITr_{i-1}$ into at most
  two new open intervals and one \singiadj\ interval.
\end{IEEEproof}

%
%
\subsection{Proof of Lemma \ref{lem:ubeq}}

\begin{relemma}{\ref{lem:ubeq}}
\UbeqText
\end{relemma}
\begin{IEEEproof}
  Recall that
  $\ITrAt{\Index}{\Offset} \Models (\Formula \SUI{\AnyIneq\Const} \AnotherFormula)$
  iff
  $\exists \Point{\Index'}{\Offset'} \in \LaterPoints{\ITr}{\Index}{\Offset} :
   ({\Offset' - \Offset} \AnyIneq \Const) \land 
   (\ITrAt{\Index'}{\Offset'} \Models \AnotherFormula)
    \land
    \big(\forall \Point{\Index''}{\Offset''} \in \LaterPoints{\ITr}{\Index}{\Offset}:
         \Point{\Index''}{\Offset''} \Earlier \Point{\Index'}{\Offset'}
         \Implies
         (\ITrAt{\Index''}{\Offset''} \Models \Formula)\big)$.
  \begin{itemize}
  \item
    The ``$\Rightarrow$'' part.

    As is easy to see from the semantics,
    $\ITrAt{\Index}{\Offset} \Models (\Formula \SUI{\AnyIneq\Const} \AnotherFormula)$ implies
    both 
    (i) $\ITrAt{\Index}{\Offset} \Models (\Formula \SU \AnotherFormula)$
    and
    (ii) $\ITrAt{\Index}{\Offset} \Models (\True \SUI{\AnyIneq\Const} \AnotherFormula)$
    corresponding to $\ITrAt{\Index}{\Offset} \Models \SFI{\AnyIneq\Const}\AnotherFormula$.
  \item
    The ``$\Leftarrow$'' part.

    By the semantics,
    if $\ITrAt{\Index}{\Offset} \Models (\Formula \SU \AnotherFormula)$
    we can pick a
    $\Point{\Index'}{\Offset'} \in \LaterPoints{\ITr}{\Index}{\Offset}$
    such that
    $
      ({\Offset' - \Offset} \ge 0) \land 
      (\ITrAt{\Index'}{\Offset'} \Models \AnotherFormula)
      \land
      \big(\forall \Point{\Index''}{\Offset''} \in \LaterPoints{\ITr}{\Index}{\Offset}:
        \Point{\Index''}{\Offset''} \Earlier \Point{\Index'}{\Offset'}
        \Implies
        (\ITrAt{\Index''}{\Offset''} \Models \Formula)\big)
    $.
    We have two cases now:
    \begin{itemize}
    \item
      If ${\Offset' - \Offset} \AnyUpper\Const$,
      then we immediately have
      $\ITrAt{\Index}{\Offset} \Models (\Formula \SUI{\AnyUpper\Const} \AnotherFormula)$.
    \item
      Otherwise,
      $\ITrAt{\Index}{\Offset} \Models \SFI{\AnyUpper\Const} \AnotherFormula$
      allows us to pick
      $\Point{\AnotherIndex'}{\AnotherOffset'} \in \LaterPoints{\ITr}{\Index}{\Offset}$
      such that
      $
      (\AnotherOffset'-\Offset \AnyUpper \Const) \land 
      (\ITrAt{\AnotherIndex'}{\AnotherOffset'} \Models \AnotherFormula)$.
      As $\AnotherOffset'-\Offset \AnyUpper \Const$,
      we know that $\Point{\AnotherIndex'}{\AnotherOffset'} \Earlier \Point{\Index'}{\Offset'}$,
      which in turn implies that
      $\forall \Point{\Index''}{\Offset''} \in \LaterPoints{\ITr}{\Index}{\Offset}:
      \Point{\Index''}{\Offset''} \Earlier \Point{\AnotherIndex'}{\AnotherOffset'}
      \Implies
      (\ITrAt{\Index''}{\Offset''} \Models \Formula)$.
      Thus we obtain
      $\ITrAt{\Index}{\Offset} \Models (\Formula \SUI{\AnyUpper\Const} \AnotherFormula)$.
    \end{itemize}
  \end{itemize}
\end{IEEEproof}

\subsection{Soundness proofs}

\begin{relemma}{\ref{Lemma:constraints_enforce_formula}}
\CenfFText
\end{relemma}

Recall, that we call an encoding $\STTSf{\Formula}$ is sound
if the following are satisfied:
\SoundDef

\newcommand{\SEncO}{\STTSf{\Op \LSF}}
\newcommand{\SEncT}{\STTSf{\LSF \Op \RSF}}
\newcommand{\Op}{\AnyTemporalOp}
\newcommand{\EL}{\Enc{\LSF}}
\newcommand{\ER}{\Enc{\RSF}}

We will now prove Lemma~\ref{Lemma:constraints_enforce_formula} separately for each operator.
Note, that as we assumed $\STTS$ to be sound for $\LSF$ and $\RSF$, we know that any point on a run where $\EL$ holds satisfies $\LSF$ and any point where $\ER$ holds satisfies $\RSF$, which will be used in the proofs without being mentioned explicitly every single time.

\newcommand{\Holds}[2]{\St_{#2}(#1)=\True}
\newcommand{\HoldsNot}[2]{\St_{#2}(#1)=\False}
\newcommand{\Hold}[3]{\St_{#3}(#1)=\St_{#3}(#2)=\True}
\newcommand{\ThreeHoldAt}[4]{\St_{#4}(#1)=\St_{#4}(#2)=\St_{#4}(#3)=\True}
\newcommand{\HoldNot}[3]{\St_{#3}(#1)=\St_{#3}(#2)=\False}
\newcommand{\HoldsFromTo}[3]{\St_{#2}(#1)=\ldots=\St_{#3}(#1)=\True}
\newcommand{\HoldsNotFromTo}[3]{\St_{#2}(#1)=\ldots=\St_{#3}(#1)=\False}
\newcommand{\HoldsFrom}[2]{\St_{#2}(#1)=\St_{#2+1}(#1)=\ldots=\True}
\newcommand{\HoldsNotFrom}[2]{\St_{#2}(#1)=\St_{#2+1}(#1)=\ldots=\False}
\newcommand{\FOUT}{\LSF\Op\RSF}
\renewcommand{\OUT}{\Enc{\FOUT}}
\begin{IEEEproof}
\emph{For ${\Op} \in\Set{{\land},{\lor}}$.} Clearly, all runs are preserved. Also, by the constraint that $\OUT\Iff\EL\Op\ER$, it immediately follows that for any $\Index\in\Naturals$ we have $\forall \Offset \in \IntervalI{\Index} : \St_{\Index}(\OUT)\Implies ( \ITrAt{\Index}{\Offset} \Models \FOUT)$.
\end{IEEEproof}

\renewcommand{\Op}{\SU}
\newcommand{\SEnc}{\SEncT}
\newcommand{\SoundProofIntro}{
Now take a $\SEnc$ run $\STTSRun = \St_0 \St_1 \St_2 \ldots$ 
  with $\TraceOf{\STTSRun} = \ITr=\Tuple{\IntervalI{0}\ValuI{0}}\Tuple{\IntervalI{1},\ValuI{1}}\ldots$, $\Index\in\Naturals$ and $\Offset\in\IntervalI{\Index}$ with $\St_{\Index}(\OUT)=\True$.
It remains to show that $\ITrSuffix{\ITr}{\Index}{\Offset}\models\FOUT$.
}

\begin{IEEEproof}
\emph{For ${\AnyTemporalOp}={\SU}$.}
Clearly, all the traces are preserved in $\SEncT$ as setting $\OUT$ to $\False$ leads to both constraints being satisfied regardless of the trace.

\SoundProofIntro

If $\IntervalI{\Index}$ is open, then by Constraint~\ref{m:enc-u-open} we know that $\Holds{\EL}{\Index}$.
Furthermore, there are three possibilities (multiple of which may be applicable):
\begin{enumerate}
\item $\Holds{\ER}{\Index}$. 
In this case we can pick any future time point on the open interval $\Index$ and demonstrate that $\RSF$ holds there and $\LSF$ holds up to that point, meaning that $\ITrSuffix{\ITr}{\Index}{\Offset}\models\FOUT$.
\item $\Holds{\ER}{\Index+1}$. 
As $\IntervalI{\Index}$ is open, we know that $\IntervalI{\Index+1}$ is a \singi . Furthermore, as $\Holds{\EL}{\Index}$ we know that $\LSF$ holds up to the single time point constituting $\IntervalI{\Index+1}$. Hence, $\ITrSuffix{\ITr}{\Index}{\Offset}\models\FOUT$.
\item 
$\Holds{\EL}{\Index}$
and
$\St_{\Index}(\ER)=\St_{\Index+1}(\ER)=\False$.
By Constraint~\ref{m:enc-u-open}, then $\Holds{\OUT}{\Index+1}$.
By the fairness constraint $\EncFair_{\FOUT}$
we know that there is a future interval on which either $\ER$ holds or $\OUT$ does not hold. Pick $\AnotherIndex>\Index+1$ as small as possible, such that $\Holds{\ER}{\AnotherIndex}$ or $\HoldsNot{\OUT}{\AnotherIndex}$.
Now
$\HoldsFromTo{\OUT}{\Index}{\AnotherIndex-1}$
and
$\HoldsNotFromTo{\ER}{\Index}{\AnotherIndex-1}$.
Note that the only way to satisfy Constraints~\ref{m:enc-u-open} and~\ref{m:enc-u-singular} on intervals $\Index,\ldots,\AnotherIndex-2$ now is by $\EL'$ holding on those intervals, meaning that $\HoldsFromTo{\EL}{\Index+1}{\AnotherIndex-1}$.
Now
\begin{itemize}
\item \label{dhTUew} 
If $\IntervalI{\AnotherIndex-1}$ is open, then by Constraint~\ref{m:enc-u-open} we know that $\Holds{\ER}{\AnotherIndex}$ or $\Holds{\OUT}{\AnotherIndex}$. As we picked $\AnotherIndex$ so that $\Holds{\ER}{\AnotherIndex}$ or $\HoldsNot{\OUT}{\AnotherIndex}$, we know that $\Holds{\ER}{\AnotherIndex}$ (meaning that $\RSF$ holds at interval $\AnotherIndex$) in either case. Furthermore, as $\IntervalI{\AnotherIndex-1}$ is open we know that $\IntervalI{\AnotherIndex}$ is a \singi , implying that $\EL$ (and thus $\LSF$) holds anywhere in between $\Point{\Index}{\Offset}$ and $\IntervalI{\AnotherIndex}$.
Thus, $\ITrSuffix{\ITr}{\Index}{\Offset}\models\FOUT$.
\item 
If $\IntervalI{\AnotherIndex-1}$ is a \singi , then by Constraint~\ref{m:enc-u-singular} we know that either $\Holds{\ER}{\AnotherIndex}$ and $\IntervalI{\AnotherIndex}$ is a \singi\ or $\Hold{\OUT}{\EL}{\AnotherIndex}$.
Again, by the choice of $\AnotherIndex$ we know that $\Holds{\OUT}{\AnotherIndex}$ implies that $\Holds{\ER}{\AnotherIndex}$. Thus, there is in either case a time point in interval $\AnotherIndex$ at which $\ER$ (and thus $\RSF$) holds such that $\EL$ (and thus $\LSF$) holds anywhere in between $\Point{\Index}{\Offset}$ and that time point.
Hence, $\ITrSuffix{\ITr}{\Index}{\Offset}\models\FOUT$.
\end{itemize} 
\end{enumerate} 

If, in contrast, $\IntervalI{\Index}$ is a \singi , then by Constraint~\ref{m:enc-u-singular} there are two possibilities:
\begin{enumerate}
\item 
$\IntervalI{\Index+1}$ is a \singi\ and $\Holds{\ER}{\Index+1}$. In this case, trivially $\ITrSuffix{\ITr}{\Index}{\Offset}\models\FOUT$.
\item 
 $\Hold{\EL}{\OUT}{\Index+1}$. If additionally $\Holds{\ER}{\Index+1}$, then $\ITrSuffix{\ITr}{\Index}{\Offset}\models\FOUT$ indiscriminately of whether $\IntervalI{\Index+1}$ is a \singi\ or an open interval. If, in contrast, $\HoldsNot{\ER}{\Index+1}$, then we can, again, pick $\AnotherIndex>\Index+1$ as small as possible, such that $\Holds{\ER}{\AnotherIndex}$ or $\HoldsNot{\OUT}{\AnotherIndex}$. By proceeding precisely in the same way as in the Case~\ref{dhTUew} for open $\IntervalI{\Index}$, we can again deduce that $\ITrSuffix{\ITr}{\Index}{\Offset}\models\FOUT$.
\end{enumerate} 

Thus, $\ITrSuffix{\ITr}{\Index}{\Offset}\models\FOUT$ holds in each of the described cases.
\end{IEEEproof}

\renewcommand{\Op}{\SFI{\leq 0}}
\renewcommand{\FOUT}{\Op\LSF}
\renewcommand{\SEnc}{\SEncO}

\begin{IEEEproof}
\emph{For ${\AnyTemporalOp}={\SFI{\leq 0}}$.}

Again, the ``preservation of traces'' property follows from the fact that the constraint is trivially satisfied globally when $\OUT$ is set to $\False$ globally.

\SoundProofIntro

By Constraint~\ref{eq:f.leqz}, we now know that $\IntervalI{\Index}$ and $\IntervalI{\Index+1}$ are both \singis . This, in particular, means that
$\IntervalI{\Index+1}=[\Offset,\Offset]$.
Furthermore, by Constraint~\ref{eq:f.leqz} one of the following holds:
\begin{itemize}
\item $\Holds{\EL}{\Index+1}$. In this case, $\ITrSuffix{\ITr}{\Index}{\Offset}\models\FOUT$ trivially.
\item $\Holds{\OUT}{\Index+1}$. In this case, $\IntervalI{\Index+2}$ is a \singi\ as well and again $\Holds{\EL}{\Index+2}$ or $\Holds{\OUT}{\Index+2}$. Applying this argument repeatedly leads to the conclusion, that there needs to be an interval on which $\EL$ holds before the next open interval.
The fact that $\ITr$ is non-zeno, furthermore, implies that there is a future open interval.
Thus, we can conclude that there is a sequence of \singiadj\ intervals starting at interval $\Index$ such that $\EL$ (and thus $\LSF$) holds on the last interval in that sequence.
Thus, $\ITrSuffix{\ITr}{\Index}{\Offset}\models\FOUT$.
\end{itemize}
In both cases we were able to demonstrate that $\ITrSuffix{\ITr}{\Index}{\Offset}\models\FOUT$.
\end{IEEEproof}


\renewcommand{\Op}{\SFI{<\Const}}
\renewcommand{\FOUT}{\Op\LSF}
\renewcommand{\SEnc}{\SEncO}

\begin{IEEEproof}
\emph{For ${\AnyTemporalOp}={\SFI{<\Const}}$.}
The ``preservation of traces'' property follows from the fact that the constraints can easily be satisfied globally when $\OUT$ is set to $\False$ globally.

\SoundProofIntro

Choose $\AnotherIndex\in\Naturals$ as large as possible such that $0\leq\AnotherIndex\leq\Index$ and either $\EReset$ holds at interval $\AnotherIndex-1$ or $\AnotherIndex=0$. 
We now know that (i) $\St_\AnotherIndex(\C)=0$ and (ii) $\Holds{\LeftOpen}{\AnotherIndex}$ iff $\IntervalI{\AnotherIndex}$ is open.
Let $\Strict\AnotherIndex=\AnotherIndex+1$ if interval $\AnotherIndex$ is a \singi\ and $\Strict\AnotherIndex=\AnotherIndex$ otherwise.
If now $\AnotherIndex<\Index$, then 
we know that $\EReset$ does not hold on interval $\Index-1$ meaning that $\Holds{\OUT}{\Index-1}$, $\HoldsNot{\EL}{\Index}$ and $\HoldsNot{\EL}{\Index-1}$ if $\IntervalI{\Index-1}$ is open. Applying the same reasoning repeatedly, we can deduce that, firstly, $\HoldsFromTo{\OUT}{\AnotherIndex}{\Index}$ and, secondly, $\HoldsNotFromTo{\EL}{\Strict\AnotherIndex}{\Index}$.

Let $\Strict\Index=\Index+1$ if $\IntervalI{\Index}$ is a \singi\ and $\Strict\Index=\Index$ otherwise.
Now assume that $\HoldsNotFrom{\EL}{\Strict\Index}$.
In this case Constraints~\ref{eq:enc.f.ub.open} and~\ref{eq:enc.f.ub.singular}
imply that $\HoldsFrom{\OUT}{\Strict\Index}$.
Thus $\EReset$ is $\False$ on intervals $\Strict\Index,\Strict\Index+1,\ldots$.
As we assumed a non-zeno trace, this implies that there is no upper bound to the value of $\C$ on the intervals $\Strict\Index,\Strict\Index+1,\ldots$ implying that $\Timing$ eventually becomes $\False$ on all intervals starting from some interval after interval $\Strict\Index$. As now $\OUT$ holds globally and $\EL$ globally does not hold starting from interval $\Strict\Index$, this contradicts Constraint~\ref{eq:enc.f.ub.timing}. Thus, assuming there is no point at which $\EL$ holds after interval $\Strict\Index$ leads to a contradiction, implying that there has to be a point where $\EL$ holds.
Thus, we can
pick $\AThirdIndex\geq\Strict\Index$ as small as possible such that $\Holds{\EL}{\AThirdIndex}$. %

\begin{itemize}
\item 
Case 1: $\Index=\AThirdIndex$.
As $\AThirdIndex\geq\Strict\Index$, this implies that  $\IntervalI{\Index}$ is open.
As, furthermore, $\Holds{\EL}{\AThirdIndex}$ and thus $\Holds{\EL}{\Index}$ we know that
$\ITrSuffix{\ITr}{\Index}{\Offset}\models{\SFI{<\Const}}\EL$.
\item 
Case 2: $\Index\neq\AThirdIndex$ and $\AThirdIndex=\AnotherIndex+1$.
As $\AnotherIndex\leq\Index\leq\AThirdIndex$ we know know that $\Index=\AnotherIndex$.
Now
\begin{itemize}
\item
If $\IntervalI{\Index}$ is a \singi , then 
$\ITrSuffix{\ITr}{\Index}{\Offset}\models{\SFI{<\Const}}\EL$ trivially holds.
\item
If $\IntervalI{\Index}$ is open, then
$\Strict\Index=\Index$.
Thus, by the choice of $\AThirdIndex$ we know that $\HoldsNot{\EL}{\Index}$. 
 As $\Holds{\OUT}{\Index}$, Constraint~\ref{eq:enc.f.ub.timing} implies that $\Timing$ holds at interval $\Index$.
As $\AnotherIndex=\Index$, we know that $\St_\Index(\C)=0$.
By the definition of $\Timing$, we now know that the value of $\D$ at interval $\Index$ is less than or equal to $\Const$. This together with the fact that $\IntervalI{\Index}$ is open implies we can pick a point in $\IntervalI{\AThirdIndex}$ that is less than $\Const$ time units away from $\Point{\Index}{\Offset}$ and, ultimately, that $\ITrSuffix{\ITr}{\Index}{\Offset}\models{\SFI{<\Const}}\EL$.
\end{itemize}
\item 
Case 3: $\Index\neq\AThirdIndex$ and $\AThirdIndex\geq\AnotherIndex+2$.
By $\HoldsNotFromTo{\EL}{\Strict\Index}{\AThirdIndex-1}$ and
by Constraints~\ref{eq:enc.f.ub.open} and~\ref{eq:enc.f.ub.singular}, we now know that $\HoldsFromTo{\OUT}{\Index+1}{\AThirdIndex-1}$.
Together with our previous observations 
we now know that $\HoldsNotFromTo{\EL}{\Strict\AnotherIndex}{\AThirdIndex-1}$ and $\HoldsFromTo{\OUT}{\AnotherIndex}{\AThirdIndex-1}$.
This implies that $\EReset$ does not hold on intervals $\AnotherIndex,\ldots,\AThirdIndex-2$, in turn implying that $\C$ and $\LeftOpen$ are updated according to Constraint~\ref{eq:enc.f.ub.clk.update} on the transitions from intervals $\AnotherIndex,\ldots,\AThirdIndex-2$ to the respective following interval. Therefore, $\St_{\AThirdIndex-1}(\C)$ is the difference between the left bound of $\IntervalI{\AnotherIndex}$ and the left bound of $\IntervalI{\AThirdIndex-1}$.
Thus, $\St_{\AThirdIndex-1}(\C)+\St_{\AThirdIndex-1}(\D)$ is the difference between the left bound of $\IntervalI{\AThirdIndex}$ and the left bound of $\IntervalI{\AnotherIndex}$.
As $\Holds{\OUT}{\AThirdIndex-1}$ and $\HoldsNot{\EL}{\AThirdIndex-1}$, Constraint~\ref{eq:enc.f.ub.timing} implies that $\Timing$ holds at interval $\AThirdIndex-1$. Now
\begin{itemize}
\item If $\IntervalI{\AnotherIndex}$ is open, then the difference between the left bounds of $\IntervalI{\AThirdIndex}$ and $\IntervalI{\AnotherIndex}$ is less than or equal to $\Const$. Then we can for every point in $\IntervalI{\AnotherIndex}$ pick another point in $\IntervalI{\AThirdIndex}$ that is less than $\Const$ time units away from the point in $\IntervalI{\AnotherIndex}$.
\item If $\IntervalI{\AnotherIndex}$ is a \singi , then the difference between the left bounds of $\IntervalI{\AThirdIndex}$ and $\IntervalI{\AnotherIndex}$ is less than $\Const$. Again, this means that we can for every point in $\IntervalI{\AnotherIndex}$ pick another point in $\IntervalI{\AThirdIndex}$ that is less than $\Const$ time units away from the point in $\IntervalI{\AnotherIndex}$.
\end{itemize}
Finally, as $\AnotherIndex\leq\Index\leq\AThirdIndex$,  we can also for $\Point{\Index}{\Offset}$ pick a point in $\IntervalI{\AThirdIndex}$ that is less than $\Const$ time units away. As $\Holds{\EL}{\AThirdIndex}$, this implies that
$\ITrSuffix{\ITr}{\Index}{\Offset}\models\FOUT$.
\end{itemize}
In each case we were able to demonstrate that
$\ITrSuffix{\ITr}{\Index}{\Offset}\models\FOUT$.
\end{IEEEproof}


\renewcommand{\Op}{\SFI{\leq\Const}}
\renewcommand{\FOUT}{\Op\LSF}
\renewcommand{\SEnc}{\SEncO}

\begin{IEEEproof}
\emph{For ${\AnyTemporalOp}={\SFI{\leq\Const}}$.}
The proof for ${\AnyTemporalOp}={\Op}$ proceeds precisely as the proof for ${\AnyTemporalOp}={\SFI{<\Const}}$, except for arguing that there are time points $\leq\Const$ time units apart in $\IntervalI{\AnotherIndex}$ and $\IntervalI{\AThirdIndex}$ in Case 3.

By the definition of $\Timing$ for ${\AnyUpper}={<}$ and as $\Timing$ holds at interval $\AThirdIndex-1$ we know that in Case 3 one of the following:
\begin{itemize}
\item The difference between the left bound of $\IntervalI{\AnotherIndex}$ is less than $\Const$ time units. In this case, we can trivially pick for any point in in $\IntervalI{\AnotherIndex}$ a point that is $\leq\Const$ time units away in $\IntervalI{\AThirdIndex}$.
\item The difference between the left bound of $\IntervalI{\AnotherIndex}$ is $\Const$ time units and $\IntervalI{\AnotherIndex}$ open or $\IntervalI{\AThirdIndex}$ is a \singi . Again we can pick for any point in in $\IntervalI{\AnotherIndex}$ a point that is $\leq\Const$ time units away in $\IntervalI{\AThirdIndex}$.
\end{itemize}
\end{IEEEproof}


\renewcommand{\Op}{\SUI{>\Const}}
\renewcommand{\FOUT}{\LSF\Op\RSF}
\renewcommand{\SEnc}{\SEncT}


\begin{IEEEproof}
\emph{For ${\AnyTemporalOp}={\SUI{>\Const}}$.}
The ``preservation of traces'' property follows from the fact that the constraints can easily be satisfied globally when $\OUT$ and $\Obligation$ are set to $\False$ globally.

\SoundProofIntro

By the fact that $\Holds{\OUT}{\Index}$ and Constraint~\ref{m:enc.u.lb.out.obl} we know that  $\Holds{\Obligation}{\Index}$.
Let $\Strict\Index=\Index$ if $\IntervalI{\Index}$ is open and $\Strict\Index=\Index+1$ otherwise.
Now
\begin{itemize}
\item 
Case 1: $\TRight$ holds on interval $\Index+1$.
Then $\Timing$  and $\ER$ hold on interval $\Index+1$.
As  $\Holds{\OUT}{\Index}$, we know that $\St_{\Index+1}(\C)=0$ and $\Holds{\RightOpen}{\Index+1}$ iff $\IntervalI{\Index}$ is open.
Now
\begin{itemize}
\item If interval $\IntervalI{\Index+1}$ is a \singi , then $\St_{\Index+1}(\C)+\St_{\Index+1}(\D)=0$. Because  $\Timing$ holds on interval $\Index+1$, we then know that $\Const=0$ and $\Holds{\RightOpen}{\Index+1}$, the latter implying that $\IntervalI{\Index}$ is open. Now Constraint~\ref{m:enc.u.lb.obl.open} implies that $\Holds{\EL}{\Index}$. Furthermore, as $\IntervalI{\Index}$ is open, we can pick a point that is more than $0$ time units away from $\Point{\Index}{\Offset}$ in $\IntervalI{\Index+1}$. Thus, $\ITrSuffix{\ITr}{\Index}{\Offset}\models\FOUT$.
\item
If, in contrast, $\IntervalI{\Index+1}$ is open then $\IntervalI{\Index}$ is a \singi\ and $\HoldsNot{\RightOpen}{\Index+1}$.
Thus, the fact that $\Timing$ is satisfied on interval $\Index+1$ implies that $\St_{\Index+1}(\D)>\Const$. Thus we can pick a point that is more than $\Const$ time units away from $\Point{\Index}{\Offset}$ in $\IntervalI{\Index+1}$.
As $\IntervalI{\Index}$ is a \singi\ and $\IntervalI{\Index+1}$ an open interval, Constraint~\ref{m:enc.u.lb.obl.singular} implies that
$\Holds{\EL}{\Index+1}$, meaning that $\ITrSuffix{\ITr}{\Index}{\Offset}\models\FOUT$.
\end{itemize}
\item 
Case 2: $\TRight$ does not hold on interval $\Index+1$ and
$\HoldsNot{\Obligation}{\Index+1}$.
Now based on
Constraint~\ref{m:enc.u.lb.out.xobl} this means that
 $\Const=0$.
Furthermore, the only way to satisfy Constraints~\ref{m:enc.u.lb.obl.open} and~\ref{m:enc.u.lb.obl.singular} is
if  $\IntervalI{\Index}$ is open and $\TRight$ and $\EL$ hold at interval $\Index$. In this case, $\Hold{\EL}{\ER}{\Index}$, implying that $\ITrSuffix{\ITr}{\Index}{\Offset}\models\LSF\SUI{>0}\RSF$. 
\item 
Case 3:
$\TRight$ does not hold on interval $\Index+1$ and $\HoldsNot{\EL}{\Index+1}$.
Again, Constraint~\ref{m:enc.u.lb.out.xobl} implies that $\Const=0$. Furthermore, 
Constraints~\ref{m:enc.u.lb.obl.open} and~\ref{m:enc.u.lb.obl.singular} can only be satisfied if $\IntervalI{\Index}$ is open and both $\LSF$ and $\TRight$ hold on interval $\Index$. Clearly,
$\ITrSuffix{\ITr}{\Index}{\Offset}\models\LSF\SUI{>0}\RSF$.
\item 
Case 4: 
$\TRight$ does not hold on interval $\Index+1$, $\Hold{\Obligation}{\EL}{\Index+1}$ and
$\TRight$ holds on interval $\Index+2$ or any later interval.
Pick $\AnotherIndex$ as small as possible such that $\AnotherIndex\geq\Index+2$ and $\TRight$ holds at interval $\AnotherIndex$.
Let  $\Strict\AnotherIndex=\AnotherIndex$ if $\IntervalI{\AnotherIndex}$ is open and $\Strict\AnotherIndex=\AnotherIndex-1$ if $\IntervalI{\AnotherIndex}$ is a \singi .
Note that Constraints~\ref{m:enc.u.lb.obl.open} and~\ref{m:enc.u.lb.obl.singular} correspond to Constraints~\ref{m:enc-u-open} and~\ref{m:enc-u-singular} in the $\SU$-encoding, except that $\OUT$ has been replaced by $\Obligation$ and $\ER$ has been replaced by $\TRight$.
This correspondence allows us to conclude that $\LSF\SU\TRight$ is satisfied everywhere on interval $\Index+1$.
By the fact that
$\LSF\SU\TRight$ holds on interval $\Index+1$
and the choice of $\AnotherIndex$
we now know that $\HoldsFromTo{\EL}{\Index+2}{\Strict\AnotherIndex}$.
Furthermore, we assumed $\Holds{\EL}{\Index+1}$. Also, if $\IntervalI{\Index}$ is open, the fact that $\Holds{\Obligation}{\Index}$ and Constraint~\ref{m:enc.u.lb.obl.open} imply that $\Holds{\EL}{\Index}$. Thus, we know that $\HoldsFromTo{\EL}{\Strict\Index}{\Strict\AnotherIndex}$.

Now choose $\AThirdIndex$ as large as possible such that $\Index\leq\AThirdIndex<\AnotherIndex$ and $\Holds{\OUT}{\AThirdIndex}$.
Now the values of $\C$ and $\RightOpen$ are set according to 
Constraint~\ref{m:enc.u.lb.clk.reset}
on interval $\AThirdIndex+1$ and 
and according to
Constraint~\ref{m:enc.u.lb.clk.update}
on intervals $\AThirdIndex+2,\ldots\AnotherIndex$.
This implies that $\St_{\AnotherIndex}(\C)$ is the difference between the left bound of $\IntervalI{\AnotherIndex}$ and the right bound of $\IntervalI{\AThirdIndex}$.
Thus, $\St_{\AnotherIndex}(\C)+\St_{\AnotherIndex}(\D)$ is the difference between the right bounds of $\IntervalI{\AnotherIndex}$ and $\IntervalI{\AThirdIndex}$.
Furthermore, $\Holds{\RightOpen}{\AnotherIndex}$ iff $\IntervalI{\AThirdIndex}$ is open.
As $\TRight$ holds on interval $\AnotherIndex$, we know by the definition of $\TRight$ that $\Timing$ holds on interval $\AnotherIndex$.
Thus, the difference between the right bounds of $\IntervalI{\AnotherIndex}$  and $\IntervalI{\AThirdIndex}$is greater or equal to $\Const$ and greater than $\Const$ if $\IntervalI{\AThirdIndex}$ is a \singi . This implies, that for every point in $\IntervalI{\AThirdIndex}$ we can pick a point in $\IntervalI{\AnotherIndex}$ that is more than $\Const$ time units away.
As $\Index\leq\AThirdIndex$, we can also pick a point in $\IntervalI{\AnotherIndex}$ that is more than $\Const$ time units away from $\Offset$.

Furthermore, by $\TRight$ holding on interval $\AnotherIndex$ we know that $\Holds{\ER}{\AnotherIndex}$. Together with the fact that $\HoldsFromTo{\EL}{\Strict\Index}{\Strict\AnotherIndex}$, this implies that $\ITrSuffix{\ITr}{\Index}{\Offset}\models\FOUT$.
\item 
Case 5: 
$\TRight$ does not hold on intervals $\Index+1,\Index+2,\ldots$ and $\Hold{\Obligation}{\EL}{\Index+1}$. 
Now
by
Constraints~\ref{m:enc.u.lb.obl.open} and~\ref{m:enc.u.lb.obl.singular}
we know that $\HoldsFrom{\EL}{\Strict\Index}$ and $\HoldsFrom{\Obligation}{\Strict\Index}$, i.e. $\EL$ and $\Obligation$ hold globally starting from interval $\Strict\Index$.
Thus, $\neg\Obligation$ holds only on finitely many intervals. By fairness constraint~$\EncFair_{\FOUT}$, this implies that $\ER$ holds on infinitely many intervals. As $\ITr$ is non-zeno, this implies that we can pick an interval that contains a point more than $\Const$ time units away from $\Point{\Index}{\Offset}$ and on which $\ER$ holds. As $\EL$ holds globally starting from interval $\Strict\Index$, this implies that $\ITrSuffix{\ITr}{\Index}{\Offset}\models\FOUT$.
\end{itemize}
In each case, we were able to demonstrate that $\ITrSuffix{\ITr}{\Index}{\Offset}\models\FOUT$.
\end{IEEEproof}


\renewcommand{\Op}{\SUI{\geq\Const}}
\renewcommand{\FOUT}{\LSF\Op\RSF}
\renewcommand{\SEnc}{\SEncT}


\begin{IEEEproof}
\emph{For ${\AnyTemporalOp}={\SUI{\geq\Const}}$.}
As a first observation, we note that in case of ${\AnyTemporalOp}={\SUI{\geq\Const}}$ we know that $\Const>0$, as we use the $\SU$ encoding to encode $\LSF \SUI{\geq 0} \RSF$.

We obtain the ${\AnyTemporalOp}={\SUI{\geq\Const}}$ proof from the ${\AnyTemporalOp}={\SUI{>\Const}}$ proof by the following modifications:
\begin{itemize}
\item In Case 1, assuming $\IntervalI{\Index+1}$ to be a \singi\ contradicts our observation that $\Const>0$, meaning that  $\IntervalI{\Index+1}$ is open, $\IntervalI{\Index}$ is a \singi . Now the fact that $\Timing$ holds on interval $\Index+1$ implies that $\St_{\Index+1}(\D)>\Const$. This allows us to to pick a point in $\IntervalI{\Index}$ that is $\geq\Const$ time units away from $\Point{\Index}{\Offset}$.
Analogously to the  ${\AnyTemporalOp}={\SUI{>\Const}}$ proof this leads to
$\ITrSuffix{\ITr}{\Index}{\Offset}\models\FOUT$.
\item Cases 2 and 3 contradict $\Const>0$.
\item In Case 4, the fact that $\Timing$ holds on interval $\AnotherIndex$ implies that either (i) the difference between the right bounds of $\IntervalI{\AnotherIndex}$ and $\IntervalI{\AThirdIndex}$ is greater than $\Const$ or (ii) the right bounds of $\IntervalI{\AnotherIndex}$ and $\IntervalI{\AThirdIndex}$ equals $\Const$ and $\IntervalI{\AThirdIndex}$ is open or $\IntervalI{\AnotherIndex}$ is a \singi . Thus, we can for every point in $\IntervalI{\AThirdIndex}$ pick a point in $\IntervalI{\AnotherIndex}$ that is $\geq\Const$ time units away.
\item Case 5 does not need modification.
\end{itemize}
\end{IEEEproof}


\renewcommand{\Op}{\SR}
\renewcommand{\FOUT}{\LSF\Op\RSF}
\renewcommand{\SEnc}{\SEncT}


\begin{IEEEproof}
\emph{For ${\AnyTemporalOp}={\SR}$.}
The ``preservation of traces'' property follows from the fact that the constraints can easily be satisfied globally when $\OUT$ and $\Obligation$ are set to $\False$ globally.

\SoundProofIntro
Let $\Strict\Index=\Index$ if $\IntervalI{\Index}$ is open and $\Strict\Index=\Index+1$ if $\IntervalI{\Index}$ is a \singi .

As a \textbf{first case} assume $\HoldsNotFrom{\EL}{\Strict\Index}$. Then by Constraints~\ref{eq:r.unt.out-open}, \ref{eq:r.unt.out-sing} and~\ref{eq:r.unt.consec} we know that $\HoldsFrom{\Obligation}{\Strict\Index}$.
Now Constraint~\ref{eq:r.unt.obl-r} implies that $\HoldsFrom{\ER}{\Strict\Index}$.
Clearly, in this case
$\ITrSuffix{\ITr}{\Index}{\Offset}\models\FOUT$.

As a \textbf{second case} assume $\EL$ holds at interval $\Strict\Index$ or any later interval. Then let $\AnotherIndex\geq\Strict\Index$ be as small as possible such that $\Holds{\EL}{\AnotherIndex}$.
Then by Constraints~\ref{eq:r.unt.out-open}, \ref{eq:r.unt.out-sing} and~\ref{eq:r.unt.consec} , we know that $\HoldsFromTo{\Obligation}{\Strict\Index}{\AnotherIndex}$.
Now
\begin{itemize}
 \item If $\IntervalI{\AnotherIndex}$ is open, then by Constraint~\ref{eq:r.unt.obl-r} we have that $\HoldsFromTo{\ER}{\Strict\Index}{\AnotherIndex-1}$. This, in turn, implies that for any time point after $\Point{\Index}{\Offset}$ at which $\ER$ does not hold there is an earlier time point in interval $\AnotherIndex$. Hence, $\ITrSuffix{\ITr}{\Index}{\Offset}\models\FOUT$.
 \item If $\IntervalI{\AnotherIndex}$ is a \singi , then by Constraint~\ref{eq:r.unt.obl-r} we have that $\HoldsFromTo{\ER}{\Strict\Index}{\AnotherIndex}$.
 Thus, the single time point in interval $\AnotherIndex$, at which $\EL$ holds, lies in between time point $\Point{\Index}{\Offset}$ and any potential future point at which $\ER$ does not hold. Again, $\ITrSuffix{\ITr}{\Index}{\Offset}\models\FOUT$.
\end{itemize}
\end{IEEEproof}


\renewcommand{\Op}{\SGI{\leq 0}}
\renewcommand{\FOUT}{\Op\LSF}
\renewcommand{\SEnc}{\SEncO}


\begin{IEEEproof}
\emph{For ${\AnyTemporalOp}={\SGI{\leq 0}}$.}
The ``preservation of traces'' property follows from the fact that the constraints can easily be satisfied globally when $\OUT$ is set to $\False$ globally.

\SoundProofIntro

Recall, that by the semantics of $\SGI{\leq\Const}$, a point on a trace satisfies $\SGI{\leq 0}\LSF$ iff all \emph{future} points that are zero time units away from that point satisfy $\LSF$.
Note that a future time point can be zero time units away only if both the current and the next interval are \singis . 

By Constraint~\ref{eq:g.leqz}, there are two possibilities:
\begin{itemize}
\item
$\IntervalI{\Index}$ or $\IntervalI{\Index+1}$ is open. In this case,
there are no future time points that are zero time units away for any point on interval $\Index$.
Hence, $\ITrSuffix{\ITr}{\Index}{\Offset}\models\FOUT$ trivially.
\item
$\IntervalI{\Index}$ and $\IntervalI{\Index+1}$ are both \singis\ and $\Hold{\EL}{\OUT}{\Index+1}$.
By the fact that $\Holds{\OUT}{\Index+1}$, we can again deduce that either $\IntervalI{\Index+2}$ is open or $\Hold{\EL}{\OUT}{\Index+2}$. Repeatedly applying this argument leads to the conclusion that $\EL$ has to hold on all future \singis\  up to the next open interval, i.e., all future intervals containing points zero time units away from $\Point{\Index}{\Offset}$.
Thus, $\ITrSuffix{\ITr}{\Index}{\Offset}\models\FOUT$.
\end{itemize}
In both cases, we were able to show that $\ITrSuffix{\ITr}{\Index}{\Offset}\models\FOUT$.
\end{IEEEproof}


\renewcommand{\Op}{\SGI{<\Const}}
\renewcommand{\FOUT}{\Op\LSF}
\renewcommand{\SEnc}{\SEncO}


\begin{IEEEproof}
\emph{For ${\AnyTemporalOp}={\SGI{<\Const}}$.}
The ``preservation of traces'' property follows from the fact that the constraints can easily be satisfied globally when $\OUT$ and $\Obligation$ are set to $\False$ globally.

\SoundProofIntro

Take an arbitrary $\AnotherIndex>\Index$ and $\SecondOffset\in\IntervalI{\AnotherIndex}$ such that $\SecondOffset-\Offset<\Const$ (if such an $\AnotherIndex$ exists). Let $\AThirdIndex$ be as large as possible such that $\Index\leq\AThirdIndex<\AnotherIndex$ and $\Holds{\OUT}{\AThirdIndex}$.
By Constraint~\ref{eq:g.ub.clk1}, we know that $\St_{\AThirdIndex+1}(\C)=0$. Furthermore, by Constraints~\ref{eq:g.ub.clk1} and~\ref{eq:g.ub.clk2}, we know that $\St_{\AnotherIndex}(\C)$ is the difference between the right bound of $\IntervalI{\AThirdIndex}$ and the left bound $\IntervalI{\AnotherIndex}$.
As $\AThirdIndex\geq\Index$ and $\SecondOffset-\Offset<\Const$, we know that this difference must be less than $\Const$ and, consequently, $\Timing$ holds on interval $\AnotherIndex$.
By Constraint~\ref{eq:g.ub.timing}, this implies that $\Holds{\EL}{\AnotherIndex}$. As we picked $\AnotherIndex$ to be an arbitrary interval containing a point less than $\Const$ time units away from $\Point{\Index}{\Offset}$, we can conclude that $\EL$ holds at any point on intervals $\Index+1,\Index+2,\ldots$ that is less than $\Const$ time units after $\Point{\Index}{\Offset}$

Additionally, Constraint~\ref{eq:g.ub.timing} ensures that $\EL$ also holds on interval $\Index$, if that interval is open. Thus, it is guaranteed that $\EL$ holds at all future points less than $\Const$ time units away and $\ITrSuffix{\ITr}{\Index}{\Offset}\models\FOUT$.
\end{IEEEproof}


\renewcommand{\Op}{\SGI{\leq\Const}}
\renewcommand{\FOUT}{\Op\LSF}
\renewcommand{\SEnc}{\SEncO}

\begin{IEEEproof}
\emph{For ${\AnyTemporalOp}={\SGI{\leq\Const}}$.}
We modify the proof for $\SGI{<\Const}$ by picking $\AnotherIndex>\Index$, $\SecondOffset\in\IntervalI{\AnotherIndex}$ with $\SecondOffset-\Offset\leq\Const$. Then we know that the difference between the left bound of $\IntervalI{\AnotherIndex}$ and the right bound of $\IntervalI{\Index}$ is less than or equal to $\Const$ and can be equal to $\Const$ only if both intervals are \singis . Now
\begin{itemize}
\item if $\St_\AnotherIndex(\C)<\Const$ then $\Timing$ is satisfied on interval $\AnotherIndex$ and we proceed as before.
\item if $\St_\AnotherIndex(\C)=\Const$ then we know that $\St_\AnotherIndex(\C)$ is precisely the difference between the left bound of $\IntervalI{\AnotherIndex}$ and the right bound of $\IntervalI{\Index}$ and $\IntervalI{\Index}$ and $\IntervalI{\AnotherIndex}$ are both \singis .
Now, $\AThirdIndex=\Index$ or there is a sequence of intervals $\IntervalI{\Index}\IntervalI{\Index+1}\ldots\IntervalI{\AThirdIndex}$ that are all \singis . In either case, $\IntervalI{\AThirdIndex}$ is a \singi\ and $\HoldsNot{\RightOpen}{\AnotherIndex}$. Thus, $\Timing$ is satisfied also in this case and we continue as before.
\end{itemize}
\end{IEEEproof}


\renewcommand{\Op}{\SRI{>\Const}}
\renewcommand{\FOUT}{\LSF\Op\RSF}
\renewcommand{\SEnc}{\SEncT}


\begin{IEEEproof}
\emph{For ${\AnyTemporalOp}={\SRI{>\Const}}$.}
The ``preservation of traces'' property follows from the fact that the constraints can easily be satisfied globally when $\OUT$ and $\Obligation$ are set to $\False$ globally.

\SoundProofIntro
As usual, let $\Strict\Index=\Index$ if interval $\Index$ is open and $\Strict\Index=\Index+1$ otherwise.

As a \textbf{first case} assume that for all $\Point{\AnotherIndex}{\SecondOffset}\in\Timepoints{\ITr}$ with $\SecondOffset - \Offset > \Const$ we have that $\Holds{\ER}{\AnotherIndex}$. In this case, we trivially have $\ITrSuffix{\ITr}{\Index}{\Offset}\models\FOUT$.

As a \textbf{second case}, assume there is
$\Point{\AnotherIndex}{\SecondOffset}\in\Timepoints{\ITr}$ with $\SecondOffset - \Offset > \Const$ and $\HoldsNot{\ER}{\AnotherIndex}$.
\begin{itemize}
\item Case 2.a:  $\Index=\AnotherIndex$:
As $\SecondOffset - \Offset > \Const$, this implies that $\D>\Const$ at interval $\Index$.
Then, Constraint~\ref{eq:nr.lb.first} implies that $\Holds{\EL}{\Index}$.
As interval $\Index$ also has to be open to allow $\D$ to be non-zero, we now know that there is point in between $\Point{\Index}{\Offset}$ and $\Point{\AnotherIndex}{\SecondOffset}$ where $\EL$ holds.
\item Case 2.b: $\Index<\AnotherIndex$ and $\HoldsNot{\Obligation}{\AnotherIndex}$.
Analogously to the proof of the lemma for the untimed $\SR$ encoding,  $\HoldsNot{\Obligation}{\AnotherIndex}$ implies that $\EL$ holds at some point in between $\Point{\Index}{\Offset}$ and $\Point{\AnotherIndex}{\SecondOffset}$.
\item Case 2.c: $\Index<\AnotherIndex$ and $\Hold{\Obligation}{\EL}{\AnotherIndex}$ and $\IntervalI{\AnotherIndex}$ is open. Then, as $\IntervalI{\AnotherIndex}$ is open and $\Holds{\EL}{\AnotherIndex}$, we know that $\EL$ holds at some point in between $\Point{\Index}{\Offset}$ and $\Point{\AnotherIndex}{\SecondOffset}$.
\item Case 2.d: $\Index<\AnotherIndex$ and $\Holds{\Obligation}{\AnotherIndex}$ and either $\IntervalI{\AnotherIndex}$ is a \singi\ or $\HoldsNot{\EL}{\AnotherIndex}$.
Then, according to Constraints~\ref{eq:nr.lb.timing} we know $\Timing$ does not hold on interval $\AnotherIndex$.
Now pick $\AThirdIndex\leq\AnotherIndex$ as large as possible such that
$\St_\AThirdIndex(\C)=0$ and one of the following:
\begin{itemize}
\item $\DelayReset$ holds at interval $\AThirdIndex-1$
\item $\EReset$ holds at interval $\AThirdIndex-1$, or
\item $\AThirdIndex=0$
\end{itemize}
Now 
Constraint~\ref{eq:nr.lb.clk.update} implies
that $\St_\AnotherIndex(\C)$ is the difference between the left bound of $\IntervalI{\AnotherIndex}$ and the left bound of $\IntervalI{\AThirdIndex}$.
Consequently, $\St_\AnotherIndex(\C)+\St_\AnotherIndex(\D)$ is the difference between the right bound of $\IntervalI{\AnotherIndex}$ and the left bound of $\IntervalI{\AThirdIndex}$. 
Now by the fact that $\Timing$ does not hold we know  that the difference between the right bound of $\IntervalI{\AnotherIndex}$ and the left bound of $\IntervalI{\AThirdIndex}$ is less than or equal to $\Const$ time units. 
Furthermore, 
$\SecondOffset - \Offset > \Const$ implies that the difference between the right bound of $\IntervalI{\AnotherIndex}$ and the left bound of $\IntervalI{\Index}$ is greater than $\Const$.
This, in turn, implies that $\AThirdIndex>\Index$ and at least one of the intervals $\IntervalI{\Index},\IntervalI{\Index+1},\ldots,\IntervalI{\AThirdIndex-1}$ is open. This implies that $\AThirdIndex>\Strict\Index>0$
Now, there are three possibilities:
\begin{itemize}
\item
$\EReset$ holds at interval $\AThirdIndex-1$ and $\Holds{\EL}{\AThirdIndex-1}$.
Then there is a point in between $\Point{\Index}{\Offset}$ and $\Point{\AnotherIndex}{\SecondOffset}$ at which $\LSF$ holds.
\item
$\EReset$ holds at interval $\AThirdIndex-1$ and $\HoldsNot{\EL}{\AThirdIndex-1}$. By the definition of $\EReset$ now $\HoldsNot{\Obligation}{\AThirdIndex-1}$. analogously to the untimed $\SR$ encoding and Case 2.b, this implies that $\EL$ holds on an interval in the range $\Strict\Index,\ldots,\AThirdIndex-1$.
\item $\DelayReset$ holds at interval $\AThirdIndex-1$. This implies that  $\Holds{\EL}{\AThirdIndex-1}$.
\end{itemize}
\end{itemize}
In each of the mentioned cases, we were able to show $\EL$ holds at some point in between $\Point{\Index}{\Offset}$ and $\Point{\AnotherIndex}{\SecondOffset}$, allowing us to conclude that $\ITrSuffix{\ITr}{\Index}{\Offset}\models\FOUT$.
\end{IEEEproof}

\renewcommand{\Op}{\SRI{\geq\Const}}
\renewcommand{\FOUT}{\LSF\Op\RSF}
\renewcommand{\SEnc}{\SEncT}


\begin{IEEEproof}
\emph{For ${\AnyTemporalOp}={\SRI{\geq\Const}}$.}
To adapt the $\SRI{>\Const}$ proof for $\SRI{\geq\Const}$ modify the second case as follows.
We pick a $\Point{\AnotherIndex}{\SecondOffset}\in\Timepoints{\ITr}$ with $\HoldsNot{\ER}{\AnotherIndex}$ and $\SecondOffset-\Offset\geq\Const$. Cases 2.a to 2.c do not require substantial changes. In Case 2.d, we pick $\AThirdIndex$ as before.

We now observe that $\SecondOffset-\Offset\geq\Const$ implies that the difference between the right bound of $\IntervalI{\AnotherIndex}$ and the left bound of $\IntervalI{\Index}$ is greater than or equal to $\Const$. Furthermore, the difference can only be equal if $\IntervalI{\Index}$ and $\IntervalI{\AnotherIndex}$ are both \singis .

The fact that $\Timing$ does not hold at interval $\AnotherIndex$ implies that $\St_{\AnotherIndex}(\C)+\St_{\AnotherIndex}(\D) \leq \Const$. Furthermore, if $\St_{\AnotherIndex}(\C)+\St_{\AnotherIndex}(\D) = \Const$ then we know that $\IntervalI{\AnotherIndex}$ is open or $\Holds{\LeftOpen}{\AnotherIndex}$.

Now
\begin{itemize}
\item If the difference between the right bound of $\IntervalI{\AnotherIndex}$ and the left bound of $\IntervalI{\Index}$ is greater than $\Const$, we again can conclude that $\AThirdIndex>\Strict\Index$ and proceed as before.
\item Likewise, if $\St_{\AnotherIndex}(\C)+\St_{\AnotherIndex}(\D) < \Const$ we conclude that $\AThirdIndex>\Strict\Index$ and proceed as before.
\item If the difference between the right bound of $\IntervalI{\AnotherIndex}$ and the left bound of $\IntervalI{\Index}$ equals $\Const$ \emph{and} $\St_{\AnotherIndex}(\C)+\St_{\AnotherIndex}(\D)=\Const$, then our previous observations tell us that (i) $\IntervalI{\Index}$ and $\IntervalI{\AnotherIndex}$ are \singis\ (ii) $\Holds{\LeftOpen}{\AnotherIndex}$. Furthermore, as $\Const>0$, $\St_{\AnotherIndex}(\C)+\St_{\AnotherIndex}(\D) = \Const$, $\St_{\AThirdIndex}(\C)=0$
and $\IntervalI{\AnotherIndex}$ is a \singi\ we know that $\AThirdIndex<\AnotherIndex$.
As the value of $\LeftOpen$ on intervals $\AThirdIndex+1,\ldots,\AnotherIndex$ set according to Constraint~\ref{eq:nr.lb.clk.update}, we know that  $\Holds{\LeftOpen}{\AThirdIndex}$. This eliminates the possibilities that $\AThirdIndex=0$ or that $\DelayReset$ holds on interval $\AThirdIndex-1$, leaving only the possibility that $\EReset$ holds on interval $\AThirdIndex-1$.
 Then by Constraint~\ref{eq:nr.lb.clk.reset}, $\IntervalI{\AThirdIndex}$ is open.
Together with our assumptions about the left-bound-to-right-bound differences and the fact that $\IntervalI{\Index}$ is a \singi , this implies that $\AThirdIndex>\Index$. Finally, as $\IntervalI{\Index}$ is a \singi\ and $\Holds{\OUT}{\Index}$, $\EReset$ does not hold on interval $\Index$, meaning that $\AThirdIndex>\Index+1=\Strict\Index$. Now we can continue as previously.
\end{itemize}
\end{IEEEproof}


\subsection{Completeness proofs}

\renewcommand{\Op}{\AnyTemporalOp}
\renewcommand{\FOUT}{\LSF\Op\RSF}
\newcommand{\OUTO}{\Op\LSF}
\renewcommand{\SEnc}{\SEncT}

\begin{relemma}{\ref{Lemma:formula_implies_constraints}}
\FenfCText
\end{relemma}

Recall, that an encoding $\STTSf{\Formula}$ is complete
if for every $\Formula$-fine trace $\ITr = \Tuple{\IntervalI{0},\ValuI{0}} \Tuple{\IntervalI{1},\ValuI{1}} \Tuple{\IntervalI{2},\ValuI{2}} \ldots$
 in $\TracesOf{\STTS}$,
there is a run $\STTSRun = \St_0 \St_1 \St_2 \ldots$ in $\STTSf{\Formula}$
  such that
  $\TraceOf{\STTSRun} = \ITr$
  and
  for all points $\Point{\Index}{\Offset}$ in $\ITr$
  it holds that
  $\ITrAt{\Index}{\Offset} \Models \Formula$
  implies
  $\ValuI{\Index}(\Enc{\Formula}) = \True$.
Thus, we can prove completeness by assuming a $\Formula$-fine trace $\ITr\in\TracesOf{\STTS}$, extending it to a run $\STTSRun$ by giving values for the auxiliary variables used in the encoding (setting $\Enc{\Formula}$ to $\True$ exactly on those intervals where $\Formula$ holds) and then arguing that all constraints of the encoding are satisfied.
Lemma~\ref{Lemma:formula_implies_constraints} is proven by structural induction. That is, it will assumed that the lemma holds for the subformulas $\LSF$ and $\RSF$. 
Like Lemma~\ref{Lemma:constraints_enforce_formula}, we will prove Lemma~\ref{Lemma:formula_implies_constraints} separately for each operator.

As the lemma assume $\Formula$-fineness, $\Formula$ either holds at all points in a given interval in $\ITr$ or $\Formula$ does not hold at any point inside the interval. We use the notation $\ITrSuffix{\ITr}{\AnotherIndex}{\cdot}\models\Formula$ to denote that $\Formula$ is satisfied by all points belonging to interval $\AnotherIndex$ and
$\ITrSuffix{\ITr}{\AnotherIndex}{\cdot}\not\models\Formula$
to denote that no point inside the interval satisfies $\Formula$.


\newcommand{\StartRulesT}{\textbf{Auxiliary variable rules:} Let $\Index\in\Naturals$. As always, we set $\St_\Index(\OUT)$ / $\St_\Index(\EL)$ / $\St_\Index(\ER)$ to $\True$ iff $\ITrSuffix{\ITr}{\Index}{\cdot}\models\FOUT$ / $\ITrSuffix{\ITr}{\Index}{\cdot}\models\LSF$ / $\ITrSuffix{\ITr}{\Index}{\cdot}\models\RSF$, respectively. }
\newcommand{\StartRulesO}{\textbf{Auxiliary variable rules:} Let $\Index\in\Naturals$. As always, we set $\St_\Index(\OUT)$ / $\St_\Index(\EL)$ to $\True$ iff $\ITrSuffix{\ITr}{\Index}{\cdot}\models\FOUT$ / $\ITrSuffix{\ITr}{\Index}{\cdot}\models\LSF$, respectively. }
\newcommand{\StartInitial}{\textbf{The initial constraint is satisfied:} }
\newcommand{\StartTransition}{\textbf{The transition constraints are satisfied:} Let $\Index\in\Naturals$. }
\newcommand{\StartFairness}{\textbf{The fairness condition is satisfied:} }

\newcommand{\StartRules}{\StartRulesT}
\renewcommand{\Op}{\SU}
\renewcommand{\FOUT}{\LSF\Op\RSF}
\renewcommand{\SEnc}{\SEncT}

\begin{IEEEproof}
\emph{For ${\AnyTemporalOp}={\SU}$.}
\StartRules


\StartTransition
Now %
\begin{itemize}
\item 
If $\HoldsNot{\OUT}{\Index}$ then both Constraints~\ref{m:enc-u-open} and~\ref{m:enc-u-singular} trivially hold at interval $\Index$.
\item 
If $\Holds{\OUT}{\Index}$ and $\IntervalI{\Index}$ is open, then Constraint~\ref{m:enc-u-singular} is trivially satisfied.
Furthermore, as $\Holds{\OUT}{\Index}$ we know that
$\ITrSuffix{\ITr}{\Index}{\cdot}\models\FOUT$.
Pick any $\FirstOffset\in\IntervalI{\Index}$.
By the semantics of $\SU$ we now know that there is a future time point
$\Point{\AnotherIndex}{\SecondOffset}\in\LaterPoints{\ITr}{\Index}{\Offset}$
such that
$\Holds{\ER}{\AnotherIndex}$
and
$\EL$ holds anywhere in between
$\Point{\Index}{\FirstOffset}$ and $\Point{\AnotherIndex}{\SecondOffset}$.
Note that due to the fact that $\IntervalI{\Index}$ is open, there is a guarantee that interval $\Index$ contains time points lying in between $\Point{\Index}{\FirstOffset}$ and $\Point{\AnotherIndex}{\SecondOffset}$, implying that $\Holds{\EL}{\Index}$.
Now if $\AnotherIndex=\Index$ or $\AnotherIndex=\Index+1$, then Constraint~\ref{m:enc-u-open} is satisfied.
Furthermore, if $\AnotherIndex>\Index+1$, then also $\Holds{\EL}{\Index+1}$ and $\ITrSuffix{\ITr}{\Index+1}{\cdot}\models\FOUT$. Thus, $\Holds{\OUT}{\Index+1}$, implying that Constraint~\ref{m:enc-u-singular} is satisfied in this case as well.
\item 
If $\Holds{\OUT}{\Index}$ and $\IntervalI{\Index}$ is a \singi , then Constraint~\ref{m:enc-u-open} is trivially satisfied.
Let $\Set{\FirstOffset}=\IntervalI{\Index}$.
Again,
we pick a time point
$\Point{\AnotherIndex}{\SecondOffset}\in\LaterPoints{\ITr}{\Index}{\SecondOffset}$
such that
$\Holds{\ER}{\AnotherIndex}$ 
and
$\EL$ holds anywhere in between
$\Point{\Index}{\FirstOffset}$ and $\Point{\AnotherIndex}{\Offset}$.
Now if $\AnotherIndex=\Index+1$ and $\IntervalI{\AnotherIndex}$ is a \singi , then Constraint~\ref{m:enc-u-singular} is satisfied. 
If $\AnotherIndex=\Index+1$ and $\IntervalI{\AnotherIndex}$ is open, then $\Hold{\EL}{\ER}{\AnotherIndex}$ meaning that $\ITrSuffix{\ITr}{\AnotherIndex}{\cdot}\models\FOUT$. Then $\Holds{\OUT}{\AnotherIndex}$ and Constraint~\ref{m:enc-u-singular} is satisfied. 
If $\AnotherIndex>\Index+1$ then we observe that $\Holds{\EL}{\Index+1}$ and $\ITrSuffix{\ITr}{\Index+1}{\cdot}\models\FOUT$, meaning that $\Holds{\OUT}{\Index+1}$ and Constraint~\ref{m:enc-u-singular} is satisfied in this case as well.
\end{itemize}

\StartFairness
It is easy to see that now the fairness constraint $\EncFair_{\FOUT}$ holds as well. We set $\OUT$ to $\True$ precisely on those intervals on which $\FOUT$ holds, implying that there is a future point at which $\RSF$ holds. Thus, if $\OUT$ holds on all intervals starting from some point, then $\FOUT$ holds globally starting from that point. Hence, there is always a future interval at which $\RSF$ holds, meaning that $\RSF$ (and thus $\ER$) holds infinitely often.
\end{IEEEproof}


\renewcommand{\StartRules}{\StartRulesO}
\renewcommand{\Op}{\SFI{\leq0}}
\renewcommand{\FOUT}{\Op\LSF}
\renewcommand{\SEnc}{\SEncO}

\begin{IEEEproof}
\emph{For ${\AnyTemporalOp}={\SFI{\leq0}}$.}
\StartRules

\StartTransition
If $\HoldsNot{\OUT}{\Index}$, then 
Constraint~\ref{eq:f.leqz}
is trivially satisfied.
If, in contrast, $\Holds{\OUT}{\Index}$, then 
we know that
$\ITrSuffix{\ITr}{\Index}{\cdot}\models\FOUT$.
Thus, there is a $\AnotherIndex>\Index$ such that $\Holds{\EL}{\AnotherIndex}$ and for every point in $\IntervalI{\Index+1}$ there is a point in $\IntervalI{\AnotherIndex}$ that is at most 0 time units away. The latter implies that intervals $\IntervalI{\Index},\ldots,\IntervalI{\AnotherIndex}$ are all \singis . Thus, in particular, $\IntervalI{\Index}$ and $\IntervalI{\Index+1}$ are \singis . Furthermore, if $\AnotherIndex=\Index+1$, then $\Holds{\EL}{\Index+1}$, implying that 
Constraint~\ref{eq:f.leqz}
is satisfied at interval $\Index$.
If, in contrast $\AnotherIndex>\Index+1$ then $\ITrSuffix{\ITr}{\Index+1}{\cdot}\models\FOUT$ as well, implying that $\Holds{\OUT}{\Index+1}$ and, ultimately, that 
Constraint~\ref{eq:f.leqz}
is satisfied at interval $\Index$ also in this case.
\end{IEEEproof}


\renewcommand{\StartRules}{\StartRulesO}
\renewcommand{\Op}{\SFI{<\Const}}
\renewcommand{\FOUT}{\Op\LSF}
\renewcommand{\SEnc}{\SEncO}
\newcommand{\ClkUpdateDesc}[3]{For $\Index>0$ we set $\St_\Index(\C)$ and $\St_\Index(#1)$ according to Constraints~\ref{#2} and~\ref{#3} and $\St_{\Index-1}$.}
\newcommand{\ClkUpdateDescT}[4]{For $\Index>0$ we set $\St_\Index(\C)$ and $\St_\Index(#1)$ according to Constraints~\ref{#2}, \ref{#3} and~\ref{#4} and $\St_{\Index-1}$.}

\begin{IEEEproof}
\emph{For ${\AnyTemporalOp}={\SFI{<\Const}}$.}
\StartRules
Furthermore, we set $\St_0(\C)=0$ and $\St_0(\LeftOpen)=\False$.
\ClkUpdateDesc{\LeftOpen}{eq:enc.f.ub.clk.reset}{eq:enc.f.ub.clk.update}

\StartInitial
Initial constraint~$\EncInit_{\FOUT}$ is trivially satisfied as $\St_0(\C)=0$ and $\St_0(\LeftOpen)=\False$.

\StartTransition
Let $\Strict\Index=\Index$ if $\IntervalI{\Index}$ is open and $\Strict\Index=\Index+1$ if $\IntervalI{\Index}$ is a \singi .
Now we will show that interval $\Index$ satisfies Constraints~\ref{eq:enc.f.ub.open}
and~\ref{eq:enc.f.ub.singular} using a case distinction.
\begin{itemize}
\item 
If $\HoldsNot{\OUT}{\Index}$ or $\Holds{\OUT}{\Index+1}$ then
Constraints~\ref{eq:enc.f.ub.open}
and~\ref{eq:enc.f.ub.singular}
are trivially satisfied.
\item 
If $\Holds{\OUT}{\Index}$ and $\HoldsNot{\OUT}{\Index+1}$, then 
we know that
$\ITrSuffix{\ITr}{\Index}{\cdot}\models\FOUT$
and
$\ITrSuffix{\ITr}{\Index+1}{\cdot}\not\models\FOUT$.
Now
\begin{itemize}
\item If $\IntervalI{\Index}$ is open then Constraint~\ref{eq:enc.f.ub.singular} is trivially satisfied.
Furthermore, by semantics of $\SFI{<\Const}$ we know that
$\ITrSuffix{\ITr}{\Index}{\cdot}\models\FOUT$
and
$\ITrSuffix{\ITr}{\Index+1}{\cdot}\not\models\FOUT$
implies that $\Holds{\EL}{\Index}$ or $\Holds{\EL}{\Index}$. In either case, Constraint~\ref{eq:enc.f.ub.open} is satisfied.
\item If $\IntervalI{\Index}$ is a \singi\ then Constraint~\ref{eq:enc.f.ub.open} is trivially satisfied.
Furthermore, by semantics of $\SFI{<\Const}$ we know that
$\ITrSuffix{\ITr}{\Index}{\cdot}\models\FOUT$
and
$\ITrSuffix{\ITr}{\Index+1}{\cdot}\not\models\FOUT$
implies that (i) $\IntervalI{\Index+1}$ is a \singi\ as well and (ii) $\Holds{\EL}{\Index+1}$. Thus, Constraint~\ref{eq:enc.f.ub.singular} is satisfied as  well.
\end{itemize}
\end{itemize} 

It remains to be shown that Constraint~\ref{eq:enc.f.ub.timing} holds on interval $\Index$, which will be done by contradiction.
Assume, Constraint~\ref{eq:enc.f.ub.timing} does not hold on interval $\Index$.
Then, $\Timing$ does not  hold on interval $\Index$, $\Holds{\OUT}{\Index}$ and $\IntervalI{\Index}$ is a \singi\ or $\HoldsNot{\EL}{\Index}$.
As $\Holds{\OUT}{\Index}$, we know that $\ITrSuffix{\ITr}{\Index}{\cdot}\models\FOUT$.
Pick $\AnotherIndex\leq\Index$ as large as possible such that $\EReset$ holds at interval $\AnotherIndex-1$ is such $\AnotherIndex$ exists and set $\AnotherIndex=0$ otherwise. Now we know that $\St_\AnotherIndex(\C)=0$ and $\Holds{\LeftOpen}{\AnotherIndex}$ iff $\IntervalI{\AnotherIndex}$ is open.
\begin{itemize}
\item 
If $\AnotherIndex=\Index$, the we know that the value of $\St_\Index(\C)=0$.
Now
\begin{itemize}
\item 
If $\IntervalI{\Index}$ is a \singi , then the value of $\D$ at interval $\Index$ is 0 as well, which implies that $\Timing$ contrary to our assumption satisfied on interval $\Index$.
\item 
If $\IntervalI{\Index}$ is open, then our assumption that Constraint~\ref{eq:enc.f.ub.timing} does not hold implies that $\HoldsNot{\EL}{\Index}$.
As $\ITrSuffix{\ITr}{\Index}{\cdot}\models\FOUT$, we now know that $\D$ at interval $\Index$ can be at most $\Const$, implying that $\Timing$ is satisfied and contradicting our assumption that Constraint~\ref{eq:enc.f.ub.timing} does not hold.
\end{itemize}
\item 
If $\AnotherIndex<\Index$, then 
we know by the fact that $\EReset$ does not hold on intervals $\AnotherIndex,\ldots,\Index-1$ that the value of $\C$ at intervals $\AnotherIndex+1,\ldots,\Index$ has been set according  to Constraint~\ref{eq:enc.f.ub.clk.update}. Hence, $\St_\Index(\C)$ is the difference between the left bound of $\IntervalI{\Index}$ and the left bound of $\IntervalI{\AnotherIndex}$. Furthermore, $\St_\Index(\C)+\St_\Index(\D)$ is the difference between the right bound of $\IntervalI{\Index}$ and the left bound of $\IntervalI{\AnotherIndex}$ and $\Holds{\LeftOpen}{\Index}$ iff $\IntervalI{\AnotherIndex}$ is open.
By the fact that $\EReset$ does not hold on intervals $\AnotherIndex,\ldots,\Index-1$ and the fact that $\Holds{\OUT}{\Index}$ we know that $\HoldsFromTo{\OUT}{\AnotherIndex}{\Index}$.
Let $\Strict\AnotherIndex=\AnotherIndex$ if $\IntervalI{\AnotherIndex}$ is open and $\Strict\AnotherIndex=\AnotherIndex+1$ if $\IntervalI{\AnotherIndex}$ is a \singi .
As $\EReset$ does not hold on intervals $\AnotherIndex,\ldots,\Index-1$ and $\HoldsFromTo{\OUT}{\AnotherIndex}{\Index}$ we know that $\HoldsNotFromTo{\EL}{\Strict\AnotherIndex}{\Index}$.
As $\Holds{\OUT}{\AnotherIndex}$, we know that $\ITrSuffix{\ITr}{\AnotherIndex}{\cdot}\models\FOUT$. Thus, for each point in $\IntervalI{\AnotherIndex}$ there is a future point at which $\LSF$ holds and that is less than $\Const$ time units away. As $\HoldsNotFromTo{\EL}{\Strict\AnotherIndex}{\Index}$, this implies that every point in $\IntervalI{\AnotherIndex}$ is less than $\Const$ time units away from a time point in $\IntervalI{\Index+1}$. Thus, the difference between the right bound of $\IntervalI{\Index}$ and the left bound of $\IntervalI{\AnotherIndex}$ is less than $\Const$ time units if $\IntervalI{\AnotherIndex}$ is a \singi\ and less than or equal to $\Const$ time units if $\IntervalI{\AnotherIndex}$ is open. Recalling that the value of $\C+\D$ at interval $\Index$ is precisely said difference and $\Holds{\LeftOpen}{\Index}$ if $\IntervalI{\AnotherIndex}$ is open, we conclude that $\Timing$ is satisfied at interval $\Index$, contradicting our assumption.
\end{itemize}
In each case, we were able to show that the assumption that Constraint~\ref{eq:enc.f.ub.timing} does not hold leads to a contradiction.
\end{IEEEproof}


\renewcommand{\StartRules}{\StartRulesO}
\renewcommand{\Op}{\SFI{\leq\Const}}
\renewcommand{\FOUT}{\Op\LSF}
\renewcommand{\SEnc}{\SEncO}

\begin{IEEEproof}
\emph{For ${\AnyTemporalOp}={\SFI{\leq\Const}}$.} The only difference between the proof for $\SFI{\leq\Const}$ and the proof for $\SFI{<\Const}$ is in arguing that Constraint~\ref{eq:enc.f.ub.timing} is satisfied.
\begin{itemize}
\item
The case where $\Index=\AnotherIndex$ and $\IntervalI{\Index}$ is a \singi\ does not need modification.
\item
In the case where $\Index=\AnotherIndex$ and $\IntervalI{\Index}$ is open, we observe that $\ITrSuffix{\ITr}{\Index}{\cdot}\models\FOUT$ and $\HoldsNot{\EL}{\Index}$. Thus, we know that the value of $\D$ at interval $\Index$ is at most $\Const$. Furthermore, as $\IntervalI{\Index}$ is open, we know that $\Holds{\LeftOpen}{\Index}$, contradicting the assumption that $\Timing$ is not satisfied.
\item
In the case where $\AnotherIndex<\Index$, we observe that $\HoldsNotFromTo{\EL}{\Strict\AnotherIndex}{\Index}$ and $\ITrSuffix{\ITr}{\AnotherIndex}{\cdot}\models\FOUT$. Thus, the difference between the left bound of $\IntervalI{\AnotherIndex}$ and the right bound of $\IntervalI{\Index}$ is less than or equal to $\Const$ and less than $\Const$ if $\IntervalI{\AnotherIndex}$ is a \singi\ and $\IntervalI{\Index+1}$ is open, again leading to a contradiction based on the fact that $\Timing$ is satisfied.
\end{itemize}
\end{IEEEproof}


\renewcommand{\StartRules}{\StartRulesT}
\renewcommand{\Op}{\SUI{>\Const}}
\renewcommand{\FOUT}{\LSF\Op\RSF}
\renewcommand{\SEnc}{\SEncT}
\begin{IEEEproof}
\emph{For ${\AnyTemporalOp}={\SUI{>\Const}}$.}
\StartRules
We set $\St_0(\C)=0$ and $\St_0(\RightOpen)=\False$.
\ClkUpdateDesc{\RightOpen}{m:enc.u.lb.clk.reset}{m:enc.u.lb.clk.update}
Furthermore, we set $\Holds{\Obligation}{\Index}$ iff at least one of the following cases holds:
\begin{enumerate}
\item $\ITrSuffix{\ITr}{\Index}{\cdot}\models\FOUT$
\item $\Const>0$, $\Index>0$ and $\ITrSuffix{\ITr}{\Index-1}{\cdot}\models\FOUT$,\label{swjrew1}
\item \label{swjrew2}
$\Index>0$,
$\IntervalI{\Index-1}$ is open, $\Holds{\Obligation}{\Index-1}$ and $\TRight$ neither holds on interval $\Index-1$ nor on interval $\Index$.
\item \label{swjrew3}
$\Index>0$,
$\IntervalI{\Index-1}$ is a \singi , $\Holds{\Obligation}{\Index-1}$ and $\TRight$ does not hold on interval $\Index$ or $\IntervalI{\Index}$ is open.
\end{enumerate}

\StartTransition
Note, that the rules for setting the value of $\Obligation$ ensure that 
Constraint~\ref{m:enc.u.lb.out.obl}
is satisfied on interval $\Index$.

Assume $\ITrSuffix{\ITr}{\Index}{\cdot}\models\FOUT$ and $\Const>0$. Take $\Offset\in\IntervalI{\Index}$ such that the difference between the right bound of $\IntervalI{\Index}$ and $\Offset$ is less than $\Const$. Now there is a future time point  more than $\Const$ time units from $\Point{\Index}{\Offset}$ at which $\RSF$ holds and up to which $\LSF$ holds. The fact that this point is more than $\Const$ time units away implies that either $\IntervalI{\Index+1}$ is open or the time point is on interval $\Index+2$ or a later interval. In either case $\Holds{\EL}{\Index+1}$. Together with the rules for setting the value of $\Obligation$, this implies that Constraint~\ref{m:enc.u.lb.out.xobl} is satisfied.

If, in contrast, $\ITrSuffix{\ITr}{\Index}{\cdot}\not\models\FOUT$ then Constraint~\ref{m:enc.u.lb.out.xobl} is trivially satisfied and if $\Const=0$, Constraint~\ref{m:enc.u.lb.out.xobl} is not used at all.

Next, we show that Constraints~\ref{m:enc.u.lb.obl.open} and~\ref{m:enc.u.lb.obl.singular} are satisfied on interval $\Index$.
\begin{itemize}
\item 
Assume $\HoldsNot{\Obligation}{\Index}$.
In this case, Constraints~\ref{m:enc.u.lb.obl.open} and~\ref{m:enc.u.lb.obl.singular} are trivially satisfied.
\item 
Assume $\Hold{\Obligation}{\OUT}{\Index}$. 
Now if $\IntervalI{\Index}$ is open, then $\Holds{\EL}{\Index}$.
Furthermore, by Constraint~\ref{m:enc.u.lb.clk.reset} we know
$\St_{\Index+1}(\C)=0$ and $\Holds{\RightOpen}{\Index+1}$ iff $\IntervalI{\Index}$ is open.
Now
\begin{itemize}
\item If $\Hold{\Obligation}{\EL}{\Index+1}$, then Constraints~\ref{m:enc.u.lb.obl.open} and~\ref{m:enc.u.lb.obl.singular} are trivially satisfied.
\item If $\HoldsNot{\EL}{\Index+1}$, then the fact that $\ITrSuffix{\ITr}{\Index}{\cdot}\models\FOUT$ implies that $\Const=0$ and one of the following:
\begin{itemize}
\item $\Holds{\ER}{\Index}$ and $\IntervalI{\Index}$ is open. In this case, $\St_\Index(\D)>0$, meaning that $\Timing$ and $\TRight$ are satisfied on interval $\Index$ and ultimately that  Constraints~\ref{m:enc.u.lb.obl.open} and~\ref{m:enc.u.lb.obl.singular} are satisfied.
\item $\Holds{\ER}{\Index+1}$, $\IntervalI{\Index}$ is open and $\IntervalI{\Index+1}$ is a \singi . In this case, $\Holds{\RightOpen}{\Index+1}$, implying that $\Timing$ and $\TRight$ are satisfied on interval $\Index+1$ and ultimately that Constraints~\ref{m:enc.u.lb.obl.open} and~\ref{m:enc.u.lb.obl.singular} are satisfied.
\end{itemize}
\item If $\HoldsNot{\Obligation}{\Index+1}$ then by the fact that $\Holds{\OUT}{\Index}$ and Rule~\ref{swjrew1} for setting the value of $\Obligation$ we know that $\Const=0$. 
If we now assume that $\IntervalI{\Index}$ is a \singi , then by Rule~\ref{swjrew3} for setting the value of $\Obligation$ we know that $\TRight$ holds on interval $\Index+1$ and $\IntervalI{\Index+1}$ is a \singi . The latter, however, implies that $\St_{\Index+1}(\C)+\St_{\Index+1}(\D)=0$ and, thus, that $\TRight$ does not hold on interval $\Index+1$. Thus, assuming $\IntervalI{\Index}$ to be a \singi\ leads to a contradiction and we know that $\IntervalI{\Index}$ is open.

Now $\IntervalI{\Index+1}$ is a \singi\ and we know that $\St_{\Index+1}(\C)+\St_{\Index+1}(\D)=0$ and $\TRight$ is not satisfied on interval $\Index+1$.
Now Rule~\ref{swjrew2} for setting the value of $\Obligation$ and the fact that $\HoldsNot{\Obligation}{\Index+1}$ imply that $\TRight$ holds on interval $\Index$. As $\IntervalI{\Index}$ is open and $\ITrSuffix{\ITr}{\Index}{\cdot}\models\FOUT$, now $\Holds{\EL}{\Index}$ and Constraints~\ref{m:enc.u.lb.obl.open} and~\ref{m:enc.u.lb.obl.singular} are satisfied.
\end{itemize}
\item 
Assume $\Holds{\Obligation}{\Index}$ and $\HoldsNot{\OUT}{\Index}$.
Choose $\AnotherIndex<\Index$ as large as possible such that $\Holds{\OUT}{\AnotherIndex}$. We know that a corresponding interval exists based on the fact that $\Holds{\Obligation}{\Index}$.
Let  $\Strict\AnotherIndex=\AnotherIndex$ if $\IntervalI{\AnotherIndex}$ is open and  $\Strict\AnotherIndex=\AnotherIndex+1$ otherwise.

Now by the choice of $\AnotherIndex$ we know $\HoldsFromTo{\Obligation}{\AnotherIndex}{\Index}$.
\begin{itemize}
\item Assume $\Const>0$.
Based on the rules for setting the value of $\Obligation$, we now know that $\TRight$ does not hold on intervals $\AnotherIndex+2,\ldots,\Index$ and not on $\AnotherIndex+1$ either if $\IntervalI{\AnotherIndex+1}$ is open.
Furthermore, if $\IntervalI{\AnotherIndex+1}$ is a \singi\ as $\St_{\AnotherIndex+1}(\C)=0$ we know that $\St_{\AnotherIndex+1}(\C)+\St_{\AnotherIndex+1}(\D)=0$ meaning $\Timing$ (and, thus, $\TRight$) does not hold on interval $\AnotherIndex+1$.
Thus, $\TRight$ does not hold on intervals $\AnotherIndex+1,\ldots,\Index$.
\item Assume $\Const=0$. Based on the rules for setting the value for $\Obligation$, we now immediately know that $\TRight$ does not hold on intervals $\AnotherIndex+1,\ldots,\Index$ and not on interval $\AnotherIndex$ either if $\IntervalI{\AnotherIndex}$ is open.
\end{itemize}

Now as $\ITrSuffix{\ITr}{\AnotherIndex}{\cdot}\models\FOUT$, we can pick $\AThirdIndex\geq\AnotherIndex$ such that $\Holds{\ER}{\AThirdIndex}$, for every point in $\IntervalI{\AnotherIndex}$ there is a point in $\IntervalI{\AThirdIndex}$ that is more than $\Const$ time units away and $\HoldsFromTo{\EL}{\Strict\AnotherIndex}{\Strict\AThirdIndex}$ with $\Strict\AThirdIndex=\AThirdIndex$ if $\IntervalI{\AnotherIndex}$ is open and $\Strict\AThirdIndex=\AThirdIndex-1$ if $\IntervalI{\AThirdIndex}$ is a \singi .
We observe that $\AThirdIndex=\AnotherIndex$ is possible only if $\Const=0$ and $\IntervalI{\AnotherIndex}$ is open.
In this case, however, $\Timing$ and $\ER$ would both hold on interval $\AnotherIndex$, meaning that $\TRight$ holds and contradicting our previous observation that $\TRight$ does not hold on interval $\AnotherIndex$ if $\IntervalI{\AnotherIndex}$ is open and $\Const=0$. Thus, $\AThirdIndex>\AnotherIndex$.

Take an arbitrary $\AFourthIndex\in\Naturals$ with $\AnotherIndex<\AFourthIndex\leq\Index$.
Now we note that $\Holds{\RightOpen}{\AFourthIndex}$ iff $\IntervalI{\AnotherIndex}$ is open.
Furthermore,
 $\St_\AFourthIndex(\C)$ is the difference between the left bound of $\IntervalI{\AFourthIndex}$ interval and the right bound of $\IntervalI{\AnotherIndex}$.
Correspondingly, $\St_\AFourthIndex(\C)+\St_\AFourthIndex(\D)$ is the difference between the right bounds of $\IntervalI{\AFourthIndex}$ and $\IntervalI{\AnotherIndex}$. Now as $\TRight$ does not hold on interval $\AFourthIndex$, we know that either $\HoldsNot{\ER}{\AFourthIndex}$ or $\Timing$ does not hold. By the definition of $\Timing$, the latter implies that the difference between the right bounds of $\IntervalI{\AFourthIndex}$ and $\IntervalI{\AnotherIndex}$ is less than or equal to $\Const$  and less than $\Const$ if $\IntervalI{\AnotherIndex}$ is open, meaning that there is a point in $\IntervalI{\AnotherIndex}$ for which there is no point in $\IntervalI{\AFourthIndex}$ that is $>\Const$ time units away.
This allows us to conclude that $\AThirdIndex\neq\AFourthIndex$.
As we picked an $\AnotherIndex<\AFourthIndex\leq\Index$ this means that $\AThirdIndex>\Index$ and, thus, $\Holds{\EL}{\Index}$.
\begin{itemize}
\item Now if $\TRight$ holds on interval $\Index+1$ and $\IntervalI{\Index}$ is open, then Constraints~\ref{m:enc.u.lb.obl.open} and~\ref{m:enc.u.lb.obl.singular} are satisfied on interval $\Index$.
\item If $\TRight$ holds on interval $\Index+1$, $\IntervalI{\Index}$ is a \singi\ and $\IntervalI{\Index+1}$ is a \singi , then Constraints~\ref{m:enc.u.lb.obl.open} and~\ref{m:enc.u.lb.obl.singular} are satisfied on interval $\Index$.
\item If $\TRight$ holds on interval $\Index+1$, $\IntervalI{\Index}$ is a \singi\ and $\IntervalI{\Index+1}$ is open, according to Rule~\ref{swjrew3} for setting the value of $\Obligation$ we have $\Holds{\Obligation}{\Index+1}$.
Furthermore, as $\AThirdIndex>\Index$ and $\IntervalI{\Index+1}$ is open we have $\Strict\AThirdIndex\geq\Index+1$, implying that $\Holds{\EL}{\Index+1}$. Thus,
 Constraints~\ref{m:enc.u.lb.obl.open} and~\ref{m:enc.u.lb.obl.singular} are satisfied on interval $\Index$.
\item If  $\TRight$ does not hold on interval $\Index+1$, then we can use the same argument used to show that $\AThirdIndex > \Index$ to show that, in fact, $\AThirdIndex>\Index+1$, implying that $\Holds{\EL}{\Index+1}$. Furthermore, by the Rules~\ref{swjrew2} and~\ref{swjrew3} for setting the value of $\Obligation$ and the fact that $\TRight$ does not hold on intervals $\Index$ and $\Index+1$, we also know that $\Holds{\Obligation}{\Index+1}$.
Hence,  Constraints~\ref{m:enc.u.lb.obl.open} and~\ref{m:enc.u.lb.obl.singular} are satisfied on interval $\Index$.
\end{itemize}
\end{itemize}
 
\StartFairness 
It remains to be shown that the
fairness constraint~$\EncFair_{\FOUT}$
is satisfied by our choice of values, which will be done by contradiction.
Assume the fairness constraint is not satisfied. Then there is an $\Index\in\Naturals$ such that $\HoldsFrom{\Obligation}{\Index}$ and $\HoldsNotFrom{\ER}{\Index}$. As $\HoldsNotFrom{\ER}{\Index}$, we know that intervals $\Index,\Index+1,\ldots$ do not satisfy $\FOUT$ and, thus, $\HoldsNotFrom{\OUT}{\Index}$.
By the rules for setting the value of $\Obligation$, we set $\Obligation$ to $\True$ only if $\OUT$ hold on the current or the previous interval or $\Obligation$ holds on the previous interval. Thus, the fact that $\Holds{\Obligation}{\Index}$ implies that there is an interval before interval $\Index$ on which $\FOUT$ holds.

Pick $\AnotherIndex$ as large as possible such that $\AnotherIndex<\Index$ and $\Holds{\OUT}{\Index}$. Let $\Strict\AnotherIndex=\AnotherIndex$ if $\IntervalI{\AnotherIndex}$ is open and $\Strict\AnotherIndex=\AnotherIndex+1$ if $\IntervalI{\AnotherIndex}$ is a \singi .
As $\ITrSuffix{\ITr}{\AnotherIndex}{\cdot}\models\FOUT$, there is a $\AThirdIndex\geq\AnotherIndex$ such that $\Holds{\ER}{\AThirdIndex}$, for every point in $\IntervalI{\AnotherIndex}$ there is a point in $\IntervalI{\AThirdIndex}$ that is more than $\Const$ time units away and $\HoldsFromTo{\EL}{\Strict\AnotherIndex}{\Strict\AThirdIndex}$ with $\Strict\AThirdIndex=\AThirdIndex$ if $\IntervalI{\AnotherIndex}$ is open and $\Strict\AThirdIndex=\AThirdIndex-1$ if $\IntervalI{\AThirdIndex}$ is a \singi .
As $\HoldsNotFrom{\ER}{\Index}$, we know that $\AThirdIndex<\Index$.
Recall, that we picked $\AnotherIndex$ to be the last interval at which $\OUT$ holds. Hence, we know that $\Obligation$ on any later interval can only be set to $\True$ based on Rules~\ref{swjrew2} and~\ref{swjrew3}. As both of these rules require $\Obligation$ to hold on the respective previous interval we know that $\HoldsFrom{\Obligation}{\AnotherIndex}$.
\begin{itemize}
\item Assume that $\AThirdIndex=\AnotherIndex$. By the semantics of $\SUI{>\Const}$, this implies that $\Const=0$ and $\IntervalI{\Index}$ is open. Then $\Timing$ is satisfied on interval $\AnotherIndex$. As, additionally, $\Holds{\ER}{\AThirdIndex}$ and, thus, $\TRight$ holds, the rules for setting the value of $\Obligation$ imply that $\HoldsNot{\Obligation}{\Index+1}$, contradicting our observation that $\HoldsFrom{\Obligation}{\AnotherIndex}$.
\item Assume that $\AThirdIndex=\AnotherIndex+1$ and $\Const>0$. Then $\IntervalI{\AnotherIndex+1}$ is open and $\St_{\AnotherIndex+1}(\D)>\Const$. Thus, both $\Timing$ and $\ER$ hold on interval $\AnotherIndex+1=\AThirdIndex$, implying that $\TRight$ holds. As $\IntervalI{\AnotherIndex+1}$ is open, our rules for setting the value for $\Obligation$ now imply that $\HoldsNot{\Obligation}{\AnotherIndex+2}$, contradicting our observation that $\HoldsFrom{\Obligation}{\AnotherIndex}$.
\item Assume that $\AThirdIndex>\AnotherIndex+1$ or $\AThirdIndex=\AnotherIndex+1$ and $\Const=0$.
Then $\Obligation$ was set to $\True$ on interval $\AThirdIndex$
by Rule~\ref{swjrew2} or~\ref{swjrew3}
implying that $\TRight$ does not hold on interval $\AThirdIndex$ or $\IntervalI{\AThirdIndex}$ is open.
As $\Obligation$ is set to $\True$ on interval $\AThirdIndex+1$
by Rule~\ref{swjrew2} or~\ref{swjrew3} as well,  $\IntervalI{\AThirdIndex}$ being open again implies that $\TRight$ does not hold on interval $\AThirdIndex$.
Thus, $\TRight$ does not hold on interval $\AThirdIndex$ and $\Timing$ does not hold on interval $\AThirdIndex$, due to the fact that we picked $\AThirdIndex$ so that $\Holds{\ER}{\AThirdIndex}$.
By the fact that $\IntervalI{\AnotherIndex}$ is the last interval on which $\OUT$ holds, we know that $\C$ and $\RightOpen$ are set based on 
Constraint~\ref{m:enc.u.lb.clk.reset}
on interval $\AnotherIndex+1$ and based on
Constraint~\ref{m:enc.u.lb.clk.update}
on all later intervals.
This implies that at interval $\AThirdIndex$, the value of $\C+\D$ is the difference between the right bounds of $\IntervalI{\AThirdIndex}$ and $\IntervalI{\AnotherIndex}$. Furthermore, 
$\Holds{\RightOpen}{\AThirdIndex}$ iff $\IntervalI{\AnotherIndex}$ is open.
Now
\begin{itemize}
\item
If $\IntervalI{\AnotherIndex}$ is open, then by the fact that $\Timing$ does not hold on interval $\AThirdIndex$ we know that the difference between the right bounds of $\IntervalI{\AnotherIndex}$ and $\IntervalI{\AThirdIndex}$ is less than $\Const$. This, however, contradicts the fact that $\AThirdIndex$ was chosen such that for every point in $\IntervalI{\AnotherIndex}$ there is a point in $\IntervalI{\AThirdIndex}$ that is more than $\Const$ time units away.
\item
If $\IntervalI{\AnotherIndex}$ is a \singi , then the difference between the right bounds of $\IntervalI{\AnotherIndex}$ and $\IntervalI{\AThirdIndex}$ is less than or equal to $\Const$. Again, this contradicts the fact that for every point in $\IntervalI{\AnotherIndex}$ there is a point in $\IntervalI{\AThirdIndex}$ that is more than $\Const$ time units away.
\end{itemize}
\end{itemize}
Thus, assuming that the fairness constraint $\EncFair_{\FOUT}$ is not satisfied leads to a contradiction.
\end{IEEEproof}


\renewcommand{\StartRules}{\StartRulesT}
\renewcommand{\Op}{\SUI{\geq\Const}}
\renewcommand{\FOUT}{\LSF\Op\RSF}
\renewcommand{\SEnc}{\SEncT}
\begin{IEEEproof}
\emph{For ${\AnyTemporalOp}={\SUI{\geq\Const}}$.}
To obtain the proof for
$\SUI{\geq\Const}$,
we make the following changes to the proof for
$\SUI{>\Const}$.
\begin{itemize}
\item All cases in which $\Const=0$ are now contradictions, as we encode $\SUI{\geq 0}$ by the $\SU$ encoding.
\item No substantial changes are needed to show that Constraints~\ref{m:enc.u.lb.out.obl} and~\ref{m:enc.u.lb.out.xobl} are satisfied.
\item When showing that Constraints~\ref{m:enc.u.lb.obl.open} and~\ref{m:enc.u.lb.obl.singular} are satisfied:
\begin{itemize}
\item In the case where $\Hold{\Obligation}{\OUT}{\Index}$ the only valid option is that $\Hold{\Obligation}{\EL}{\Index+1}$, as all other cases required $\Const=0$.
\item In the case where $\Holds{\Obligation}{\Index}$ and $\HoldsNot{\OUT}{\Index}$, we pick $\AThirdIndex$ so that for every point in $\AnotherIndex$ there is a point at least $\Const$ time units away in $\IntervalI{\AThirdIndex}$ (and the properties regarding $\EL$ and $\ER$ hold).

After picking $\AFourthIndex$, we note that if $\Timing$ does  not hold on interval $\AFourthIndex$ then by the definition of $\Timing$ the difference between the right bounds of $\IntervalI{\AFourthIndex}$ and $\IntervalI{\AnotherIndex}$ is less than or equal to $\Const$  and less than $\Const$ if $\IntervalI{\AnotherIndex}$ is open or $\IntervalI{\AFourthIndex}$ is a \singi .
Then, there is a point in $\IntervalI{\AnotherIndex}$ for which there is no point in $\IntervalI{\AFourthIndex}$ that is $\geq\Const$ time units away.
Again, we conclude that $\AThirdIndex\neq\AFourthIndex$. We proceed as before.
\end{itemize}
\item When showing that the fairness constraint is satisfied we pick $\AThirdIndex$ so that for every point in $\AnotherIndex$ there is a point at least $\Const$ time units away in $\IntervalI{\AThirdIndex}$ (and the properties regarding $\EL$ and $\ER$ hold).

Then, the final case distinction is replaced by
\begin{itemize}
\item
If $\IntervalI{\AnotherIndex}$ is open or $\IntervalI{\AThirdIndex}$ a \singi , then by the fact that $\Timing$ does not hold on interval $\AThirdIndex$ we know that the difference between the right bounds of $\IntervalI{\AnotherIndex}$ and $\IntervalI{\AThirdIndex}$ is less than $\Const$. This, however, contradicts the fact that $\AThirdIndex$ was chosen such that for every point in $\IntervalI{\AnotherIndex}$ there is a point in $\IntervalI{\AThirdIndex}$ that is at least $\Const$ time units away.
\item
If $\IntervalI{\AnotherIndex}$ is a \singi\ and $\IntervalI{\AThirdIndex}$ an open interval, then the difference between the right bounds of $\IntervalI{\AnotherIndex}$ and $\IntervalI{\AThirdIndex}$ is less than or equal to $\Const$. Again, this contradicts the fact that for every point in $\IntervalI{\AnotherIndex}$ there is a point in $\IntervalI{\AThirdIndex}$ that is at least $\Const$ time units away.
\end{itemize}
\end{itemize}
\end{IEEEproof}


\renewcommand{\StartRules}{\StartRulesT}
\renewcommand{\Op}{\SR}
\renewcommand{\FOUT}{\LSF\Op\RSF}
\renewcommand{\SEnc}{\SEncT}
\begin{IEEEproof}
\emph{For ${\AnyTemporalOp}={\SR}$.}
\StartRules
We set $\Holds{\Obligation}{\Index}$ iff at least one of the following holds:
\begin{enumerate}
\item \label{sWBKR}
$\IntervalI{\Index}$ is open and $\Holds{\OUT}{\Index}$,
\item \label{lhfJi}
$\IntervalI{\Index-1}$ is a \singi\ and $\Holds{\OUT}{\Index-1}$, or
\item \label{zTYUZ}
$\Holds{\Obligation}{\Index-1}$ and $\HoldsNot{\EL}{\Index-1}$.
\end{enumerate}

\StartTransition
Constraints~\ref{eq:r.unt.out-open},~\ref{eq:r.unt.out-sing} and~\ref{eq:r.unt.consec} are trivially satisfied based on the rules for setting  the value of $\Obligation$. Thus, it only remains to be shown that Constraint~\ref{eq:r.unt.obl-r} is satisfied as well.
We distinguish the following cases:
\begin{itemize}
\item 
Case 1:
$\HoldsNot{\Obligation}{\Index}$. In this case, Constraint~\ref{eq:r.unt.obl-r} is trivially satisfied.
\item 
Case 2:
$\Holds{\Obligation}{\Index}$ and Rule~\ref{sWBKR} for setting the value of $\Obligation$ applies. That is, $\IntervalI{\Index}$ is open and $\Holds{\OUT}{\Index}$. As $\Holds{\OUT}{\Index}$, we know that $\ITrSuffix{\ITr}{\Index}{\cdot}\models\FOUT$. On any open interval satisfying $\FOUT$ either $\LSF$ or $\RSF$ (or both) must hold, immediately implying that Constraint~\ref{eq:r.unt.obl-r} is satisfied.
\item 
Case 3:
$\Holds{\Obligation}{\Index}$ and Rule~\ref{sWBKR} does not apply but Rule~\ref{lhfJi} does apply.
Then $\IntervalI{\Index-1}$ is a \singi\ and $\Holds{\OUT}{\Index-1}$, implying that $\ITrSuffix{\ITr}{\Index-1}{\cdot}\models\FOUT$. If $\IntervalI{\Index}$ itself is a \singi, then there is no time point in between the single time point in $\IntervalI{\Index-1}$ and the single time point in $\IntervalI{\Index}$. Thus, $\ITrSuffix{\ITr}{\Index-1}{\cdot}\models\FOUT$ implies that $\Holds{\ER}{\Index}$, in turn implying that Constraint~\ref{eq:r.unt.obl-r} is satisfied.
If, in contrast, $\IntervalI{\Index}$ is open, then $\ITrSuffix{\ITr}{\Index-1}{\cdot}\models\FOUT$ implies that either $\Holds{\EL}{\Index}$ or $\Holds{\ER}{\Index}$ (or both). Thus, also in this case Constraint~\ref{eq:r.unt.obl-r} is satisfied.
\item 
Case 4:
$\Holds{\Obligation}{\Index}$ and Rule~\ref{sWBKR} and Rule~\ref{lhfJi} do not apply. In this case, Rule~\ref{zTYUZ} has to apply, as we would not have set $\Holds{\Obligation}{\Index}$ if no rule applied. Thus, $\Holds{\Obligation}{\Index-1}$ and $\HoldsNot{\EL}{\Index-1}$.
We now choose $\AnotherIndex<\Index$ as large as possible such that (i) $\Holds{\Obligation}{\AnotherIndex}$ and (ii) either $\AnotherIndex=0$ or $\HoldsNot{\Obligation}{\AnotherIndex-1}$.
Now we know that $\Holds{\Obligation}{\AnotherIndex}$ based on
Rule~\ref{sWBKR} or Rule~\ref{lhfJi}. Let $\Strict\AnotherIndex=\AnotherIndex$ iff Rule~\ref{sWBKR} applies to interval $\AnotherIndex$ and $\Strict\AnotherIndex=\AnotherIndex-1$ otherwise. We now know that $\Holds{\OUT}{\Strict\AnotherIndex}$ and, thus, $\ITrSuffix{\ITr}{\Strict\AnotherIndex}{\cdot}\models\FOUT$. Furthermore, as $\Obligation$ propagated up to interval $\Index$ through Rule~\ref{zTYUZ}, we know that $\HoldsNotFromTo{\EL}{\AnotherIndex}{\Index-1}$. Now take any $\FirstOffset\in\IntervalI{\Index}$ and $\SecondOffset\in\IntervalI{\AnotherIndex}$.
We now perform another case distinction based on the type of $\IntervalI{\Index}$.
\begin{itemize}
\item
If $\IntervalI{\Index}$ is a \singi , then for any $\Point{\AThirdIndex}{\Offset}\in\Timepoints{\ITr}$ with $\Point{\Strict\AnotherIndex}{\SecondOffset}\Earlier\Point{\AThirdIndex}{\Offset}\Earlier\Point{\Index}{\FirstOffset}$ it holds that $\AnotherIndex\leq\AThirdIndex<\Index$, implying that $\HoldsNot{\EL}{\AThirdIndex}$. Thus, there is no time point in between  $\Point{\Strict\AnotherIndex}{\SecondOffset}$ and $\Point{\Index}{\FirstOffset}$ at which $\LSF$ holds. As $\ITrSuffix{\ITr}{\Strict\AnotherIndex}{\SecondOffset}\models\FOUT$, this means that $\Holds{\ER}{\Index}$ and Constraint~\ref{eq:r.unt.obl-r} is satisfied.
\item
If $\IntervalI{\Index}$ is open, then for any $\Point{\AThirdIndex}{\Offset}\in\Timepoints{\ITr}$ with $\Point{\Strict\AnotherIndex}{\SecondOffset}\Earlier\Point{\AThirdIndex}{\Offset}\Earlier\Point{\Index}{\FirstOffset}$ it holds that $\AnotherIndex\leq\AThirdIndex\leq\Index$, meaning that $\Holds{\EL}{\AnotherIndex}$ implies $\AThirdIndex=\Index$. Thus, if there is a time point in between  $\Point{\Strict\AnotherIndex}{\SecondOffset}$ and $\Point{\Index}{\FirstOffset}$ at which $\LSF$ holds, that time point has to be part of interval $\Index$, meaning that $\Holds{\EL}{\Index}$. If there is no such time point, then the fact that $\ITrSuffix{\ITr}{\Strict\AnotherIndex}{\SecondOffset}\models\FOUT$ implies that $\Holds{\ER}{\Index}$. Hence, we have $\Holds{\EL}{\Index}$ or $\Holds{\ER}{\Index}$ (or both) and Constraint~\ref{eq:r.unt.obl-r} is satisfied.
\end{itemize}
\end{itemize}
Either way, we were able to demonstrate that Constraint~\ref{eq:r.unt.obl-r} is satisfied.
\end{IEEEproof}


\renewcommand{\StartRules}{\StartRulesO}
\renewcommand{\Op}{\SGI{\leq0}}
\renewcommand{\FOUT}{\Op\LSF}
\renewcommand{\SEnc}{\SEncO}

\begin{IEEEproof}
\emph{For ${\AnyTemporalOp}={\SGI{\leq 0}}$.}
\StartRules

\StartTransition
Now
\begin{itemize}
\item
If $\HoldsNot{\OUT}{\Index}$ or $\IntervalI{\Index}$ is open or $\IntervalI{\Index+1}$ is open, then 
Constraint~\ref{eq:g.leqz}
is trivially satisfied on interval $\Index$.
\item
If, $\Holds{\OUT}{\Index}$ and $\IntervalI{\Index}$ is a \singi\ and $\IntervalI{\Index+1}$ is a \singi , then pick $\AnotherIndex\geq\Index+1$ as large as possible such that  $\IntervalI{\Index},\ldots,\IntervalI{\AnotherIndex}$ are all \singis .
$\Holds{\OUT}{\Index}$ means that
$\ITrSuffix{\ITr}{\Index}{\cdot}\models\FOUT$.
Thus, we know that $\HoldsFromTo{\EL}{\Index+1}{\AnotherIndex}$.
This implies that
$\ITrSuffix{\ITr}{\Index+1}{\cdot}\models\OUT$
as well and, thus, $\Holds{\OUT}{\Index+1}$.
Thus, 
Constraint~\ref{eq:g.leqz}
is satisfied on interval $\Index$ in this case as well.
\end{itemize}
Hence, 
Constraint~\ref{eq:g.leqz}
is in each case satisfied on interval $\Index$.
\end{IEEEproof}


\renewcommand{\StartRules}{\StartRulesO}
\renewcommand{\Op}{\SG_{<\Const}}
\renewcommand{\FOUT}{\Op\LSF}
\renewcommand{\SEnc}{\SEncO}

\begin{IEEEproof}
\emph{For ${\AnyTemporalOp}={\SG_{<\Const}}$.}
\StartRules
We set $\St_0(\C)=\Const+1$ and $\Holds{\RightOpen}{0}$.
\ClkUpdateDesc{\RightOpen}{eq:g.ub.clk1}{eq:g.ub.clk2}

\StartTransition

It remains to show that Constraint~\ref{eq:g.ub.timing} is satisfied on interval $\Index$.
\begin{itemize}
\item
Case 1: $\Timing$ does not hold on interval $\Index$ and $\HoldsNot{\OUT}{\Index}$ or $\IntervalI{\Index}$ is a \singi .
In this case Constraint~\ref{eq:g.ub.timing} is trivially satisfied.
\item
Case 2: $\Holds{\OUT}{\Index}$ an $\IntervalI{\Index}$ is open. As $\ITrSuffix{\ITr}{\Index}{\cdot}\models\FOUT$ and $\IntervalI{\Index}$ is open, we now know that $\Holds{\EL}{\Index}$.
Thus, Constraint~\ref{eq:g.ub.timing} is satisfied.
\item
Case 3: $\Timing$ holds on interval $\Index$.
Then we note that there is a previous interval on which $\OUT$ holds. If no such previous interval would exist, then $\St_\Index(\C)$ would be at least the initial value of $\Const+1$, contradicting the  assumption that $\Timing$ holds. We can, thus, pick $\AnotherIndex<\Index$ is as large as possible such that $\Holds{\OUT}{\AnotherIndex}$, implying $\ITrSuffix{\ITr}{\AnotherIndex}{\cdot}\models\FOUT$.

As is easy to see, repeated application of Constraints~\ref{eq:g.ub.clk1}
and~\ref{eq:g.ub.clk2} leads to $\St_\Index(\C)$ being the difference between the left bound of $\IntervalI{\Index}$ and the right bound of $\IntervalI{\AnotherIndex}$. Furthermore, $\Holds{\RightOpen}{\Index}$ iff $\IntervalI{\AnotherIndex}$ is open.
As $\Timing$ holds at interval $\Index$ we know that $\St_\Index(\C)<\Const$. That is, the difference between left bound of $\IntervalI{\Index}$ and the right bound of $\IntervalI{\AnotherIndex}$  is less than $\Const$ time units. Indiscriminately of whether $\IntervalI{\AnotherIndex}$ and $\IntervalI{\Index}$ are open or \singis , this implies that there are $\FirstOffset\in\IntervalI{\Index}$ and $\SecondOffset\in\IntervalI{\AnotherIndex}$ with $\FirstOffset-\SecondOffset<\Const$. This implies that $\Holds{\EL}{\Index}$, as $\ITrSuffix{\ITr}{\AnotherIndex}{\cdot}\models\FOUT$. Thus,
Constraint~\ref{eq:g.ub.timing} is satisfied in this case as well.
\end{itemize}
\end{IEEEproof}


\renewcommand{\StartRules}{\StartRulesO}
\renewcommand{\Op}{\SG_{<\Const}}
\renewcommand{\FOUT}{\Op\LSF}
\renewcommand{\SEnc}{\SEncO}

\begin{IEEEproof}
\emph{For ${\AnyTemporalOp}={\SG_{\leq\Const}}$.}
The proof for ${\AnyTemporalOp}={\SG_{\leq\Const}}$ proceeds precisely as the proof for ${\AnyTemporalOp}={\SG_{<\Const}}$ up to the point in Case 2 where we observe that $\St_\Index(\C)<\Const$. For ${\AnyTemporalOp}={\SG_{\leq\Const}}$, we observe instead that $\St_\Index(\C)<\Const$ or $\St_\Index(\C)\leq\Const$ and intervals $\IntervalI{\Index}$ and $\IntervalI{\AnotherIndex}$ are both \singis .
Thus, we can pick time points in $\IntervalI{\Index}$ and $\IntervalI{\AnotherIndex}$ that are $\leq\Const$ time units apart. Hence, the fact that $\ITrSuffix{\ITr}{\AnotherIndex}{\cdot}\models\FOUT$ again implies $\Holds{\EL}{\Index}$ and Constraint~\ref{eq:g.ub.timing} is satisfied.
\end{IEEEproof}


\renewcommand{\StartRules}{\StartRulesT}
\renewcommand{\Op}{\SRI{>\Const}}
\renewcommand{\FOUT}{\LSF\Op\RSF}
\renewcommand{\SEnc}{\SEncT}
\begin{IEEEproof}
\emph{For ${\AnyTemporalOp}={\SRI{>\Const}}$.}
\StartRules
We set $\St_0(\C)=0$ and $\HoldsNot{\LeftOpen}{0}$.
\ClkUpdateDescT{\LeftOpen}{eq:nr.lb.clk.delayed}{eq:nr.lb.clk.reset}{eq:nr.lb.clk.update}
For setting the value of $\Obligation$, we use the same rules used for the untimed release $\SR$. That is, we set $\Holds{\Obligation}{\Index}$ iff at least one of the following holds:
\begin{enumerate}
\item \label{RIsWBKR}
$\IntervalI{\Index}$ is open and $\Holds{\OUT}{\Index}$,
\item \label{RIlhfJi}
$\IntervalI{\Index-1}$ is a \singi\ and $\Holds{\OUT}{\Index-1}$, or
\item \label{RIzTYUZ}
$\Holds{\Obligation}{\Index-1}$ and $\HoldsNot{\EL}{\Index-1}$.
\end{enumerate}

\StartInitial
Initial Constraint $\EncInit_{\FOUT}$ is trivially satisfied as $\St_0(\C)=0$ and $\HoldsNot{\LeftOpen}{0}$.

\StartTransition
We observe that Constraints~\ref{eq:nr.lb.left.open},~\ref{eq:nr.lb.left.singular} and~\ref{eq:nr.lb.obligation.update} correspond to Constraints~\ref{eq:r.unt.out-open},~\ref{eq:r.unt.out-sing} and~\ref{eq:r.unt.consec} in the encoding of the untimed release operator.
Consequently, Constraints~\ref{eq:nr.lb.left.open},~\ref{eq:nr.lb.left.singular} and~\ref{eq:nr.lb.obligation.update} are satisfied by the fact that we use the exact same rules for setting the value of $\St_\Index(\Obligation)$.

The left hand side of the implication in Constraint~\ref{eq:nr.lb.first} is satisfied precisely if $\ITrSuffix{\ITr}{\Index}{\cdot}\models\FOUT$ and $\IntervalI{\Index}$ is an open interval whose bounds are more than $\Const$ time units apart. Then, there are time points $\FirstOffset,\SecondOffset\in\IntervalI{\Index}$ such that $\SecondOffset-\FirstOffset>\Const$. By the semantics of $\SR$, this implies that $\RSF$ holds at $\Point{\Index}{\SecondOffset}$ or that $\LSF$ holds somewhere in between $\Point{\Index}{\FirstOffset}$ and $\Point{\Index}{\SecondOffset}$. Thus, $\Holds{\EL}{\Index}$ or $\Holds{\ER}{\Index}$ and the right hand side of the implication is satisfied as well.

We now argue that Constraint~\ref{eq:nr.lb.timing} is satisfied on interval $\Index$ using a case distinction:

\textbf{Case 1}:
$\HoldsNot{\Obligation}{\Index}$
\emph{or}
$\Timing$ is not satisfied at interval $\Index$
\emph{or}
interval $\IntervalI{\Index}$ is open and $\Holds{\EL}{\Index}$.
In this case, Constraint~\ref{eq:nr.lb.timing} is trivially satisfied at interval $\Index$.

\textbf{Case 2}:
$\Holds{\Obligation}{\Index}$, either $\IntervalI{\Index}$ is a \singi\ or $\HoldsNot{\EL}{\Index}$ and $\Timing$ is satisfied at interval $\Index$.
Note that $\Timing$ holding at interval $\Index$ implies that $\St_{\Index}(\C)>0$.
Now pick $\AnotherIndex\leq\Index$ as large as possible such that one of the following:
(i) $\AnotherIndex<\Index$ and $\DelayReset$ holds at interval $\AnotherIndex$, (ii) $\EReset$ holds at interval $\AnotherIndex-1$ or (iii) $\AnotherIndex=0$.
By the choice of $\AnotherIndex$ we know that (a) $\EReset$ does not hold at intervals $\AnotherIndex\ldots\Index-1$ (b) $\DelayReset$ does not hold at intervals $\AnotherIndex+1,\ldots,\Index-1$
and (c) $\Holds{\OUT}{\AnotherIndex}$. If $\AnotherIndex$ was chosen based on (i) or (ii), this immediately follows from the definition of $\DelayReset$ and $\EReset$. If, in contrast, $\AnotherIndex$ was chosen based on (iii) (implying $\AnotherIndex=0$) then assuming $\HoldsNot{\OUT}{\AnotherIndex}$ leads to the observation that both $\OUT$ and $\Obligation$ are $\False$ on a prefix of $\STTSRun$. Thus $\EReset$ holds on the first interval on whose successor $\OUT$ holds. As $\Holds{\OUT}{\Index}$, this the case on is interval $\Index-1$ at the latest, contradicting the assumption that we picked $\AnotherIndex$ based on (iii). Thus, $\Holds{\OUT}{\AnotherIndex}$ in each case.

Based on the update rules for $\C$ and $\LeftOpen$, we know that $\Holds{\LeftOpen}{\Index}$ iff $\IntervalI{\AnotherIndex}$ is open.
Furthermore,
$\St_{\Index}(\C)$  is the difference between the left bound of $\IntervalI{\Index}$ and the left bound of $\IntervalI{\AnotherIndex}$, i.e. $\St_{\Index}(\C)+\St_{\Index}(\D)$ is the difference of the right bound of $\IntervalI{\Index}$ and the left bound of $\IntervalI{\AnotherIndex}$.
By our Case 2 assumption that $\Timing$ is satisfied, this difference is greater than $\Const$. Therefore, we can indiscriminately of the type of $\IntervalI{\Index}$ and $\IntervalI{\AnotherIndex}$ pick $\FirstOffset\in\IntervalI{\Index}, \SecondOffset\in\IntervalI{\AnotherIndex}$ such that $\FirstOffset-\SecondOffset>\Const$.

Let
$\Strict\Index\Def\Index$ if $\IntervalI{\Index}$ is open and
$\Strict\Index\Def\Index-1$ otherwise.
Furthermore, let
$\Strict\AnotherIndex\Def\AnotherIndex$ if $\IntervalI{\AnotherIndex}$ is open and
$\Strict\AnotherIndex\Def\AnotherIndex+1$ otherwise.
Note that $\Strict\Index\geq\Strict\AnotherIndex$. (As $\Index>\AnotherIndex$, we know that $\Strict\Index\geq\Strict\AnotherIndex-1$. Additionally, $\Strict\Index=\Strict\AnotherIndex-1$ would require that $\Index=\AnotherIndex+1$ and that both $\IntervalI{\Index}$ and $\IntervalI{\AnotherIndex}$ to be \singis\ which contradicts $\C+\D>\Const$ at interval $\Index$)
Furthermore, note that the time points lying in between $\Point{\AnotherIndex}{\FirstOffset}$ and $\Point{\Index}{\SecondOffset}$ now all belong to $\IntervalI{\Strict\AnotherIndex},\ldots,\IntervalI{\Strict\Index}$.
We now claim that both
$\HoldsFromTo{\Obligation}{\Strict\AnotherIndex}{\Strict\Index}$
and
$\HoldsNotFromTo{\EL}{\Strict\AnotherIndex}{\Strict\Index}$.
Note, that by our Case 2 assumptions we have $\Holds{\Obligation}{\Index}$ and $\HoldsNot{\EL}{\Index}$ if $\IntervalI{\Index}$ is open (meaning $\Strict\Index=\Index$). Thus, we only have to show that
$\HoldsFromTo{\Obligation}{\Strict\AnotherIndex}{\Index-1}$
and
$\HoldsNotFromTo{\EL}{\Strict\AnotherIndex}{\Index-1}$, which will proven by induction over $\AThirdIndex=\Strict\AnotherIndex,\ldots,\Index-1$, for each $\AThirdIndex$ assuming that we established
$\HoldsFromTo{\Obligation}{\Strict\AnotherIndex}{\AThirdIndex-1}$
and
$\HoldsNotFromTo{\EL}{\Strict\AnotherIndex}{\AThirdIndex-1}$ already.

\textbf{Base case:} $\AThirdIndex=\Strict\AnotherIndex$.
$\Holds{\Obligation}{\Strict\AnotherIndex}$ follows immediately from the fact that $\Holds{\OUT}{\AnotherIndex}$ and Rules~\ref{RIsWBKR} and~\ref{RIlhfJi} for setting the value of $\Obligation$.

Now assume $\Holds{\EL}{\AThirdIndex}$.
Recall, that $\Holds{\Obligation}{\Index}$.
This allows us to pick a $\AFourthIndex$ as small as possible such that $\AThirdIndex<\AFourthIndex\leq\Index$ and $\Holds{\Obligation}{\AFourthIndex}$. Then:

\begin{itemize}
\item If $\AFourthIndex=\AThirdIndex+1$, then based on the fact that $\Holds{\EL}{\AThirdIndex}$, we know that $\Obligation$ was set to $\True$ on interval $\AFourthIndex$ not based on Rule~\ref{RIzTYUZ} but based on Rule~\ref{RIsWBKR} or~\ref{RIlhfJi}.
\item If $\AFourthIndex>\AThirdIndex+1$, then by the choice of $\AFourthIndex$ we know that $\HoldsNot{\Obligation}{\AFourthIndex-1}$. Thus, we deduct that, again, $\Obligation$ was not set to $\True$ on interval $\AFourthIndex$ based on Rule~\ref{RIzTYUZ} but based on Rule~\ref{RIsWBKR} or~\ref{RIlhfJi}.
\end{itemize}

Now we split based on the rule by which $\Obligation$ was set to $\True$ on interval $\AFourthIndex$.
\begin{itemize}
\item 
If $\Obligation$ was set to $\True$ based on Rule~\ref{RIsWBKR} and Rule~\ref{RIlhfJi} does not apply,
then $\Holds{\OUT}{\AFourthIndex}$ and $\HoldsNot{\OUT}{\AFourthIndex-1}$ (as otherwise Rule~\ref{RIlhfJi} would apply).
Furthermore, $\IntervalI{\AFourthIndex}$ is open, implying that  $\IntervalI{\AFourthIndex-1}$ is a\singi .
\begin{itemize}
\item If $\AFourthIndex=\AThirdIndex+1$, then
as $\Holds{\EL}{\AThirdIndex}$, $\HoldsNot{\OUT}{\AThirdIndex}$ and $\Holds{\OUT}{\AThirdIndex+1}$ we conclude that $\EReset$ holds at interval $\AFourthIndex-1=\AThirdIndex>\AnotherIndex$, contradicting observation (a).
\item If $\AFourthIndex>\AThirdIndex+1$, then $\HoldsNot{\Obligation}{\AFourthIndex-1}$. As additionally $\Holds{\OUT}{\AFourthIndex}$, $\HoldsNot{\OUT}{\AThirdIndex}$, we know that $\EReset$ holds on interval $\AFourthIndex-1>\AThirdIndex\geq\AnotherIndex$, again contradicting observation (a).
\end{itemize}
\item 
If $\Obligation$ was set to $\True$ based on Rule~\ref{RIlhfJi}, then $\IntervalI{\AFourthIndex-1}$ is a \singi\ and $\Holds{\OUT}{\AFourthIndex-1}$.
Now:
\begin{itemize}
\item 
If $\AFourthIndex=\AThirdIndex+1$, then $\IntervalI{\AThirdIndex}$ is a \singi , implying that $\AThirdIndex>\AnotherIndex$ (as $\AThirdIndex\geq\Strict\AnotherIndex$ it is not possible that $\AThirdIndex=\AnotherIndex$ when $\IntervalI{\AThirdIndex}$ is a \singi ).
Now $\ThreeHoldAt{\EL}{\OUT}{\Obligation}{\AThirdIndex}$,
$\IntervalI{\AThirdIndex}$ is a \singi
and
$\DelayReset$ holds at interval $\AThirdIndex>\AnotherIndex$. This contradicts observation (b).
\item 
If $\AFourthIndex=\AThirdIndex+2$, we further split cases based on the type of $\IntervalI{\AThirdIndex}$.
\begin{itemize}
\item If $\IntervalI{\AThirdIndex}$ is open, then $\Holds{\EL}{\AThirdIndex}$ and $\Holds{\OUT}{\AThirdIndex+1}$, meaning that $\EReset$ is satisfied at interval $\AThirdIndex$, again contradicting (a).
\item If $\IntervalI{\AThirdIndex}$ is a \singi , then we again observe that $\AThirdIndex>\AnotherIndex$. Now one last split is necessary:
\begin{itemize}
\item If $\Holds{\OUT}{\AThirdIndex}$, then $\DelayReset$ holds at interval $\AThirdIndex>\AnotherIndex$, contradicting (b).
\item If $\HoldsNot{\OUT}{\AThirdIndex}$, then $\EReset$ holds on interval $\AThirdIndex$ based on the fact that, additionally, $\Holds{\EL}{\AThirdIndex}$ and $\Holds{\OUT}{\AThirdIndex+1}$, contradicting (a).
\end{itemize}
\end{itemize}
\item 
Assume $\AFourthIndex>\AThirdIndex+2$.
By the choice of $\AFourthIndex$ we know that
$\HoldsNot{\Obligation}{\AFourthIndex-2}$ and $\HoldsNot{\Obligation}{\AFourthIndex-1}$, implying that $\HoldsNot{\OUT}{\AFourthIndex-2}$
Now $\EReset$ holds at interval $\AFourthIndex-2>\AThirdIndex\geq\AnotherIndex$ due to the fact that additionally $\Holds{\OUT}{\AFourthIndex-1}$, again contradicting (a).
\end{itemize}
\end{itemize}
As each case ended in a contradiction we conclude that $\HoldsNot{\EL}{\AThirdIndex}$.

\textbf{Inductive step:} $\Strict\AnotherIndex<\AThirdIndex<\Index$.
Now
$\Holds{\Obligation}{\AThirdIndex}$ follows from the inductive hypothesis that
$\Holds{\Obligation}{\AThirdIndex-1}$
and
$\HoldsNot{\EL}{\AThirdIndex-1}$
and Rule~\ref{RIzTYUZ} for setting the value of $\Obligation$.
Furthermore, we can derive that $\HoldsNot{\EL}{\AThirdIndex}$ by the same arguments used in the base case.
Thus, we have shown inductively that
$\HoldsFromTo{\Obligation}{\Strict\AnotherIndex}{\Index-1}$
and
$\HoldsNotFromTo{\EL}{\Strict\AnotherIndex}{\Index-1}$.

Now as $\HoldsNotFromTo{\EL}{\Strict\AnotherIndex}{\Strict\Index}$ we know that
there is no time point in between $\Point{\AnotherIndex}{\SecondOffset}$ and $\Point{\Index}{\FirstOffset}$ at which $\LSF$ holds.
As $\ITrSuffix{\ITr}{\AnotherIndex}{\SecondOffset}\models\FOUT$ and $\FirstOffset-\SecondOffset>\Const$ we can conclude that $\RSF$ holds at time point $\Point{\Index}{\FirstOffset}$, meaning that $\Holds{\ER}{\Index}$ and ultimately implying that Constraint~\ref{eq:nr.lb.timing} is satisfied at interval $\Index$.
\end{IEEEproof}


\renewcommand{\StartRules}{\StartRulesT}
\renewcommand{\Op}{\SRI{\geq\Const}}
\renewcommand{\FOUT}{\LSF\Op\RSF}
\renewcommand{\SEnc}{\SEncT}
\begin{IEEEproof}
\emph{For ${\AnyTemporalOp}={\Op}$.}
To adapt the  proof for ${\AnyTemporalOp}={\SRI{>\Const}}$ to ${\AnyTemporalOp}={\Op}$ we only need to argue that in Case 2, based on the fact that $\Timing$ holds we can pick $\FirstOffset\in\IntervalI{\Index}$, $\SecondOffset\in\IntervalI{\AnotherIndex}$ with $\FirstOffset-\SecondOffset\geq\Const$. As $\Timing$ holds, we know that the difference between the right bound of $\IntervalI{\Index}$ and the left bound of $\IntervalI{\AnotherIndex}$ is $>\Const$ or the difference is $\geq\Const$ and both $\IntervalI{\Index}$ and $\IntervalI{\AnotherIndex}$ are \singis . In both cases, we can pick $\FirstOffset$, $\SecondOffset$ with $\FirstOffset-\SecondOffset\geq\Const$.
\end{IEEEproof}

%
%
\subsection{Proof of Lemma~\ref{lem:sddfuiet}}

\begin{relemma}{\ref{lem:sddfuiet}}
  Assume two states, $\St$ and $\AnotherSt$,
  such that $\St \REquiv \AnotherSt$.
  It holds that
  (i) $\St \Models \EncInit$ iff $\AnotherSt \Models \EncInit$, and
  (ii) $\St \Models \EncInv$ iff $\AnotherSt \Models \EncInv$.
  Furthermore,
  if there is
  a $\Delay_\St \in \RealsNonNeg$ and
  a state $\St'$
  such that
  ${\St \cup \Set{\Delay \mapsto \Delay_\St} \cup \Setdef{y' \mapsto \St'(y)}{y \in {\Clocks \cup \NonClocks}}} \Models \EncTr$,
  then
  there is a $\Delay_\AnotherSt \in \RealsNonNeg$
  and a state $\AnotherSt'$
  such that
  ${\AnotherSt \cup \Set{\Delay \mapsto \Delay_\AnotherSt} \cup \Setdef{y' \mapsto \AnotherSt'(y)}{y \in {\Clocks \cup \NonClocks}}} \Models \EncTr$
  and
  $\St' \REquiv \AnotherSt'$.
\end{relemma}
\begin{IEEEproof}
  As
  $\St \REquiv \AnotherSt$ and
  the only atoms involving clock variables in $\EncInit$ and $\EncInv$
  are of form $\Clock \AnyIneq \Const$,
  the definitions of $\ClockMax{\Clock}$ and $\REquiv$
  directly imply that
  (i) $\St \Models \EncInit$ iff $\AnotherSt \Models \EncInit$, and
  (ii) $\St \Models \EncInv$ iff $\AnotherSt \Models \EncInv$.

  To prove the remaining claim,
  consider the state $\St''$ such that
  (i) $\St''(\Clock) = \St(\Clock)+\Delay_\St$ for each clock $\Clock \in \Clocks$, and
  (ii) $\St''(\NonClock) = \St'(\NonClock)$ for each non-clock $\NonClock \in \NonClocks$.
  As ${\St \cup \Set{\Delay \mapsto \Delay_\St} \cup \Setdef{y' \mapsto \St'(y)}{y \in {\Clocks \cup \NonClocks}}} \Models \EncTr$,
  we have $\Delay_\St \ge 0$ and
  for each $\Clock \in \Clocks$ either
  $\St'(\Clock) = 0$ or
  $\St'(\Clock) = \St(\Clock)+\Delay_\St = \St''(\Clock)$.
  That is,
  intuitively $\St''$ is obtained from $\St'$
  by ``unresetting'' the reset clocks.

  Next,
  take any $\Delay_\AnotherSt \in \Reals$
  and
  state $\AnotherSt''$
  such that
  (i) $\Delay_\AnotherSt \ge 0$,
  (ii) $\Delay_\AnotherSt = 0 \Iff \Delay_\St = 0$,
  (iii) $\AnotherSt''(\Clock) = \AnotherSt(\Clock)+\Delay_\AnotherSt$ for each clock $\Clock \in \Clocks$,
  (iv) $\AnotherSt''(\NonClock) = \St'(\NonClock)$ for each non-clock $\NonClock \in \NonClocks$,
  and
  (v) $\St'' \REquiv \AnotherSt''$.
  Such $\Delay_\AnotherSt$ and $\AnotherSt''$ exists
  because $\St \REquiv \AnotherSt$ and
  of the fact that clock valuations in the same region have
  time successors in same regions \cite{AlurDill:TCS1994}.
  Let $\AnotherSt'$ be the state such that
  (i) for each $\Clock \in \Clocks$,
  $\AnotherSt'(\Clock) = 0$ if $\St'(\Clock)=0$ and
  $\AnotherSt'(\Clock) = \AnotherSt''(\Clock) =\AnotherSt(\Clock)+\Delay_\AnotherSt$ otherwise,
  and
  (ii) $\AnotherSt'(\NonClock) = \AnotherSt''(\NonClock) = \St'(\NonClock)$ for each non-clock $\NonClock \in \NonClocks$.
  Now $\St' \REquiv \AnotherSt'$.
  As a summary, intuitively $\AnotherSt'$ is a state in the region that is obtained by letting time pass in the similar manner as when moving from $\St$ to $\St'$ and then resetting the same clocks.

  Now we only have to show that ${\AnotherSt \cup \Set{\Delay \mapsto \Delay_\AnotherSt} \cup \Setdef{y' \mapsto \AnotherSt'(y)}{y \in {\Clocks \cup \NonClocks}}} \Models \EncTr$.
  We do this by showing that the atoms in $\EncTr$ evaluate to the same boolean value under both
  $\St \cup \Set{\Delay \mapsto \Delay_\St} \cup \Setdef{y' \mapsto \St'(y)}{y \in {\Clocks \cup \NonClocks}}$
  and
  $\AnotherSt \cup \Set{\Delay \mapsto \Delay_\AnotherSt} \cup \Setdef{y' \mapsto \AnotherSt'(y)}{y \in {\Clocks \cup \NonClocks}}$.
  \begin{itemize}
  \item
    Case: the atom does not involve variables in $\Clocks \cup \ClocksNext \cup \Set{\delta}$.

    In this case the atom evaluates to true under
    $\St \cup \Setdef{y' \mapsto \St'(y)}{y \in {\Clocks \cup \NonClocks}}$
    if and only if
    it does under $\AnotherSt \cup \Setdef{y' \mapsto \AnotherSt'(y)}{y \in {\Clocks \cup \NonClocks}}$
    because
    $\St \REquiv \AnotherSt$
    and
    $\St' \REquiv \AnotherSt'$.

  \item
    Case: the atom is of form $\Clock' = 0$. 

    Because $\St' \REquiv \AnotherSt'$,
    the atom evaluates to true under $\Setdef{y' \mapsto \St'(y)}{y \in {\Clocks \cup \NonClocks}}$
    if and only if
    it does under $\Setdef{y' \mapsto \AnotherSt'(y)}{y \in {\Clocks \cup \NonClocks}}$.

  \item
    Case: the atom is of form $\Clock' = \Clock+\delta$. 

    We have to consider the following:
    \begin{enumerate}
    \item
      Sub-case $\St'(\Clock) = 0$.
   
      Thus $\AnotherSt'(\Clock) = 0$ as well because $\St' \REquiv \AnotherSt'$.
      Now $\Clock' = \Clock+\delta$ evaluates to true under 
      $\St \cup \Set{\Delay \mapsto \Delay_\St} \cup \Setdef{y' \mapsto \St'(y)}{y \in {\Clocks \cup \NonClocks}}$
      if and only if
      $\St(\Clock) = 0$ and $\Delay_\St = 0$
      (as $\Clock$ and $\delta$ always have non-negative values).
      \begin{enumerate}
      \item 
        If $\St(\Clock) = 0$ and $\Delay_\St = 0$,
        then $\AnotherSt(\Clock) = 0$ and $\Delay_\AnotherSt = 0$ as well
        because $\St \REquiv \AnotherSt$ and
        $\St'' \REquiv \AnotherSt''$
        (forcing that $\St(\Clock)+\Delay_\St = 0$
        if and only if
        $\AnotherSt(\Clock)+\Delay_\AnotherSt = 0$).

      \item 
        If $\St(\Clock) > 0$,
        then $\AnotherSt(\Clock) > 0$ as $\St \REquiv \AnotherSt$,
        and thus
        $\St'(\Clock) \neq \St(\Clock)+\Delay_\St$
        and
        $\AnotherSt'(\Clock) \neq \AnotherSt(\Clock)+\Delay_\AnotherSt$.
        %
      
      \item 
        If $\St(\Clock) = 0$ and $\Delay_\St > 0$,
        then $\AnotherSt(\Clock) = 0$
        as
\newpage 
        $\St \REquiv \AnotherSt$ and
        $\Delay_\AnotherSt > 0$
        as $\St'' \REquiv \AnotherSt''$ and
        $\St''(\Clock) = \St(\Clock)+\Delay_\St \ge 0$,
        implying that
        $\St'(\Clock) \neq \St(\Clock)+\Delay_\St$
        and
        $\AnotherSt'(\Clock) \neq \AnotherSt(\Clock)+\Delay_\AnotherSt$.
        %
    \end{enumerate}

    \item
      Sub-case $\St'(\Clock) > 0$.

      Now also $\AnotherSt'(\Clock) > 0$ as $\St' \REquiv \AnotherSt'$.
      As $\St'(\Clock) > 0$,
      it must be that $\St'(\Clock) = \St(\Clock) + \Delay_\St$
      of the restriction imposed on $\EncTr$.
      By the construction of $\AnotherSt''$ and $\AnotherSt'$,
      $\AnotherSt'(\Clock) = \AnotherSt''(\Clock) = \AnotherSt(\Clock) + \Delay_\AnotherSt$.
    \end{enumerate}

  \item
    Case: the atom is of form $\Clock \AnyIneq \Const$.

    Because $\St \REquiv \AnotherSt$,
    the atom evaluates to true under $\St$
    if and only if
    it does under $\AnotherSt$.

  \item
    Case: the atom is of form $\Clock + \delta \AnyIneq \Const$.

    By the construction of $\St''$ and $\AnotherSt''$,
    and the fact that $\St'' \REquiv \AnotherSt''$,
    we have that
    $\St''(\Clock) = \St(\Clock) + \Delay_\St \AnyIneq \Const$ if and only if
    $\AnotherSt''(\Clock) = \AnotherSt(\Clock) + \Delay_\AnotherSt \AnyIneq \Const$.

  \item
    Case: the atom is of form $\Delay \AnyIneq 0$.

    Because
    $\Delay_\St \ge 0$,
    $\Delay_\AnotherSt \ge 0$, and
    $\Delay_\St = 0 \Iff \Delay_\AnotherSt = 0$,
    the atom evaluates to true under $\Set{\Delay \mapsto \Delay_\St}$
    if and only if
    it does under $\Set{\Delay \mapsto \Delay_\AnotherSt}$.

  \end{itemize}%
\end{IEEEproof}

\fi

\end{document}